\documentclass[11pt]{article}
\usepackage{jheppub}
\usepackage[T1]{fontenc}
\usepackage[latin9]{inputenc}
\usepackage{xcolor}
\usepackage{amsmath}
\usepackage{amssymb}
\usepackage{graphicx}
\usepackage{float}
\usepackage{graphicx}
\usepackage{graphbox}
\usepackage{tikz}
\usepackage{tikz-cd} 
\usetikzlibrary{shapes.geometric, arrows}
\usepackage{mathtools}
\usepackage{verbatim}
\usepackage{dsfont}
\usepackage{empheq}
\usepackage{cancel}
\usepackage{stackengine}
\usepackage{mathrsfs}
\usepackage{comment}
\usepackage{pifont}
\usepackage{multirow}

\newcommand{\cmark}{\ding{51}}%

\renewcommand{\sec}[1]{sec.~\ref{#1}}

\renewcommand{\d}{\text{d}}

\newcommand{\nn}{\nonumber}
\newcommand{\be}{\begin{equation}}
\newcommand{\ee}{\end{equation}}
\newcommand{\ba}{\begin{eqnarray}}
\newcommand{\ea}{\end{eqnarray}}

\newcommand{\la}{\langle}
\newcommand{\ra}{\rangle}
\newcommand{\braket}[2]{\big\langle{#1}|{#2}\big\rangle}
\newcommand{\bra}[1]{\big\langle{#1}|}
\newcommand{\ket}[1]{|{#1}\big\rangle}

\newcommand{\Space}{\Cbb^n\setminus \{\ell_\perp^2=0\} \cup \{D=0\}}
\newcommand{\Spacedual}{\Cbb^n\setminus \{\ell_\perp^2=0\} }
\renewcommand{\div}{\{D=0\}}

\newcommand{\dual}{\vee}

\newcommand{\xbig}{{x\text{-big}}}
\newcommand{\xsmall}{{x\text{-small}}}

\newcommand{\trfive}{{\text{tr}_5}}

\newcommand{\dint}{d_\text{int}}

\newcommand{\F}{\mathcal{F}}
\renewcommand{\L}{\mathcal{L}}
\renewcommand{\O}{\mathcal{O}}
\newcommand{\B}{\mathcal{B}}

\newcommand{\C}{\mathcal{C}}
\newcommand{\A}{\mathcal{A}}

\newcommand{\bk}{{\rm bk}}
\newcommand{\bd}{{\rm bd}}

\newcommand{\CP}{\mathbb{CP}}

\newcommand{\Cbb}{\mathbb{C}}

\newcommand{\im}{{\rm Im}}
\newcommand{\Li}{{\rm Li}}

\newcommand{\res}{{\rm Res}}

\renewcommand{\ker}{{\rm ker}}

\newcommand{\alg}{{\rm alg}}
\newcommand{\dr}{{\rm dR}}
\newcommand{\drc}{{\rm dR,c}}

\newcommand{\pent}{{\rm pent}}

\newcommand{\vphi}{\varphi}

\newcommand{\vep}{\varepsilon}

\newcommand{\eps}{\varepsilon}

\renewcommand{\th}{\theta}

\newcommand{\disc}{{\rm Disc}}

\newcommand{\ab}[1]{\langle#1\rangle}
\renewcommand{\sb}[1]{[#1]}
\newcommand{\keta}[1]{|#1\rangle}
\newcommand{\kets}[1]{|#1]}
\newcommand{\braa}[1]{\langle#1|}
\newcommand{\bras}[1]{[#1|}
\newcommand{\pslash}{p\!\!\!/}

\newcommand{\bs}[1]{\boldsymbol{#1}} 
\newcommand{\mat}[1]{\underline{\boldsymbol{#1}}}

\newcommand{\ipaa}[2]{\left\langle #1 \vert #2 \right\rangle}

\newcommand{\hooklongrightarrow}{\lhook\joinrel\longrightarrow}

\title{Duals of Feynman Integrals, 2: Generalized Unitarity}
\author[a]{Simon Caron-Huot}
\author[a]{Andrzej Pokraka}
\affiliation[a]{
	Department of Physics, McGill University, 
	3600 Rue University, 
	Montr\'eal, Canada
}
\emailAdd{schuot@physics.mcgill.ca}
\emailAdd{andrzej.pokraka@mail.mcgill.ca}

\abstract{
The first paper of this series introduced objects (elements of twisted relative cohomology) that are Poincar\'e dual to Feynman integrals. 
We show how to use the pairing between these spaces -- an algebraic invariant called the intersection number -- to express a scattering
amplitude over a minimal basis of integrals, bypassing the generation of integration-by-parts identities.
The initial information is the integrand on cuts of various topologies, computable as products of on-shell trees,
providing a systematic approach to generalized unitarity.  We give two algorithms for computing the multi-variate intersection number.
As a first example, we compute 4- and 5-point gluon amplitudes in generic spacetime dimension. 
We also examine the 4-dimensional limit of our formalism and provide prescriptions for extracting rational terms.
}

\begin{document}

\maketitle

\section{Introduction}

Generalized unitarity \cite{Bern:1994cg, Bern:1995db, Bern:2004ky, Britto:2004nc, Buchbinder:2005wp, Anastasiou:2006jv, Britto:2006fc, 
Britto:2007tt, Forde:2007mi, Badger:2008cm, Giele:2008ve,  
Britto:2008vq, Cachazo:2008vp, 
Bern:2010qa, Bourjaily:2017wjl, Bourjaily:2019iqr, Feng:2021spv} is a powerful technique for constructing loop-level scattering amplitudes.
It is a generic procedure that applies to scattering amplitudes in any theory (planar or not) and has played an important role in obtaining state of the art predictions relevant for LHC phenomenology (see \cite{
Abreu:2017xsl, Chicherin:2018mue, Jin:2019nya} for a very non-exhaustive list 
of recent work); for example,
5-point amplitudes at NNLO accuracy in both QCD and  $\mathcal{N}=4$ sYM have been recently obtained \cite{
Abreu:2018aqd, Chicherin:2018yne, Badger:2019djh, Abreu:2020cwb, Kallweit:2020gcp, Abreu:2021oya}.

Unitarity has a long history in quantum field theory beginning with the optical theorem and Cutkosky rules \cite{Cutkosky:1960sp}. Unlike unitarity cuts, generalized unitarity places multiple internal lines on shell, which subdivides an amplitude into more than two pieces. 
The amplitude can be constructed as a linear combination of a known basis of Feynman integrals
with coefficients that are rational functions in the kinematic data (momentum and polarizations),
determined by comparing the generalized unitarity cuts of the amplitude with those of a suitable ansatz.

At one stage or another,  Feynman integrals are typically reduced to a minimal basis, called master integrals,
using integration-by-parts (IBP) identities. IBP identities are linear relations among Feynman integrals that come from the fact that  total derivatives integrate to zero in dimensional regularization \cite{Chetyrkin:1981qh, Tkachov:1981wb}.
While these strategies have been extremely successful \cite{Laporta:2001dd, vonManteuffel:2012np, vonManteuffel:2014ixa, Maierhoefer:2017hyi, Smirnov:2019qkx}, one is left with little understanding for when or how cancellations occur.  This motivates exploring other strategies.

The end result of projecting an amplitude onto master integrals takes the generic form
\be
 \mathcal{A}^{(L)}(\{p_k\},\eps)  = \sum_i c_i(\{p_k\},\eps) F_i(\{p_k\},\eps)\,, \label{generalized unitarity}
\ee
where the set of master integrals $F_i$ is finite for any process \cite{Smirnov:2010hn, Lee:2013hzt, Bitoun:2018afx}.
In general, the master integrals are complicated transcendental functions which depend on the process but not on the theory.  On the other hand, the coefficients $c_i$ depend on the theory but are algebraic functions of the kinematic data (i.e., may contain square root normalization factors but not more complicated branch cut). 
Our focus here will be on extracting the coefficients $c_i$ of master integrals, starting from a given representation of the loop integrand, or more simply, of its cuts.

Recently, an option for extracting the $c_i$ that bypasses the generation of IBP identities has been suggested \cite{Mastrolia:2018uzb, Frellesvig:2019kgj, Frellesvig:2019uqt,  Mizera:2019vvs, Mizera:2020wdt, Frellesvig:2020qot}. 
Working with Feynman integrals modulo IBP identities means that we are actually interested in the cohomology class of a Feynman integrand. 
These classes come with a nondegenerate pairing called the intersection number, which pair integrands with suitably defined dual forms.
Then, a Feynman integral can be projected onto a chosen basis via the intersection number
\begin{align}
	c_i = \la f^\vee_i \vert \mathcal{I} \ra
\end{align}
where $f^\vee_i$ is dual to the integrand of $F_i$ and $\mathcal{I}$ is the integrand for the process at hand.
Thus, provided that the dual space is known, the coefficients $c_i$ can be extracted.

In \cite{Caron-Huot:2021xqj}, we identified the space of dual forms as being a certain \emph{relative} twisted cohomology group.
Unlike normal loop integrands, dual forms are supported on a subset of generalized unitarity cuts and contain $\delta$-functions that force some propagators on-shell.
All other factors are essentially polynomials. That is, any propagator is either cut or absent.
For maximal cuts, the pairing becomes a series of residues as in generalized unitarity, yet, no information is lost even for non-maximal cuts. 
We then discussed the differential equations which simultaneously characterize dual forms and Feynman integrals.
This provides a mathematical basis for unitarity methods, and an alternative explanation for why scattering amplitudes $\mathcal{A}$ are determined by on-shell subprocesses.  

The (co)homological study of Feynman integrals dates back to the 1960's (a nice selection of reprints can be found in \cite{Hwa:102287}). 
However, missing the tools of dimensional regularization \cite{THOOFT1972189} progress was quickly hampered. 
Recently, there has been a resurgence of interest in (co)homological studies of Feynman integrals. 
The appearance of interesting transcendental numbers (multiple zeta values and generalizations)
established geometric and number theoretic connections to Feynman integrals (see  \cite{Broadhurst:1985vq, Broadhurst:1996ye, Broadhurst:1996yd, Bloch:2005bh, Brown:2009rc, Brown:2009ta, Bonisch:2021yfw, Broedel:2021zij} for some examples) where these ``interesting'' numbers arise as periods associated to the
geometry of a given Feynman integral.  
More recently, \cite{Mastrolia:2018uzb, Frellesvig:2019kgj, Frellesvig:2019uqt,  Mizera:2019vvs, Mizera:2020wdt, Frellesvig:2020qot} have shown how to make sense of the geometry (or (co)homology) of dimensionally regulated integrals. 

In order to exploit the factorization of cuts into products of on-shell trees, we find it compelling to work directly in the physical momentum space,
rather than in auxiliary spaces such as Feynman parameters.
Since many mathematical theorems only apply to ``generic'' situations where propagators are raised to non-integer exponents, earlier works along this line \cite{Mastrolia:2018uzb, Frellesvig:2019kgj, Frellesvig:2019uqt, Mizera:2019vvs, Mizera:2020wdt, Frellesvig:2020qot} added infinitesimal exponents to all propagators, which promotes poles to branch cuts but unfortunately seems to obscure the connection with generalized unitarity.
The main result of \cite{Caron-Huot:2021xqj} is that the physical case (propagators with integer powers) is intrinsically simpler, as noted above:
propagators become geometric boundaries (hence ``relative'' cohomology) and never appear in denominators.

The present paper focuses on the extraction of integral coefficients, $c_i$, via the intersection with dual forms. 
The key idea is to start with a canonical algebraic (meromorphic) representative of a dual form and apply
a compact support isomorphism: the c-map, to bring it to a compactly supported version, which can be used in an integral.
This allows to compute intersection integrals in a sequence of algebraic steps.  
The c-map is unique modulo exact forms, and different choices
can produce various equivalent formulas for the same coefficient.

Section \ref{sec:dual form review} reviews the main features of dual forms. 
Section \ref{sec:intersection intro} provides an example driven introduction to intersection theory as well as the compactly supported versions of dual pentagons, boxes and triangles. 
The c-map for $(p>2)$-dual forms is presented in section \ref{sec:higher form c-map}.   
In section \ref{sec:cbub4} we provide a detailed example of the formalism of \ref{sec:higher form c-map} by computing the compactly supported bubble dual form at 4-points. 
As an application, we compute the  one-loop 4- and 5-point gluon helicity amplitudes starting from cut representations of the integrands in section \ref{sec:gluon amplitudes}. 
We also examine the 4-dimensional limit of our formalism and provide an algorithm to extract the rational terms of 4-point amplitudes (higher multiplicity is left for future work). 
We conclude in section \ref{sec:conclusion}.

\section{Review of dual forms \label{sec:dual form review}}

In this section, we review the basics of one-loop dual forms. The first step is to define the differential form $\d^d\ell$ when $d$ is continuous. Then, we explain how Feynman integrals are naturally elements of a twisted cohomology group and elucidate the associated dual-cohomology group. 

\subsection{Feynman integrals and twisted cohomology}

To define the measure $\d^d\ell$ for continuous $d$, we set $d=\dint{-}2\vep$ where $\dint \in \mathbb{N}$ is a non-negative integer, and split
split momenta into a physical (integer dimensional) subspace and an extra $(-2\vep)$-dimensional subspace. All external momentum $p_i$ live (by definition)
in the physical $\dint$-dimensional subspace, while the loop momenta $\ell_i$ can have a non-trivial $\ell_\perp$ component
\begin{align}
	p^\mu_i = (p_i^0, p_i^1, \dots, p_i^{\dint}, 0_\perp), 
	\qquad 
	\ell^\mu = (\ell^0, \ell^1, \dots, \ell^{\dint},\ell_{\perp}).
\end{align}
The measure then follows from the product rule for dimensional regularization \cite{THOOFT1972189}, and from the fact that
Feynman integrals depend only on the combination $\ell_\perp^2$:
\begin{align} \label{eq:one-loop measure}
  \boxed{	
\frac{d^d\ell}{\pi^{d/2}}
		= \C_{\dint} u \frac{\d^{\dint}\ell \wedge d(\ell_\perp^2)}{\ell_\perp^2},
	\qquad 
	u = (\ell_\perp^2)^{-\vep},
	\qquad 
	\C_{\dint} =  \frac{1}{\pi^{\dint/2}\ \Gamma(-\vep)}.
	}
\end{align}
Note that, viewed as a function of complex momenta,
the measure is multivalued: the ``twist'' $u$ has a non-integer exponent whenever $\vep\neq0$. 
We thus treat a one-loop integrand in dimensional regularization as a $(n=\dint{+}1)$-form.

The generalization to $L$ loops includes $L(L+1)/2$ dot products $\ell_{i\perp}{\cdot}\ell_{j\perp}$ \cite{Caron-Huot:2021xqj}.
The standard integration contour for a Feynman integral is the $\mathbb{R}^{n}$ subspace consisting of real Minkowski momenta (with the usual Feynman $i0$ prescription), times the region over which the Gram matrix $\ell_{i\perp}{\cdot}\ell_{j\perp}$ is positive definite.
However, this contour plays no role in the integral reduction problem that is the focus of this paper:
the intersection pairing to be defined shortly involves a $(2n)$-dimensional integral over all of $\mathbb{C}^n$.

To more easily keep track of the multivaluedness of a Feynman form, it is convenient to factor out the twist $u$ and work only with single-valued forms $\vphi$:   
\be
 \d^{4-2\eps}\ell\ f(\ell,p) \equiv u\ \vphi.
\ee
We can completely ``forget'' about the multivaluedness by further introducing a covariant derivative
\be
	\d( u\ \vphi ) = u \nabla_{\omega} \vphi 
\ee
where $\nabla_\omega = \d +\omega \wedge$ with $\omega = \d\log(u)=-\vep \d\log(\ell_\perp^2)$. The connection is
curvature-free and keeps track of the multi-valuedness of the original integrand, somewhat analogously to the gauge potential in the Aharonov-Bohm effect.

One property of the physical integration contour will be important to us: total derivatives integrate to zero.
This means that Feynman integrands are only defined modulo total derivatives or equivalently up to integration-by-parts identities: $u \vphi \simeq u \vphi + \d(u \psi)$ where $\psi$ is an $(n{-}1)$-form if $\vphi$ is an $n$-form.  
Note that we could equivalently shift the single-valued form $\vphi$ by a covariant derivative: $\vphi\simeq\vphi+\nabla_{\omega}\psi$. This allows us to talk about integration by parts in a single-valued framework.

The modding out by total (covariant) derivatives is precisely the idea behind de Rham cohomology groups, whose elements
are equivalence classes of \emph{closed} differential forms (those with vanishing total derivative) modulo \emph{exact} differential forms (those that are a total (covariant) derivative). Intuitively, the integral of a closed form is unchanged under small contour deformations, while exact forms are integration-by-parts identities. 
Thus, Feynman integrands are elements of twisted de Rham cohomology groups: 
\begin{align}
	\vphi \in H^n_\dr(X,\nabla_\omega) 
	\equiv \frac{
		\ker \nabla_\omega: \Omega^n_\dr(X) \to \Omega^{n+1}_\dr(X)
	}{
		\im \nabla_\omega: \Omega^{n-1}_\dr(X) \to \Omega^{n}_\dr(X)
	}
\end{align}
where $\Omega^p_\dr(X)$ is the set of all smooth $p$-forms on the manifold $X$. For Feynman integrands, the manifold is simply complex space with all singularities excised: $X = \Space$. Points on the zero locus of the twist $\{\ell_\perp^2=0\}$ and infinity $\{\ell_\perp^2=\infty\}$ are called \emph{twisted singularities} or \emph{twisted boundaries} \cite{Mastrolia:2018uzb, Frellesvig:2019kgj, Frellesvig:2019uqt, Mizera:2019vvs, Mizera:2020wdt, Frellesvig:2020qot, Caron-Huot:2021xqj}.
These singularities are regulated by $\vep$ in the sense that the covariant derivative is locally invertible. On the other hand, \emph{unregulated singularities} or \emph{poles} appear whenever a propagator vanishes $\{D=0\}$ \cite{Matsumoto:2018aa, Mizera:2020wdt,Caron-Huot:2021xqj}. Near a pole the covariant derivative can only be inverted in the absence of a residue, and is only invertible up to an integration constant. 

\subsection{The intersection pairing and dual forms}

We review the intersection with dual forms following \cite{Caron-Huot:2021xqj}.
By Poincar\'e duality in (2n)-dimensional space, $p$-forms are dual to $(2n-p)$-forms where we obtain a number by integrating their wedge product over the full $n$-dimensional complex space. For this product to make sense, the pairing must be single-valued. Thus, the dual forms come with the opposite twisting: $u^\vee = u^{-1}$. 
The intersection pairing is then \cite{aomoto2011theory, yoshida2013hypergeometric}:
\be
 	\braket{\vphi^\dual}{\vphi} 
 	\equiv \frac{(-1)^{ \frac{p(p-1)}{2}} }{(2\pi i)^n} \int_X \vphi^\dual_c \wedge \vphi
	= \frac{1}{(2\pi i)^n} \int_X (\vphi^\dual_c)^T \wedge \vphi . 
 \label{pairing}
\ee
where $T$ denotes the transpose of the wedge product.\footnote{This way, if we integrate one variable at a time, anti-holomorphic and holomorphic pairs of differentials are always adjacent: $\int d\bar{z}_n \wedge \cdots \wedge d\bar{z}_1 \wedge dz_1 \wedge \cdots \wedge dz_n \to \int d\bar{z}_n \wedge \cdots \wedge d\bar{z}_2 \wedge dz_2 \wedge \cdots \wedge dz_n \to \cdots \to \int d\bar{z}_n \wedge dz_n$.}
Since $X$ is non-compact, $\vphi^\dual_c$ must have compact support so that the integral is well-defined for any representative
$\vphi$ ({\it ie.} converge near poles and branch points of the Feynman integral).\footnote{All that is required is that the \emph{product} $\vphi^\vee\wedge\vphi$ has compact support. Thus, one could instead try to compactify $\vphi$. However, one cannot make $\vphi$ with compact support about the propagator poles. This is one reason why previous works deformed the Feynman integrals by raising all propagators to some non-integer exponent.}

Demanding that \eqref{pairing} is invariant under changes of representative $\vphi$ implies that $\vphi^\dual$ is closed,
and since $\vphi$ is closed it is easy to see that $\vphi^\dual$ is defined modulo exact forms.
Thus the duals of Feynman integrals are also elements of a cohomology. 
However, it is a different cohomology of \emph{compactly supported} forms with the opposite twist:
\begin{empheq}[box=\fbox]{align*}
		\mbox{dual forms } 
		&\in H^n_\drc(\Space;\nabla_{-\omega}) \quad &\mbox{(compact)}.
\end{empheq}
While smooth compact forms make it easy to see that the above pairing is well-defined and non-degenerate, they are cumbersome to work with.
The main idea of \cite{Caron-Huot:2021xqj} is to populate the space using algebraic representatives.
Intuitively, we can account for the compact support property by adding small boundaries around the singularities.  Integration by parts on a manifold with boundaries produces surface terms,
and relative cohomology offers a simple bookkeeping scheme to track those:
\emph{relative} twisted de Rham cohomology is isomorphic to \emph{compactly supported} twisted de Rham cohomology.
As all our boundaries satisfy simple algebraic equations,
the smooth relative space is further isomorphic to an algebraic version:
\begin{empheq}[box=\fbox]{align*}
		\mbox{dual integrands } 
		& \in H^n_\text{alg} (\Spacedual,\div;\nabla_{-\omega})
			&\mbox{(alg./holo. forms)},\\
		&\simeq H^n_\dr(\Spacedual,\div;\nabla_{-\omega}) \quad 
			&\mbox{(smooth relative)},\\
		&\simeq H^n_\drc(\Space;\nabla_{-\omega}) \quad &\mbox{(compact)}.
\end{empheq}
The first space is where dual forms can be most simply written down; in \cite{Caron-Huot:2021xqj} the differential equations
satisfied by dual forms were verified to match with the (transpose of the) ones for integrals, thus verifying the isomorphism. The third space is where the intersection \refeq{pairing} is calculated in practice.
Describing the isomorphisms will be the main topic of this paper.

Elements of relative cohomology are simply formal sums
where each term has support on one of the boundaries, which we denote as:
\be
 	\th\ \psi
 	+ \delta_{1}(\th\ \psi_1)
	+ \delta_{2}(\th\ \psi_2)
 	+ \delta_{12} (\th\ \psi_{12}) 
	+ \cdots 
	\label{delta notation}
\ee
where the first term is a bulk form while the remaining terms are supported on boundaries indicated by the symbol $\delta_I$. Here $I$ is a set indexing a given boundary: $\delta_1$ is the $D_1=0$ boundary, $\delta_{12}$ is the $D_1=D_2=0$ boundary and so on. It is also important to note that the $\delta_I$ are totally antisymmetric in the index $I$: $\delta_{ij} = -\delta_{ji}$. Furthermore, each form is multiplied by the symbol $\th$ which reminds us to keep track of boundary terms when taking a derivative 
\be \label{eq:boundary terms}
	\nabla \delta_I (\theta\ \psi_I) 
	= (-1)^{|I|} \delta_I \left(\theta\ \nabla\vert_I\ \psi_I + \sum_{j \notin I} \delta_j \left( \theta\ \psi_I\big|_{j} \right) \right).
\ee 
The rules are easy to remember if one views $\theta$ as a literal product of step functions, each of which
vanish in the neighbourhood of one boundary (see above eq.~\eqref{eq:dtheta norm} below): $\theta=\prod_i \theta_i$.
Then, the notation naturally suggests:
\begin{align} \label{eq: dtheta nabla delta}
	\d \theta  =  \sum_{j} \delta_j\ \theta, 
	\qquad 
	\nabla \delta_{I}(\bullet) = (-1)^{|I|} \delta_I(\nabla\vert_{I}\ \bullet).
\end{align}
In the absence of twist, the isomorphism between smooth relative and smooth compact cohomologies would essentially replace the combinatorial symbols $\th$ and $\delta$ by literal Heaviside and delta functions \cite{Caron-Huot:2021xqj}, which satisfy equivalent rules.

Since generic smooth forms are cumbersome to manipulate, we work primarily with elements of algebraic cohomology, which is naturally embedded as a subgroup. 
To compute intersection numbers we thus use a series of isomorphisms\footnote{In a situation where the definitions of relative algebraic and relative de Rham cohomology would become ambiguous, for example if twisted and untwisted boundaries intersect,
the isomorphism with $H_{\rm dR,c}$ should be taken as the primary definition of the dual space.}
\begin{equation*}
	H^p_\alg  
	\hooklongrightarrow
	H^p_\dr
	\overset{c}{\longrightarrow}       
	H^p_\drc                   \,.
\end{equation*}
Algebraic forms are, by definition, holomorphic and the $c$-map
produces representatives in the compactly supported cohomology whose anti-holomorphic dependence enters \emph{exclusively} in a very simple manner: through $\delta$-functions supported on products of small circles. Thus, intersection numbers are computed algebraically via residues!

\subsection{A basis of one-loop dual forms}

The geometry of one-loop integrals is particularly simple due to the fact that the cohomology of the bulk space is trivial and all non-trivial elements are supported on at least one boundary (for multiloop integrals, at least one propagator must be cut in \emph{every} loop). Cutting a propagator sets $\ell_\perp^2$ to some quadric in the physical loop momentum variables while linearizing all other propagators. This means that the twist locus
is a sphere while the propagator/boundaries are hyperplanes. Moreover, further cuts do not change the geometry because the intersection of a sphere and a hyperplane is just a lower-dimensional sphere. 

Near 4-dimensions ($\dint=4$), one-loop
dual forms with six or more cut legs vanish since the hyperplanes have vanishing intersection;  this reflects the well known reducibility of integrals with six or more legs.
The basis thus consists of a pentagon, boxes, triangles, bubbles and tadpoles.
A uniform transcendental (UT) dual-basis (for $\dint=4$) is \cite{Caron-Huot:2021xqj}
\begin{eqnarray}
	\label{eq:tadpoles}
	&\text{Tadpoles:} \quad
	&\vphi^\vee_{a} = c_3\, \delta_{a} \left( \th\ \frac{ r_{a}\ \d^4 k_{a} }{ (r_{a}^2 + k^2_{a})^2 } \right),
\\
	\label{eq:bubbles}
	&\text{Bubbles:} \quad
	&\vphi^\vee_{ab} = c_{2}\,\ \delta_{ab} \left( \th\ \frac{ r_{ab}\ \d^3 k_{ab} }{ (r_{ab}^2 + k^2_{ab})^2 } \right),
\\
	\label{eq:triangles}\
	&\text{Triangles:} \quad
	&\vphi^\vee_{abc} = c_1\, \delta_{abc} \left( \th\ \frac{ r_{abc}\ \d^2 k_{abc} }{ r_{abc}^2 + k^2_{abc} } \right),
\\
	\label{eq:boxes}
	&\text{Boxes:} \quad
	&\vphi^\vee_{abcd} = c_0\, \delta_{abcd} \left( \th\ \frac{ r_{abcd}\ \d^1 k_{abcd} }{ r_{abcd}^2 + k^2_{abcd} } \right),
\\
	\label{eq:pentagon}
	&\text{Pentagon:} \quad
	&\vphi^\vee_{abcde} = c_{-1}\, \delta_{abcde} \left( \th\ r_{abcde}\ \d^0 k_{abcde} \right) 
\,.
\end{eqnarray}
where the radii $r_\bullet$ are ratios of kinematic determinants and the normalizations $c_i$ are such that the associated differential equation is pure.\footnote{Here, the $c_l$ are given by $c_{l} = -16(2\vep)^{\lfloor\frac{l+1}{2}\rfloor-2}\ (-\vep)_{\lfloor\frac{l}{2}\rfloor+1}$. Note that this differs from \cite{Caron-Huot:2021xqj} by an overall $-4/\vep^2$ since we normalize the intersection matrix to identity here.} 
Here, the $k_I^\mu$ is the loop momentum in the directions transverse to the cut $I$.
Note that the integration measure
$\d^0k_{abcde} = \d^{4}\ell\wedge\d\ell^2_\perp\vert_{abcde}/(\d D_a \wedge \cdots \wedge \d D_e)$ is a sign factor that preserves the orientation of the original measure and keeps the dual pentagon independent of the order in which the cuts are taken. Similarly, the anti-symmetry in the $\d^\bullet k_I$ compensates for the anti-symmetry of the $\delta_I$ so that the basis forms are independent of the order in which cuts are taken. 

For example, for a planar 5-point amplitude with generic masses, the basis contains a unique pentagon, 5 boxes, 10 triangles, 10 bubbles and 5 tadpoles for a total basis size of 31. Taking various kinematic degenerations can cause the cohomology on various cuts to become trivial decreasing the size of the basis. In the total massless degeneration, all tadpole- and triangle-cut cohomologies as well as half of the bubble-cut cohomologies become trivial decreasing the basis size to 11. 

Each basis dual form has singularities on the zero locus of a (cut dependent) sphere. In the case of the tadpole and bubble, the basis forms have double poles at the zero locus of the sphere. These higher order poles act as dimension shifts so that the tadpole and bubble dual forms extract the coefficient of the 2-dimensional tadpole and bubble coefficients. The dimension shifted tadpoles and bubbles are naturally pure functions in 2-dimensions. 

The purity of the integrated expressions can be seen from the $\vep$-form of the dual differential equation \cite{Caron-Huot:2021xqj}
\begin{align}
	\nabla^\vee_{\text{kin}} \vphi^\vee_J 
	\simeq \Omega_{JJ^\prime}^\vee \wedge \vphi^\vee_{J^\prime}. 
\end{align}
Here, $J$ is a multi-index labeling the dual forms and $\Omega^\vee$ is the dual kinematic connection
\begin{align} \label{eq:oddOmega}
\left.\begin{array}{ll} \displaystyle
	\Omega_{J;J}^\vee 	&= \vep\ d\log r_J^2
	\\[0.1em]\displaystyle
	\Omega_{J;Ja}^\vee &=
		2\vep\ d\tanh^{-1}(i_J a_J)
	\\[0.1em]\displaystyle
	\Omega_{J;Jab}^\vee  &= \vep\ d\cos^{-1}(a_{J0}b_{J0})
\end{array}
\quad\right\}\quad	\text{for } \dint-|J| \text{ odd}\,,
\end{align}
and 
\begin{align} \label{eq:evenOmega}
\left.\begin{array}{ll} \displaystyle
	\Omega_{J;J}^\vee 
		&=\vep\ d\log r_J^2
	\\[0.1em]\displaystyle
	\Omega_{J;Ja}^\vee 
		&= \vep\ d\cos^{-1}(i_{J}a_{J})
	\\[0.1em]\displaystyle
	\Omega_{J;Jab}^\vee
	&=\vep\ d\cos^{-1}(a_{J}b_{J})
\end{array}
\quad\right\}\quad \text{for } \dint-|J| \text{ even}\,,	
\end{align}
where the brackets $(a_J b_J)$ are dot products of embedding space vectors (see \cite{Caron-Huot:2021xqj} for the details). We will always choose a basis of Feynman forms that is dual to the basis \eqref{eq:tadpoles}-\eqref{eq:pentagon}: $\la \vphi^\vee_J \vert \vphi_{J^\prime} \ra \propto \delta_{JJ^\prime}$. This ensures that the kinematic connection for the basis of Feynman forms is the minus transpose of the dual kinematic connection: $\Omega_{JJ^\prime} = -\Omega^\vee_{J^\prime J}$. For generic masses, the basis of Feynman forms is simply the standard UT basis of scalar one-loop integrals. However, for degenerate kinematics such as in the complete massless limit, the basis of Feynman forms changes discontinuously and, in this case, are reorganized into linear combinations of the old basis such that the box (pentagon) integrals are finite (vanishing) at $d=4$. 
In the following, we will use the standard $G$-notation to denote the (near 4-dimensional) Feynman integrals
\begin{align}
	G_{\nu_1,\dots,\nu_n} 
	= \C_4 \int  u
	\frac{
		\d\ell_\perp^2 \wedge \d^4\ell
	}{
		\ell_\perp^2 D_1^{\nu_1} \cdots D_n^{\nu_n}
	}
\end{align}
where $\C_4$ and $u$ are given by \eqref{eq:one-loop measure}. 

Below we review the massless 4- and 5-point degenerate limits from \cite{Caron-Huot:2021xqj}.
At 4-points, a standard basis of UT Feynman \emph{forms} is \cite{Henn:2014qga}
\begin{align}
	\vphi_{\text{bub}_{s}} = \vep (2\vep{-}1) 
	\frac{ \d\ell_\perp^2 \wedge \d^4\ell}{\ell_\perp^2 D_1 D_3},
	\quad
	\vphi_{\text{bub}_{t}} = \vep (2\vep{-}1)
	\frac{ \d\ell_\perp^2 \wedge \d^4\ell}{\ell_\perp^2 D_2 D_4},
	\quad
	\vphi_{\text{box}} = \vep^2 st 
	\frac{ \d\ell_\perp^2 \wedge \d^4\ell}{\ell_\perp^2 D_1 \cdots D_4}.
\end{align}
Integrating these forms yields the basis of UT Feynman \emph{integrals}
\begin{align}
	\mathscr{I}[\vphi_{\text{bub}_s}] = \vep (2\vep{-}1) G_{1,0,1,0},
	\quad 
	\mathscr{I}[\vphi_{\text{bub}_t}] = \vep (2\vep{-}1) G_{0,1,0,1},
	\quad 
	\mathscr{I}[\vphi_{\text{box}}] = \vep^2 st  G_{1,1,1,1},
\end{align}
where 
\begin{align}
	\mathscr{I}[\vphi] = \mathcal{C}_4 \int u\ \vphi.
\end{align}
While this basis is not orthonormal to our dual basis, the following linear combinations are:
\begin{align}
	\vphi_{13} = 2 \vphi_{\text{bub}_s},
	\qquad
	\vphi_{24} = 2 \vphi_{\text{bub}_t},
	\qquad
	\label{eq:4pt box}
	\vphi_{1234} =	\vphi_{\text{box}} + 2 \vphi_{\text{bub}_s} + 2 \vphi_{\text{bub}_t}. 
\end{align}
Consequently, they satisfy the following differential equation 
\begin{align} 
	\nabla_\text{kin} \bs{\vphi} \simeq (-\mat{\Omega}^\vee_{4\text{pt-massless}} )^T \wedge \bs{\vphi}
\end{align}
where $\bs{\vphi} = ( \vphi_{13}\ \vphi_{24}\ \vphi_{1234} )^T$ and $\simeq$ means equality modulo total (covariant) derivatives (IPBs). 
The corresponding kinematic connection 
\begin{align} \label{eq:4pt kinematic connection}
	(\mat{\Omega}^\vee_{4\text{pt-massless}})^T
	= \vep 
	\begin{pmatrix}
		d\log(s) & 0 & 0 \\
		0 & d\log(t) & 0 \\
	 	d\log(s/t) & d\log(t/s) & d\log(st/u)
	\end{pmatrix}
\end{align}
can be obtained from the degeneration of \eqref{eq:oddOmega} and \eqref{eq:evenOmega}. 
Of course, since the physical integration contour is closed\footnote{For Feynman integrals, the canonical contour $\mathbb{R}^4\times\mathbb{R}^+$ is closed since the twist $(\ell^2_\perp)^{-\vep}$ regulates singularities at the boundary $\ell_\perp^2=0,\infty$. Note that the integrals for generic $\vep$ are defined by analytic continuation from a region of convergence. Therefore, there is no boundary term at $\ell_\perp^2=0,\infty$ regardless of the sign of $\vep$. This contour is an element of the twisted homology group $H_5(\mathbb{C}^5\setminus\{\ell_\perp^2=0,\infty\}\cup\{D=0\},\mathcal{L}_\omega)$.}, integrals satisfy the same differential equation as the cohomology class of forms:
\begin{align} \label{eq:4pt integral DEQ}
	\d_\text{kin} \ \mathscr{I}[\vphi_i] 
	= \mathcal{C}_4 \int u\ \nabla_\text{kin} \vphi_i
	=  (-\mat{\Omega}^\vee)^T_{ij}\ \mathscr{I}[\vphi_j].
\end{align}
Thus the $\mathscr{I}[\vphi_i]$ can be obtained by integrating \eqref{eq:4pt integral DEQ} order by order in $\vep$ where the integration constants are fixed by requiring the absence of spurious singularities at $u=s_{13}=0$ \cite{Henn:2014qga}.
Explicitly, one  finds
\begin{align}
\label{eq:Ibub}
	\mathscr{I}[\vphi_{i,i+2}] 
	&= - 2 r_\Gamma(-s_{i,i+1})^{-\vep}
	\nn\\
	&= -2 + 2\vep \log(-s_{i,i+1}) 
	+ \vep^2 (\zeta_2{-}\log^2(-s_{i,i+1}))
	+ \O(\vep^3),
	\\
	\label{eq:Ibox massless}
	\mathscr{I}[\vphi_{1234}] & = - \vep^2 \left(\pi^2 + \log^2(t/s) \right) + \mathcal{O}(\vep^3),
\end{align}
where $r_\Gamma = e^{\vep\gamma_E}\Gamma^2(1-\vep)\Gamma(1+\vep)/\Gamma(1-2\vep)$ 
and $s_{i,i+1} = -(p_i + p_{i+1})^2$ are the Mandelstam invariants (Lorentz invariant scalar products of particle momenta).
For the 4-point integrals above, $s = s_{12}$, $t = s_{23}$ and $u=s_{13}$.

The basis of Feynman integrals picked out by the dual basis \eqref{eq:4pt box} is particularly nice since the box integral $\mathscr{I}[\vphi_{1234}]$ is finite. In fact, 
$\mathscr{I}[\vphi_{1234}] = \vep^2 st G_{1,1,1,1} + 2\vep(2\vep-1) (G_{1,0,1,0}+G_{0,1,0,1})$ 
is simply the 6-dimensional box integral written in terms of  4-dimensional integrals (up to an overall normalization).

For massless five point scattering, the basis of Feynman integrals consists of 5 bubbles \eqref{eq:Ibub}, 5 boxes with a massive corner and a single pentagon.
The boxes with one massive corner have a distinct expression from \eqref{eq:Ibox massless} and are defined by 
\begin{align} 
	\mathscr{I}[\vphi_{1234}]^{\rm 5-pt}	
		&= \vep^2 s_{12} s_{23} G_{1,1,1,1,0} 
		+ \mathscr{I}[\vphi_{13}]
		+ \mathscr{I}[\vphi_{24}] 
		- \mathscr{I}[\vphi_{14}]
	\label{eq:5ptFI boxes}
	\\ 
	&= -2\vep^2
	\bigg(
		\zeta_2 
		+ \frac12 \log^2\left(\frac{s_{12}}{s_{23}}\right)
		+ \text{Li}_2\left(1-\frac{s_{45}}{s_{12}}\right)
		+ \text{Li}_2\left(1-\frac{s_{45}}{s_{23}}\right)
	\bigg)
	+ \O(\vep^3).
	%
	\nn
\end{align}
and cyclic permutations thereof.
Equation \eqref{eq:5ptFI boxes} is simply the 6-dimensional 1-mass box integral written in terms of 4-dimensional integrals. It is finite at $d=4$ and should be contrasted with its divergent counterpart
\begin{align}
	G_{1,1,1,1,0}
	&= \frac{2r_\Gamma}{s_{12} s_{23}} \frac{(- s_{12})^{-\vep}(- s_{23})^{-\vep}}{(- s_{45})^{-\vep}}
	\bigg( 
		\frac{1}{\vep^2} 
		+ \text{Li}_{2}\left(1-\frac{s_{12}}{s_{45}}\right) 
		+ \text{Li}_{2}\left(1-\frac{ s_{23}}{s_{45}}\right) 
		- \zeta_2 +\O(\vep)
	\bigg).
\end{align}

The last element of the basis is the 6-dimensional pentagon, which is the combination of 4-dimensional pentagons and boxes that vanishes at $d=4$ \cite{vanNeerven:1983vr}
\begin{align} \label{eq:5ptFI pentagon}
	\mathscr{I}[\vphi_{12345}] &= \frac{2\vep^2}{\sqrt{-(012345)^2}} \bigg(
		2(12345)^2 G_{1,1,1,1,1}
	\nn\\&\qquad
	- (01234)\cdot(12345)\ G_{1,1,1,1,0}
	- (01235)\cdot(12354)\ G_{1,1,1,0,1}
	\nn\\&\qquad
	- (01245)\cdot(12343)\ G_{1,1,0,1,1}
	- (01345)\cdot(12342)\ G_{1,0,1,1,1}
	\nn\\&\qquad
	- (02345)\cdot(12341)\ G_{0,1,1,1,1}
	\bigg) = 0 + \O(\vep^3).
\end{align}
Here, the symbols $(\bullet)^2$ and $(\bullet)\cdot(\bullet)$ correspond to minors of a kinematic Gram matrix (see section 2.5 of \cite{Caron-Huot:2021xqj}). Since this pentagon integral vanishes like $\vep^3$, it does not contribute to the physical part of one-loop amplitudes and is only included for completeness. 

Equations \eqref{eq:Ibub}, \eqref{eq:Ibox massless} and \eqref{eq:5ptFI boxes} 
complete the list of Feynman integrals that will appear in the examples considered in the following sections.

\section{Introduction to intersection theory, and pentagons, boxes and triangles \label{sec:intersection intro}}

In this section we will illustrate how to use equation \eqref{pairing} and provide an explicit realization of the map from holomorphic (algebraic) forms to their compactly supported versions.

Section \ref{sec:distributions} introduces the required distributions for the compactifying procedure. 
We then work through examples on both compact and non-compact manifolds in sections \ref{sec:compact ex} and \ref{sec:non-compact ex}. 
The compact supported versions of the dual pentagon, box and triangle are constructed in sections \ref{sec:cpent}, \ref{sec:cbox} and \ref{sec:ctri}.

\subsection{The $\theta$ and $d\theta$ distributions \label{sec:distributions}}

\begin{figure}[t]
\centering
\includegraphics[scale=.4]{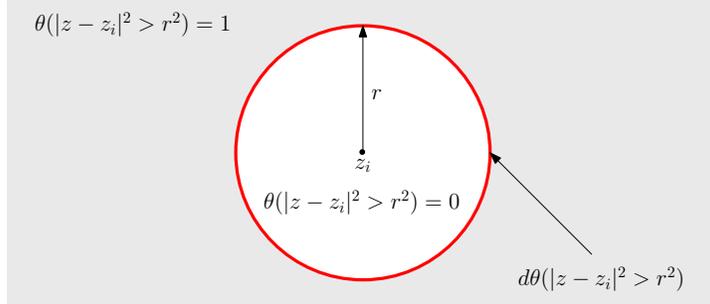}
\caption{
	\label{fig:dtheta}
 The regulating step function $\theta$: equal to unity outside a small disc, zero inside.
Its derivative $d\theta$ is essentially a $\delta$ function supported on the red contour and is dual to a residue contour
-- that is, the integral of $\d\theta(z-z_i)\wedge f(z) \d z$ takes the residue of $f$ at $z_i$.
}
\end{figure}

In practice, we will need the following two basic distributions:
\begin{itemize}
\item $\th(z)\equiv \theta(|z|{-}\eta)$: a 0-form that vanishes inside a disc of radius $\eta$ around $z=0$, and is unity outside.
\item $\d\th(z)$: a 1-form supported on a circle of radius $\eta$ around $z=0$.
\end{itemize}
The later satisfies the following useful property (see figure \ref{fig:dtheta}):
\be \label{eq:dtheta norm}
 \boxed{\frac{1}{2\pi i} \int \d\th(z) \wedge \frac{\d z}{z} = 1.} 
\ee
or more simply, $\langle \d\th(z),\frac{\d z}{z}\rangle=1$.
One could take $\theta$ to be a smooth function but in practice the Heaviside distribution works just as well.

Boundary-supported forms in relative cohomology, with the boundary labeled by the combinatorial symbol $\delta_I$, are mapped to
ambient forms in $H_{\dr,c}$ via the Leray coboundary
\be \label{eq:c-leray}
\boxed{
	\delta_{i_1 \cdots i_{|I|}}(\phi^\vee) = \frac{u}{u\vert_I}
	d\theta_{i_{1}} \wedge \cdots \wedge d\theta_{i_{|I|}} \wedge \phi^\vee\vert_I.
	}
\ee
The formal symbol $\theta$ becomes a (minimal) product of Heaviside step functions enforcing compact support.
The restriction $\vert_I$ is technically the pullback of a projection map from a product of small circles to the codimension-$|I|$ boundary.
The formal rules of the algebraic dual forms \eqref{eq: dtheta nabla delta} follow from the definition \eqref{eq:c-leray} and the chain rule. Using \eqref{eq:dtheta norm}, the intersection with a dual form supported on a cut simply becomes
\be \label{eq:action of delta in intersection numbers}
	\boxed{
	\bigg\la \delta_{1 \cdots p} ( \phi^\vee ) \bigg\vert \vphi \bigg\ra
	= \bigg\la \phi^\vee \bigg\vert \res_{1 \cdots p} \left(\frac{u}{u\vert_{1 \cdots p}} \vphi \right) \bigg\ra.
	}
\ee
The above equation demonstrates the power of relative cohomology: intersections localize to cut surfaces simplifying the calculation!

Equations \eqref{eq:dtheta norm} and \eqref{eq:action of delta in intersection numbers} form the basis of all calculations in this paper,
and we now proceed to explain the general theory by means of examples.

\subsection{Compact examples: Riemann sphere and elliptic curves \label{sec:compact ex}}


As a trivial example, consider the Riemann sphere $\CP^1$ (without any twisting).
The Betti numbers (dimensions of homology groups) are $\{b_0,b_1,b_2\}= \{1,0,1\}$: the nonzero
entries represent a point and the whole sphere.
Algebraic representatives could be defined using \v{C}ech cohomology; we omit these
and directly define algebraic-like objects that play the same role in the smooth compact world
(see appendix D of \cite{Caron-Huot:2021xqj} for its connection with the standard $H^2_{\rm alg}(\CP^1)$).

Algebraic-like generators of $H_\drc(\CP^1)$ are
\be
 g_0 = 1, \quad g_2 = \d\theta(z-a)\wedge\frac{\d z}{z-a} \equiv \delta^2(z-a) \label{sphere}
\ee
where $a$ is an arbitrary point on the sphere. It is obvious that these are respectively closed 0- and 2-forms: $d g_i=0$.
Geometrically, a two-form represents a density on the sphere, and from eq.~\eqref{eq:dtheta norm} the second form
can be regarded as a unit mass (times $2\pi i$) concentrated near the point $z=a$.
It is defined modulo total derivatives of smooth compact 1-form, and we restrict to algebraic like forms:
meromorphic one-form multiplied by $\theta$ or $d\theta$ to mask their singularities.
Since a meromorphic one-form on a sphere has at least two singular points (including infinity),
the class $g_2$ does not depend on the choice of point $a$:
\be
0\simeq \d\left( \frac{\th(z-a)\d z}{z-a} - \frac{\th(z-b)\d z}{z-b} \right) = \delta^2(z-a) -\delta^2(z-b). \label{holo anomaly}
\ee
This is effectively the familiar holomorphic anomaly equation, which technically corresponds to omitting
the $\th$ factors on the left-hand-side, which would replace the right-hand-side by an equivalent distribution.
In this compact example, the cohomology $H_{\rm c}^n(\CP^1)$ is its own dual,
and integration yields a non-degenerate intersection pairing between 0 and 2 forms:
\be
\braket{g_0}{g_2} = \braket{g_2}{g_0}\equiv \frac{1}{2\pi i}\int_{\CP^1} \delta^2(z-a)\wedge (1) = 1.
\ee

As a second compact example, consider an elliptic curve: $y^2=x(x-1)(x-a)$, topologically a torus.
It is well-known that $H^1$ is now two-dimensional (a torus has two basic 1-cycles),
and we again have one-dimensional $H^0$ and $H^2$ with representatives similar to eq.~(\ref{sphere}).
A standard basis of algebraic 1-forms spanning $H^1$ is:
\be
 g_{\alg,1} = \frac{\d x}{y}, \qquad g_{\alg,1}^{\prime} = \frac{x \d x}{y}.
\ee
While it is easy to see that both forms are naively closed (they depend only on holomorphic variables),
the second one is not a well-defined smooth-compact form due to its singularity near infinity.
However, it is readily patched up to define a well-defined closed form.
To see this, a good local coordinate there is $w=\frac{x}{y} \sim \frac{1}{\sqrt{x}}\to 0$, where
\be
 \lim_{w\to 0} g_{\alg,1} = -2 \d w, \qquad \lim_{w\to 0} g_{\alg,1}^{\prime}  = \frac{-2 \d w}{w^2}  + 0 \frac{\d w}{w} + \O(w^2 \d w).
\label{eq:naive singular limit}
\ee
The fact that the residue vanishes is significant: otherwise the second form couldn't be closed due to a holomorphic anomaly.
However, the form is \emph{still} not closed since $\d(\d w/w^2)\neq 0$ is a derivative of the holomorphic anomaly.
The following general procedure will be used throughout the paper to fix this.
We excise a small neighborhood of the singularity by slapping on a step function $\theta(w)$,
and then we add a $\d\theta$ term to ``patch it up'' and make the result closed:
\be
g_1^\prime \equiv \left[ g_{\alg,1}^\prime \right]_c \equiv \frac{x\ \d x}{y}\theta(w) + \psi(w) \d\theta(w)\ .
\ee
This is closed provided $\psi$ is a local 0-form primitive: $\d\psi = \frac{x}{y} \d x$.
Because $\d\theta(w)$ is concentrated on a small circle around $w=0$, in practice, $\psi$ need only be computed as a Laurent series in $w$.
From eq.~(\ref{eq:naive singular limit}),
\be
\d\psi = \frac{-2 \d w}{w^2} + 0\frac{\d w}{w} + \O(w^2 \d w)\quad\Rightarrow\quad \psi = \frac{2}{w} + C + \O(w^3)
\label{eq:1st c-map example}
\ee
where $C$ is an arbitrary integration constant, which has no effect on the cohomology class of $g_1^\prime$
(the form $\d\theta(w)$ is cohomologous to zero).
Note that a primitive $\psi$ only exists when the residue vanishes ($\log w$ would not be a valid single-valued primitive).
Thus we defined a valid de Rham class $g_1^\prime$ canonically associated to $x\ \d x/y$:
it ``looks'' like it everywhere except near $w=0$, where its singularity is regulated in a unique way.
The upshot of this method is that it is easy to compute intersections, because regularization has brought in an antiholomorphic differential $\d\bar{z}$ (through $\d\theta$):\footnote{
Upon writing the volume 2-cycle on the torus as the product of geometric $A$ and $B$ 1-cycles,
this essentially gives Legendre's famous period relation:
\be \int_{T^2=A\wedge B} g_1\wedge g_1^\prime
= \det\left(\begin{array}{cc} \int_A g_1 & \int_A g_1^\prime \\ \int_B g_1 & \int_B g_1^\prime\end{array}\right) = 4\times 2\pi i\ .
\ee}
\be
 \langle g_1|g_1^\prime\rangle = \frac{1}{2\pi i} \int_{T^2}  (-2 \d w) \wedge \left(\frac{2}{w} \d\theta(w)\right)  = 4.
\ee

There are two important lessons from these examples.  First, using the ``building block" distributions
$\theta$ and $\d\theta$ it is straightforward to write algebraic-like representatives of de Rham cohomology classes,
where all the ``antihomolorphic" dependence is hidden inside $\delta$-functions.
Second, thanks to these $\delta$-functions, intersections of such forms can always be computed algebraically (by residues),
even on topologically nontrivial spaces such as an elliptic curve.

\subsection{Non-compact examples: complex plane and degenerate ${}_2F_1$ \label{sec:non-compact ex}}

The simplest non-compact complex space is perhaps the complex plane $\Cbb=\CP^1\setminus\{\infty\}$.
Because of non-compactness, its de Rham cohomology $H^n_{\rm dR}(\Cbb)$ is not isomorphic to its dual $H^n_{\rm dR,c}(\Cbb)$.
Indeed, $H^2_{\rm dR}(\Cbb)=0$ (any two-form density can be pushed out to infinity),
while $H^0_{\rm dR,c}(\Cbb)=0$ (there exists no compactly-supported constant function).
However, the remaining $\CP^1$ generators in eq.~(\ref{sphere}) survive, and the 
the cohomology and its dual admit the following generators:
\be
 H^{0,1,2}_{\rm dR}(\Cbb) = \{1, \emptyset, \emptyset\},\qquad
 H^{0,1,2}_{\rm dR,c}(\Cbb) = \{\emptyset,\emptyset,\delta^2(z)\} .
\ee
Although these are now clearly non-isomorphic, the non-degenerate pairing between them has survived.
This is the most general form of Poincar\'e duality that we will use in this paper:
that integration gives a nondegenerate pairing between $H_{\rm dR,c}^{2n-k}$ and $H^{k}_{\rm dR}$.

We finally turn to a realistic model for Feynman integrals: a family of ${}_2F_1$ hypergeometric integral which contains
both poles and branch points (arguably the two essential features of a Feynman integral).
Consider the family, for (possibly negative) integers $m,n,p$:
\be\begin{aligned}
I_{m,n,p}(x) &\equiv \int_0^1 \d z\; \frac{z^{\alpha_0-m} (1-z)^{\alpha_1-n}}{(z-x)^p}
\\ &= B(1{+}\alpha_0{-}m,1{+}\alpha_1{-}n)\times (-x)^{-p} {_2F_1}(p,1{+}\alpha_0{-}m;2{+}\alpha_0{+}\alpha_1{-}m{-}n;x^{-1}).
\end{aligned}\label{2F1 family}\ee
The exponents $\alpha_i\in\mathbb{C}$ are assumed noninteger, and we will write this as $\int u\ \vphi$
where $u=z^{\alpha_0} (1-z)^{\alpha_1}$ denotes the multi-valed part; $x \in\mathbb C \backslash [0,1]$ and $B$ is the Euler beta function.

\begin{figure} \centering
\includegraphics[align=c,width=.46\textwidth,]{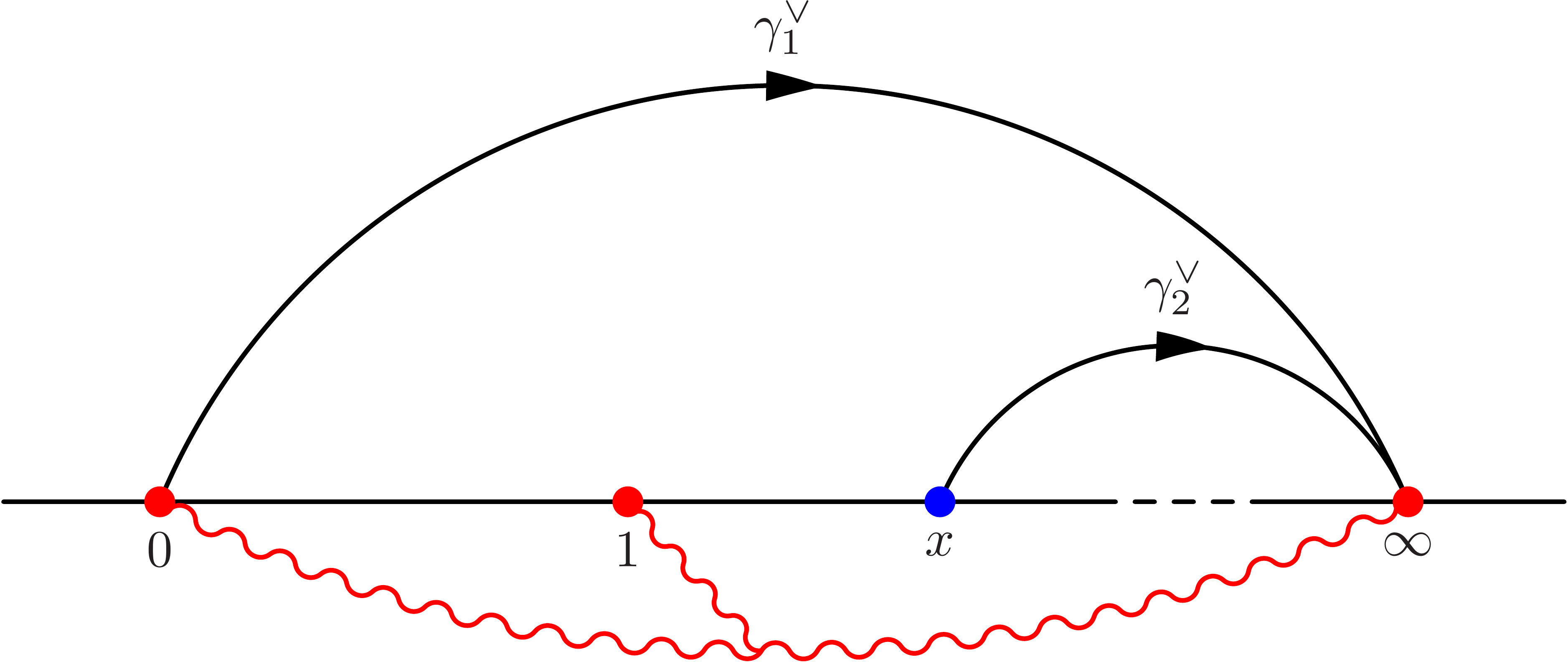}\qquad\includegraphics[align=c,width=.46\textwidth,]{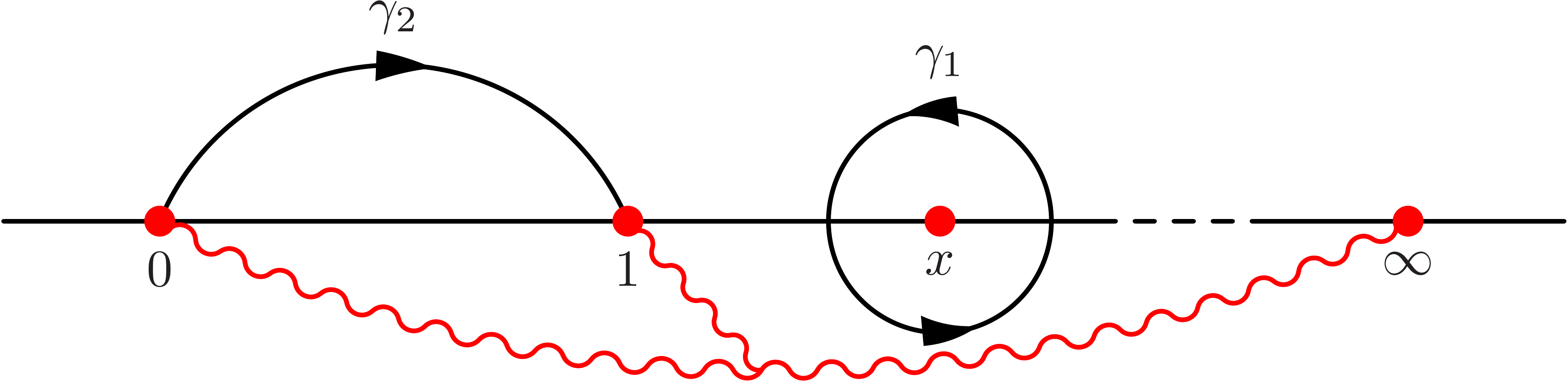}
\caption{\label{fig:2F1 cycles}	Homology cycles that illustrate the basis choices \eqref{2F1 basis} and \eqref{2F1 dual basis} respectively.
The figures show respectively element of $H_1(X^\vee,\{x\};\mathcal L_{-\omega})$ (left) and $H_1(X;\mathcal L_\omega)$ (right), for $X$ in \eqref{2F1 H1}.
Here the wavy red lines denote branch cuts.	Red dots are singular point and blue dots are boundaries.}
\end{figure}

For any closed contour, integration-by-parts in $z$ produces valid identities between these functions since surface terms vanish (due to the non-integer exponents at the boundaries 0, 1 and $\infty$)\footnote{In this example, closed contours  can have the endpoints $0,1$ and $\infty$. The endpoint $x$ is special and its treatment depends on which space we are considering: $X$ or $X^\vee$ (see fig.~\ref{fig:2F1 cycles}.). On $X$, the endpoint $x$ can only be encircled since there is no non-integer twisting at $x$ and the forms may be singular at $x$. On the other hand, $x$ is a valid endpoint on $X^\vee$ since the forms are non-singular there. However, taking $x$ as an endpoint generates boundary terms when integrating-by-parts. }. 
The space of independent integrals is thus the cohomology
\be
 H^1_{\rm dR}(X=\CP^1\setminus\{0,1,x,\infty\};\nabla_\omega) \label{2F1 H1}
\ee
where we introduced the covariant derivative $\nabla_\omega=d+\omega$ with connection
\be
 \omega= u^{-1}du = \alpha_0 \d\log z + \alpha_1 \d\log(1-z).
\ee

We will be interested in the problem of integral reduction:
what is a basis of the cohomology (\ref{2F1 H1}), and how do we express a given integral $I_{m,n,p}$ in this basis?
For hypergeometric functions with generic indices, this problem has been considered in many references \cite{aomoto1975,cho_matsumoto_1995,AOMOTO1997119,matsumoto1998,aomoto2011theory,aomoto2012,aomoto2015,Aomoto:2017npl};
in the generic setup, $\omega$ contains an extra term $\alpha_2 \d\log(z-x)$ which turns that pole into a branch point.
The cohomology is two-dimensional (reflecting the fact that ${}_2F_1$ satisfies a second-order differential equation)
and is dual to essentially the same space with connection $-\omega$.
Their non-degenerate pairing allows for efficient extraction of coefficients. 

We now detail the analogous story in the degenerate case (\ref{2F1 family}).
The key is to use the appropriate space of dual forms, which is now spanned by compactly supported forms that vanish near the boundary $z=x$:
\be
 H^1_{\rm dR}(X;\nabla_\omega)^\dual =  H^1_{\rm dR,c}(X;\nabla_{-\omega})\simeq H^1_{\rm alg}(X,\{x\},-\omega)\,.
\ee
For concreteness, we choose the following convenient representatives of $H^1_{\rm dR}$ \cite{Matsumoto:2018aa}:
\be
 	\ket{\vphi_1} = \d\log z=\frac{\d z}{z},
 	\qquad 
	\ket{\vphi_2} = \d\log(z-x) = \frac{\d z}{z-x}. 
	\label{2F1 basis}
\ee
It is easy to prove directly that any form in (\ref{2F1 H1}) can be reduced 
to this basis (establishing that the cohomology is at most two-dimensional),
using the following observations.\footnote{
There is no need to make either of these forms compact support, because their singularities are \emph{outside} the space $=\CP^1\setminus\{0,1,x,\infty\}$. For generic twist $\alpha_i$, one can also show that $H^0_{\rm dR}(X;\nabla_\omega)=0=H^2_{\rm dR}(X;\nabla_\omega)$.
}
First, forms with singularities stronger than a single pole at $0,1,z,\infty$ 
can be eliminated, by adding $\nabla_\omega \psi$ of a suitable 0-form $\psi$
that has a singularity at only one of these points. This leaves the three forms 
$\d\log(z), \d\log(1-z)$ and $\d\log(z-x)$ as a potential basis, however
one combination is cohomologous  to zero, $\nabla_\omega 1 = \alpha_0 \d\log z + \alpha_1 \d\log(1-z)$, which removes one.
That the two classes in eq.~(\ref{2F1 basis}) 
are indeed linearly independent will be confirmed by the non-degenerate 
intersection pairing computed shortly.

We may depict graphically the two forms in eq.~(\ref{2F1 basis}) as paths  from $0$ to infinity, and $x$ to infinity, respectively, as shown in 
figure~\ref{fig:2F1 cycles}(a).  One might then guess, geometrically, ``orthonormal'' 
dual cycles in figure~\ref{fig:2F1 cycles}(b) which intersect the cycles of 
figure~\ref{fig:2F1 cycles}(a) only once.
We will now find explicit dual forms which precisely reflect this picture.

The new feature of having the pole at $z=x$ is that the form $d\theta(z-x)$ is 
\emph{not} exact in $H^1_{\rm dR,c}$ (because neither the step function 
$\theta(z-x)\equiv \theta(|z-x|-\eta)$ nor its complement form compactly 
supported 0-form in $X=\CP^1\setminus\{0,1,x,\infty\}$).
This will be a very convenient choice for our second dual basis element:
\be
 \bra{\vphi_1^\dual} = \left[\d\log\frac{z}{1-z}\right]_c,
 \qquad 
 \bra{\vphi_2^\dual} = \delta_{z-x}(1) = \frac{u(z)}{u(x)} \d\theta(z-x). 
 \label{2F1 dual basis}
\ee
The factor $u(z)/u(x)$ is required to make the second form closed;
it is single-valued \emph{along the small circle on which $\d\theta$ is supported}.
Let us now elaborate on the first form, which is a ``compactly supported version'' 
of the algebraic form $\d\log\frac{z}{1-z}$. It is constructed by a procedure analogous to that used 
in eq.~(\ref{eq:1st c-map example}): we slap a step functions then add patch-up terms to make the result closed.
In general:
\be \label{eq:c-map}
	\left[\vphi^\dual\right]_c \equiv \vphi^\dual\prod_{i} \theta(z-z_i)\ + \sum_{i}  \psi_i \d\theta(z-z_i) 
\ee
where $z_i\in \{0,1,x,\infty\}$ and $\psi_i= \nabla_{-\omega}^{-1} \vphi^\dual$ 
is a local primitive of $\vphi^\dual$ around $z_i$. This primitive is unique around 
each point with a twist, and can be obtained as a Laurent series. For example, near 
the origin, we have
\be
\nabla_{-\omega} (z^n) = (n-\alpha_0)z^{n-1}\d z+\mbox{less singular}
\ee
which implies that $\nabla_{-\omega}$ can be uniquely inverted order by order in 
a Laurent series. The procedure generates denominators of the form 
$1/(n -\alpha_0)$ where $n$ is an integer. For $\vphi^\dual=\frac{\d z}{z(1-z)}$, 
in particular, we get
\be\begin{aligned}
 \psi_0 &= -\frac{1}{\alpha_0} &+\quad& \frac{\alpha_0+\alpha_1}{\alpha_0(1-\alpha_0)}z &+\quad& O(z^2), \\
 \psi_1 &= \phantom{-}\frac{1}{\alpha_1} &-\quad& \frac{\alpha_0+\alpha_1}{\alpha_1(1-\alpha_1)}(1-z) &+\quad& O((1-z)^2),\\
 \psi_\infty &= \phantom{-1}0 &+\quad& \frac{1}{1+\alpha_0+\alpha_1}\frac{1}{z} &+\quad& O(1/z^2). \label{2F1 primitives}
\end{aligned}\ee
The untwisted point is no different: 
\be\begin{aligned}
 \psi_{x} &= \phantom{-1} 0 &+\quad& \phantom{-1} \frac{z-x}{x(1-x)} \phantom{-1} &+\quad& O((z-x)^2),
\end{aligned}\ee
but notice that the first term is now an integration constant, not determined by the differential equation.
We set to zero the constants around untwisted boundaries as a canonical definition of the c-map.
However, note that representatives of the relative cohomology cannot have poles at untwisted points:
if $\vphi^\vee$ had a reside, one would not be able to invert the covariant derivative.

Instead of series solutions, one could formally write the primitive $\psi_{z_i}$ as the integral
\be
	\psi_{z_i} [\vphi^\vee]
	= \nabla_{-\omega}^{-1} \vphi^\vee \;\; (\text{near } z=z_i)
	= \int_{z_i}^z \frac{u(z,\{s_a\})}{u(z^\prime,\{s_a\})} \vphi^\vee(z^\prime,\{s_a\})
\ee
where $\{s_a\}$ is a set of ``\textit{kinematic}'' variables (in the present case $\{s_a\} = \{x\}$). While evaluation of this integral can be challenging (it is essentially a twisted period), it does provide primitives valid to all orders in $(z-z_i)$, which is sometimes useful. 


Given a holomorphic form $\vphi$, its intersection with the dual basis (\ref{2F1 dual basis}),
defined by integration as usual (see eq.~(\ref{pairing})),
can now be explicitly computed:
\be
 \braket{\vphi_1^\dual}{\vphi} = \sum_{z_i=0,1,x,\infty} {\rm Res}_{z=z_i} (\psi_i \vphi)\,, \qquad
 \braket{\vphi_2^\dual}{\vphi} = {\rm Res}_{z=x} \left(\frac{u(z)}{u(x)} \vphi\right). \label{2F1 residues}
\ee
Notice that if $\vphi$ only has logarithmic singularities, only the leading term of the primitives (\ref{2F1 primitives})
is required. Similarly, for logarithmic forms, the factor $\frac{u(z)}{u(x)}$ is irrelevant. Then the intersections simplify to
\be
 \braket{\vphi_1^\dual}{\vphi} = \tfrac{1}{\alpha_1}{\rm Res}_{z=1} \vphi-\tfrac{1}{\alpha_0}{\rm Res}_{z=0}\vphi,\qquad
 \braket{\vphi_2^\dual}{\vphi} = {\rm Res}_{z=x} \vphi \qquad\mbox{($\vphi$ logarithmic)}. \label{2F1 log}
\ee
The full primitives $\psi_i$ (and $u(z)/u(x)$) are however very important
if $\vphi$ has higher-order poles. The number of required terms is dictated by the order of the poles in $\vphi$.
Eq.~(\ref{2F1 log}) immediately yields the intersection with the basis $\ket{\vphi_j} = \{ \d\log z,\d\log(z-w)\}$ defined
in eq.~(\ref{2F1 basis}):
\be
 C_{ij} \equiv \braket{\vphi^\dual_i}{\vphi_j} =
 \left(	\begin{array}{c@{\ \ }c} -\frac{1}{\alpha_0} & 0\\0 & 1\end{array} \right)\ . \label{2F1 Cij}
\ee
Notice that it is diagonal, as anticipated from the geometric picture in figure \ref{fig:2F1 cycles}.

As a simple application, consider the reduction of the integral $I_{1,2,2}$ (defined in eq.~(\ref{2F1 family})),
corresponding to the form $\vphi_{1,2,2} = \frac{\d z}{z(1-z)^2(z-x)^2}$.
Since it has double poles at $1$ and $x$, all the terms in two of the primitives (\ref{2F1 primitives}) are needed, and we find
\be\begin{aligned}
 \braket{\vphi_1^\dual}{\vphi_{1,2,2}} &= \frac{-1}{\alpha_0 x^2} + \frac{\alpha_0+4\alpha_1-3-x(\alpha_0+2\alpha_1-1)}{\alpha_1(1-\alpha_1)(1-x)^3}+\frac{1}{x^2(1-x)^3} \equiv A,
\\
 \braket{\vphi_2^\dual}{\vphi_{1,2,2}} &= \frac{\alpha_0 - 1 - x (\alpha_0 + \alpha_1 - 3)}{x^2 (1 - x)^3} \equiv B,
\end{aligned}
\ee
which implies the following identity between hypergeometric functions:
\be
 I_{1,2,2}(x) = -\alpha_0A I_{1,0,0} + B I_{0,0,1}(x),
\ee
which may be tested directly via numerical integration or series expansion in $1/x$.
Note also that here $I_{1,0,0}=B(\alpha_0,1+\alpha_1)$ is a constant independent of $x$.
The $-\alpha_0$ factor originates from inverting the intersection (\ref{2F1 Cij}).

\subsubsection{Differential equation}

As a further application, let us obtain the differential equation satisfied by the integral $I_{0,0,1}$.
At the level of cohomology, we would like the Picard-Fuchs (or Gauss-Manin) connection: 
\be
 \nabla_\omega^{\rm full} \ket{\vphi_j} = \Omega_{ij} \ket{\vphi_i} + (\nabla_\omega^{z}\mbox{-exact})
\ee
where $\d^{\rm full} \equiv \d x \partial_x + \d z \partial_z$ includes both internal variables and external parameters, $\nabla_\omega^z$ only $z$-derivatives, and where $\Omega_{ij}$ does not depend on $z$. The calculation is particularly straightforward if we keep the basis elements in their $\d\log$ form, interpreting the total derivative $d$ as $\d^{\rm full}$ -- we now drop the ``full'' superscript. (The form $\d\log(z-x)$ then includes a term $\frac{\d x}{x-z}$, which is a 0-form with respect to $z$ that does not contribute to the calculation below \cite{Herrmann:2019upk}.) Then
\be
 \Omega_{ij} = C^{-1}_{ji'}\braket{\vphi_{i'}^\dual}{\omega\wedge \vphi_i}.
\ee
The form on the right is logarithmic and we can use the simplified residue formula (\ref{2F1 log}).
For example,
\be\begin{aligned}
 \Omega_{21} &=
 -\alpha_0 \left[\tfrac{1}{\alpha_1}{\rm Res}_{z=1}-\tfrac{1}{\alpha_0}{\rm Res}_{z=0}\right] \Big( -\big(\alpha_0 \d\log z+\alpha_1 \d\log(1-z)\big)\wedge \d\log(z-x)\Big)
  \\ &= \alpha_0 \d\log \frac{1-x}{x}.
\end{aligned}\ee
Note that the large parenthesis is a 2-form proportional to $\d z\wedge \d x$; its residue at $z=0$ simply extracts the $\d\log z$ term
and puts $z=0$ in what it multiplies. (Our sign convention is that ${\rm Res}_{z=0}$ of a two-form plucks a factor $\d z/z$ from the left,
which is compatible with the way the residue appears in our formula: ${\rm Res}_{z=0}\vphi \equiv \frac{1}{2\pi i} \int \d\theta(z)\wedge \vphi$.)
It is easy to see that $\Omega_{i1}=0$ (there is no $\d x$ to be found in $\omega\wedge \d\log(z)$), and the remaining component
$\Omega_{22}$ is a single residue at $z=x$ of the above parenthesis. Thus altogether
\be
 \Omega = \left(\begin{array}{c@{\ \ }c} 
 	0 & 0 \\ 
	\alpha_0 \d\log \frac{1-x}{x} 
		& \alpha_0 \d\log x + \alpha_1 \d\log(1-x) \end{array}\right),
\ee
which may be compared (after integrating with respect to $z$) with the hypergeometric equation.
To be fully explicit, the differential equation is:
\be
 \frac{d}{\d x} \left(\begin{array}{c} I_{1,0,0}(x) \\ I_{0,0,1}(x)\end{array}\right)
 = \left(\begin{array}{c@{\hspace{2em}}c} 0 & 0 \\ \frac{-\alpha_0}{x(1-x)} & \frac{\alpha_0}{x} + \frac{\alpha_1}{x-1} \end{array}\right)
 \left(\begin{array}{c} I_{1,0,0}(x) \\ I_{0,0,1}(x)\end{array}\right). \label{2F1 diff eq}
\ee
The fact that $I_{1,0,0}(x)\equiv C$ is constant was observed already.

To complete the analogy with the physical problem of Feynman integrals, one would consider the twists
$\alpha_i$ as small parameters: $\alpha_i\propto \eps$, and specialize to the $\eps\to 0$ limit.
The differential equation (\ref{2F1 diff eq}) then has the canonical form of ref.~\cite{Henn:2013pwa}: proportional to $\eps$.
It can be readily integrated order by order in terms of harmonic polylogarithms,
using the boundary condition that $I_{0,0,1}(\infty)=0$ (clear from the definition (\ref{2F1 family})):
\begin{align}
 \tfrac{1}{C}I_{0,0,1}(x) 
 &= \alpha_0 \log\left(\frac{x-1}{x}\right) 
 + \alpha_0^2 \Li_2\left(\frac{1}{x}\right) -\alpha_0\alpha_1 \Li_2\left(\frac{1}{1-x}\right)
 + O(\alpha^3).
\end{align}

%
%
%

\subsection{Zero-dimensional: pentagon dual \label{sec:cpent}}

For the remainder of this section, we take $\dint=4$. 
Then, the maximal topology is the pentagon with propagators
\begin{align} \label{eq:pentprop}
	D_1 = \ell_\perp^2 + \ell^2 + m_1^2 , 
	\quad 
	D_2 = \ell_\perp^2 + (\ell + p_1)^2 + m_2^2 ,
	\quad
	D_3 = \ell_\perp^2 + (\ell + p_1 + p_2)^2 + m_3^2 ,
	\nn \\
	D_4 = \ell_\perp^2 + (\ell + p_1 + p_2 + p_3)^2 + m_4^2 ,
	\quad
	D_5 = \ell_\perp^2 + (\ell + p_1 + p_2 + p_4)^2 + m_5^2.
\end{align}
Any additional propagators will be linearly dependent on the $D_{i\leq5}$ since vector space of external momentum is 4-dimensional. 

Compactifying the pentagon dual form is trivial since 
\be
	\vphi^\vee_\pent 
	= c_{-1} \delta_{12345}(1) 
	= c_{-1} \frac{u}{u_{12345}} \bigwedge_{i=1}^5 d\th(D_i).
\ee
where $u = (\ell_\perp)^{-\vep}$, $u_{12345} = u\vert_{12345} = (r_{12345}^2)^{-\vep}$ and we have used equation \eqref{eq:c-leray} for the compactly supported Leray map. It is now straightforward to extract the pentagon coefficient of a general amplitude $\vphi$
\be
	c_\pent 
	= \frac{ \ipaa{\vphi^\vee_\pent}{\vphi} }{ \ipaa{\vphi^\vee_\pent}{\vphi_\pent}}
	= \res_{12345} \left[ \frac{u}{u_{12345}} \vphi \right] \bigg\slash \res_{12345} \left[ \frac{u}{u_{12345}} \vphi_\pent \right]. 
\ee
It is simply the max cut residue of $\vphi$ weighted by $u/u_{12345}$. Note that $u/u_{12345}$ is only needed when the propagators
in $\vphi$ have higher order poles.

\subsection{One-dimensional: box dual \label{sec:cbox}}

The algebraic box dual \eqref{eq:boxes} is a one-form supported on the box cut.
Since the Leray coboundary preserves compact support, it suffices to find the compactly supported version of this one-form
before we apply the coboundary to it.

Specializing to the 1234-box, we have 
\begin{align}
	[\vphi^\vee_{1234}]_c = \delta_{1234}([\phi_{1234}^\vee]_c),
	\quad 
	\phi_{1234}^\vee = c_0 \frac{r_{1234}\ \d\ell_{1234}}{r_{1234}^2 + \ell_{1234}^2},
	\quad 
	c_0 = \frac{4}{\vep}.
\end{align}
Here, $\ell_{1234} = \ell \cdot e_{1234}$ where $e_{1234}$ is a unit basis vector of the external kinematic space perpendicular to
the momenta at each corner of the box.

On the 1234-cut, the twist is $u_{1234}^\vee = u^\vee\vert_{1234} = \left( r_{1234}^2 + \ell_{1234}^2 \right)^{\vep}$ and there is also a boundary corresponding to the pentagon cut \be
	D_5\vert_{1234} 
	\propto \ell_{1234} - \sqrt{r_{12345}^2-r_{1234}^2}.
\ee
Thus, we will need primitives for $\phi^\vee_{1234}$ near the twisted singular points $\{ \pm r_{1234}, \infty \}$ as well as the boundary point $\ell_{0} \equiv \sqrt{r_{12345}^2-r_{1234}^2}$. These 1-form primitives are easily computed:
\begin{align}
	\psi^\vee_+ &= -\frac{c_0}{2 \vep} 
		+ \frac{c_0 (\ell_{1234} - r_{1234})}{2(1+\vep) r_{1234}} 
		+ \O\Big((\ell_{1234}-r_{1234})^2\Big),
\\
	\psi^\vee_- &= \frac{c_0}{2 \vep } 
		+ \frac{c_0 (\ell_{1234} + r_{1234})}{2(1+\vep) r_{1234}} 
		+ \O\Big((\ell_{1234}+r_{1234})^2\Big),
\\
	\psi^\vee_\infty &= \frac{c_0\ r_{1234}}{(1-2\vep)\ell_{1234}} 
		+ \frac{c_0\ r_{1234}^3}{(3-2\vep)(1-2\vep)\ell_{1234}^3}
		+ \O\Big(\frac{1}{\ell_{1234}^4}\Big),
\\
	\psi^\vee_0 &= \frac{c_0}{r_{12345}} (\ell_{1234}-\ell_0) 
		+ \frac{c_0 (1+\vep)\ell_0}{r_{12345}^3} (\ell_{1234}-\ell_0)^2 
		+ \O\Big( (\ell_{1234}-\ell_0)^3 \Big).
\end{align}
Moreover, at one-loop, closed form expressions for the primitives exist (Appendix \ref{app:closed form box prim}). Then, the 1234-box coefficient for any Feynman integral $\vphi$ is given by the intersection
\begin{align} \label{eq:boxcoeff}
	c_{1234}[\vphi] 
	&=\frac{\la \vphi^\vee_{1234} \vert \vphi \ra}{\la \vphi^\vee_{1234} \vert \vphi_{1234} \ra}
	\nn \\
	&= 
		\Big\la [\phi^\vee_{1234}]_c \Big\vert \res_{1234} \left[ \frac{u_{1234}}{u}\vphi \right] \Big\ra
	\nn \\
	&=
	\sum_\alpha \res_\alpha  \left[ \psi^\vee_\alpha\ \res_{1234} \left[ \frac{u_{1234}}{u}\vphi \right] \right]
\end{align}
where the index $\alpha$ runs over the twisted singular points and boundary. Note that if there are no $D_5$ propagators, the contribution from the $\ell_0$ residue will always vanish.

It is enlightening to take the $\vep\to 0$ limit of \eqref{eq:boxcoeff} and compare with the BCF formula for the box coefficient \cite{Britto:2004nc}. Only the two twisted singularities contribute: 
\be \lim_{\vep\to 0} \frac{\vep}{c_0}c_{1234}[\vphi]  = -\sum_\pm \pm\res_{\pm r_{1234}} \res_{1234} [\vphi]\,.
\ee
There are powers of $\vep$ on the left ($c_0=\frac{4}{\vep}$) because the box forms in our basis of Feynman integrals contain a factor of $\vep^2$.
The two singular points of the twist $\ell_{1234}=\pm r_{1234}$ correspond to the two on-shell solutions to the quad-cut in 4-dimensions,
reproducing the BCF formula.  In general, eq.~\eqref{eq:boxcoeff} contains additional $\vep$ corrections;
the residue at $D_5=0$ can contribute if some propagator is squared, and the residue at infinity is needed for $\vphi$ that have numerators.

\subsection{Two-dimensional null coordinates: triangle dual \label{sec:ctri}}

In this section, we provide an efficient algorithm for compactifying certain 2-forms
bypassing the need to introduce the machinery of fibration (section \ref{sec:higher form c-map}). 

Coordinate choices greatly influence the compactification procedure. 
For example, one can often find coordinates where the cohomology on a fibre vanishes (or is smaller than it ought to be). 
Rather than being problematic, such singular coordinate choices can actually be used to simplify the compactification process. 
As an example, we will compactify the triangle dual form in null coordinates and obtain  remarkably simple expressions for triangle dual intersection numbers.
(Related but distinct coordinates could also be considered, such as those used in \cite{Forde:2007mi,Badger:2008cm,Giele:2008ve}, where the $\ell_\perp^2$
integral is kept for the last stage. Here we simply aim to experiment with rotations of Cartesian coordinates to see what kind of formulas we can obtain.)

We start by considering a massive triangle assuming no boundaries in order to demonstrate the procedure simply (boundaries will be added after). Specializing to the 123-triangle, equation \eqref{eq:triangles} yields, 
\begin{align}
	\vphi^\vee_{123} = \delta_{123}(\phi^\vee_{123}),
	\qquad 
	\phi^\vee_{123} = c_1 \frac{\d z_1 \wedge \d z_2}{1 - z_1^2 -z_2^2},
	\qquad
	c_1 = 8,
\end{align}
where the twist on the 123-cut is $u^\vee_{123} = u^\vee\vert_{123} = (1 - z_1^2 -z_2^2)^\vep$. Note that we have rescaled the coordinates of \eqref{eq:triangles} for ease of reading (effectively setting $r_{123}=1$). 


While triangle dual can straightforwardly be compactified in these coordinates using the fibration technology of section \ref{sec:higher form c-map}
to deal with $z_1$ and $z_2$ one at a time, we can try to be more clever and linearize the twist.
Changing to null coordinates $z_1 \to \frac{1}{2} (x+y), z_2 \to \frac{i}{2} (x-y)$ the twist and triangle dual form become
\be
	u^\vee_{123} = (1 - xy)^\vep,
	\qquad
	\phi^\vee_{123} = ic_1 \frac{\d x \wedge \d y}{1 - x y}. 
\ee
Looking as a function of $y$ one sees that the $H^1$ of the fibre vanishes for generic base point $x$;
in the language of section \ref{sec:higher form c-map} such a coordinate choice leads to singular fibres.

To see what is happening, consider trying to make primitives for $\phi^\vee_{123}$:
\begin{align} \label{eq:almost global primitives}
	\psi_{\xbig}^\vee = i \frac{c_1}{\vep} \frac{\d x}{x}
	\quad\text{ and }\quad
	\psi_\xsmall^\vee = -i \frac{c_1}{\vep} \frac{\d y}{y}
\end{align}
(the subscripts will become clear later). Taking the covariant derivative of the first reveals that this primitive is \emph{almost global}
\begin{align}
	\nabla^\vee \left( i\frac{c_1}{\vep} \frac{\d x}{x} \right)
	= \phi^\vee_{123} + \delta^2(x)
\end{align}
where $\delta^2(x)=d\theta(x)\wedge \frac{dx}{x}$ is a holomorphic anomaly \eqref{holo anomaly} that is caused by the fact that the $y$-primitive
becomes singular at $x=0$.  Thus, the light-cone coordinate transformation has concentrated the entire cohomology near the singular fibre $x=0$. 
A similar formula exists for the other almost global primitive that concentrates the cohomology on the $y=0$ line instead.

One shortcut for computing the particular intersection number $\la \phi^\vee_{123} | \phi \ra$ would be to use the fact that the (non-dual) form $\phi$
also has almost global primitives.
Then, one can show that $\phi^\vee_{123}$ is cohomologous to $\# \delta^2(x)$ and $\phi$ is cohomologous to $\delta^2(y)$. Since the wedge product of these delta function forms has compact support the intersection number is well defined without further compactification and localizes on $x=y=0$. 

\begin{figure}
	\centering
	\includegraphics[scale=.45]{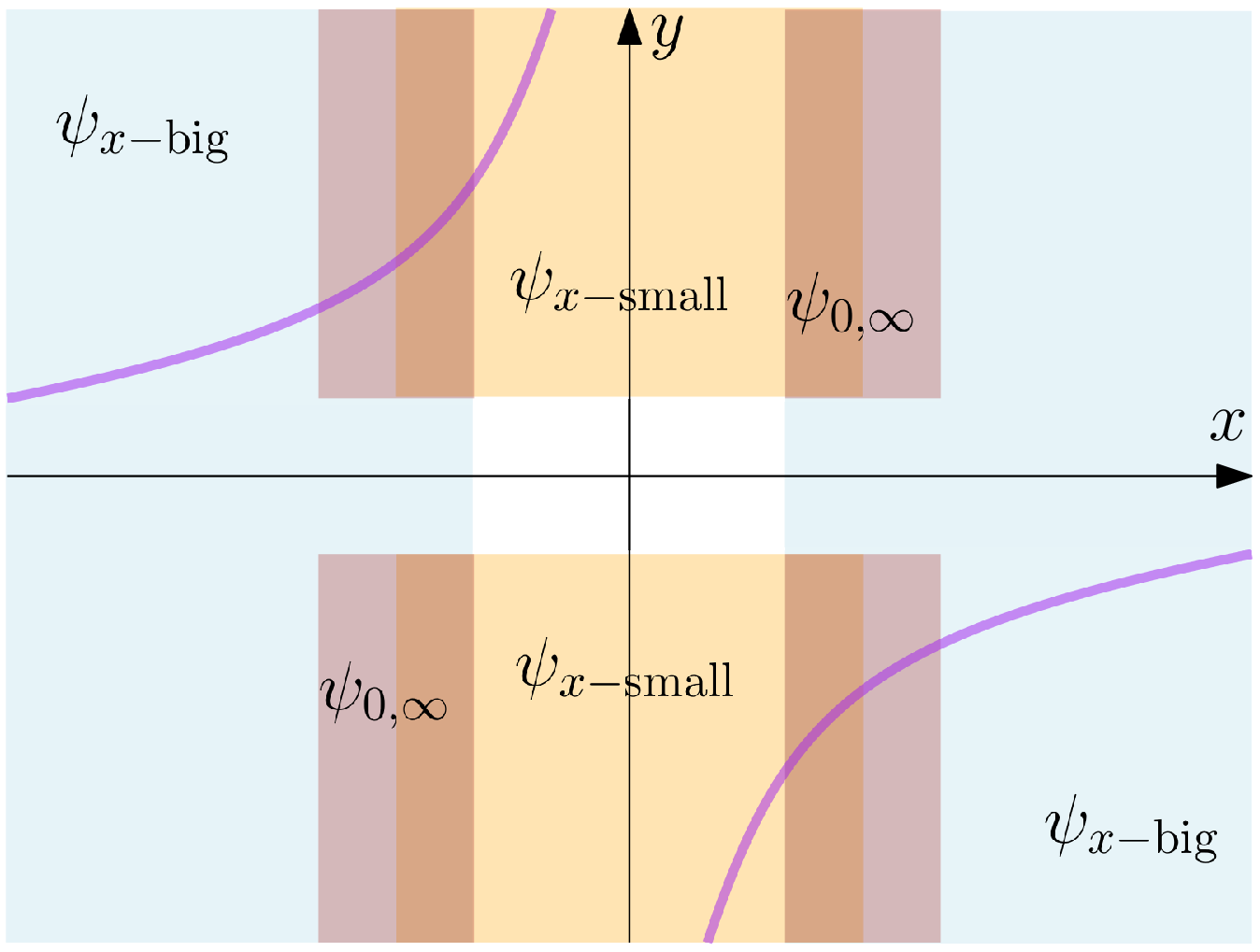}
	\qquad
	\includegraphics[scale=.45]{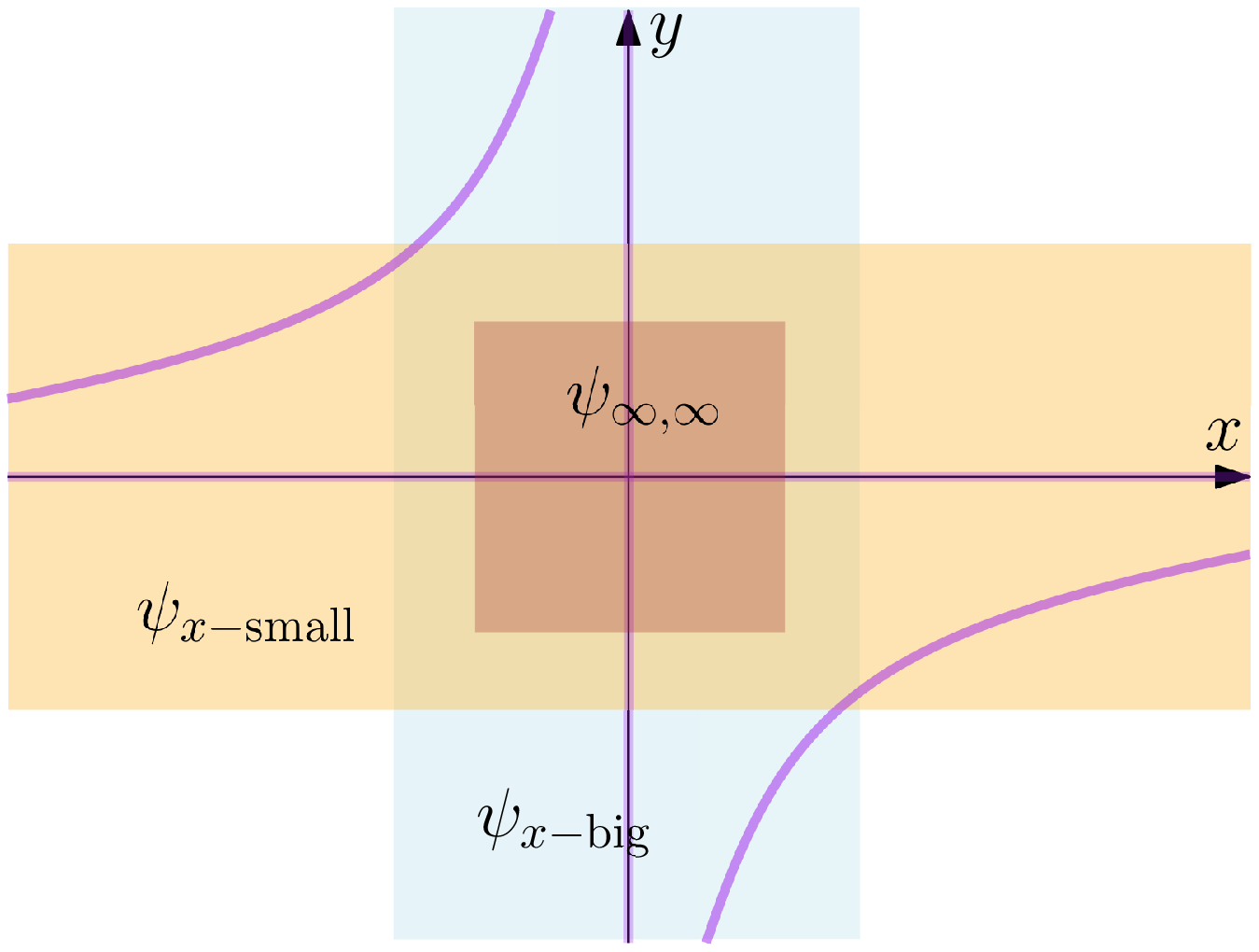}
	\caption{ \label{fig:LCprimitives}
		The removed surfaces (twist and the hyperplanes at $\infty$) are shown in purple. 
		The blue regions are where we use the primitive $\psi_\xbig$, while in the orange regions, we use the primitive $\psi_\xsmall$. 
		The red regions are the neighbourhoods on which the double primitives are supported.
		Right: normal chart containing the origin. 
		Left: chart at infinity where the origin is the point $(\infty,\infty)$.
	}
\end{figure}

However, as before, we prefer to fully compactify the dual forms and leave Feynman integrands untouched.
The trick to constructing the compactly supported version of $\phi^\vee_{123}$ will be to insist that $\psi_\xsmall$ is used when $x<\epsilon$ and $\psi_\xbig$ is used when $x>\epsilon$ (see figure~\ref{fig:LCprimitives}); this will yield a remarkably simple formula for the triangle coefficients in $d$-dimensions!

As the starting point, consider the following ansatz for $[\phi^\vee_{123}]_c$
\begin{align}
	[\phi^\vee_{123}]_c^{(0)}
	&=\theta(1-xy) \theta(1/x) \theta(1/y) \phi^\vee_{123}
	\nn \\
	&= \Big(
		1 - \bar\th(1-xy) - \bar\th (1/x) - \bar\th(1/y) 
		\nn \\ & \qquad
		- \bar\th(1-xy) \bar\th(1/x) - \bar\th(1-xy)\bar\th(1/y) - \bar\th(1/x) \bar\th(1/y) 
	\Big) \phi^\vee_{123}.
\end{align}
Obviously, $[\phi^\vee_{123}]_c^{(0)}$ has compact support but is not closed. We can remove all the terms with a single $d\bar\theta$ in $\nabla[\phi^\vee_{123}]_c^{(0)}$ by adding some patch-up terms
\begin{align}
	[\phi^\vee_{123}]_c^{(1)}
	&= [\phi^\vee_{123}]_c^{(0)} 
	+ \theta(1/y)\ d\bar{\theta}(1/x) \wedge \psi_\xbig^\vee
	+ \theta(1/x)\ d\bar{\theta}(1/y) \wedge \psi_\xsmall^\vee
	\nn \\ & \qquad
	{\color{purple} 
		- (1-\bar{\theta}(x)) \ d\bar{\theta}(1{-}xy)\wedge\psi_\xbig^\vee 
	}
	{\color{purple} 
		- \bar{\theta}(x) \ d\bar{\theta}(1{-}xy) \wedge\psi_\xsmall^\vee
	}
	\nn \\ & \qquad
	{\color{purple} 
		+ \bar{\theta}(1{-}xy) \ d\bar{\theta}(1/x) \wedge \psi_\xbig^\vee 
	}
	{\color{purple} 
		+ \bar{\theta}(1{-}xy) \ d\bar{\theta}(1/y) \wedge \psi_\xsmall^\vee
	}.
%
\end{align}
The terms highlighted in red showcase how the primitive in the neighbourhood of $1-xy=0$ is split between the regions where $x>\epsilon$ and $x<\epsilon$. This splitting introduces an additional $\theta$ function at $x=0$ and means that the intersection number can potentially localize on $x=0$ even though it is not a singular point of the twist! 

Testing $[\phi^\vee_{123}]_c^{(1)}$ for closure, we find 
\begin{align}
	\nabla[\phi^\vee_{123}]_c^{(1)} 
	= d\bar\theta(x) \wedge d\bar\theta(1-xy) 
		\wedge (\psi_\xbig^\vee - \psi_\xsmall^\vee) 
	\nn \\
	+ d\bar\theta(1/y) \wedge d\bar\theta(1/x) 
		\wedge (\psi_\xbig^\vee - \psi_\xsmall^\vee).
\end{align}
The double $d\bar\theta$ terms can be removed by constructing primitives for the difference $\psi_\xbig^\vee - \psi_\xsmall^\vee$ in the regions where the $d\bar\th$'s overlap
\begin{align}
	\nabla \psi_{0,\infty}^\vee 
	&= \psi_\xbig^\vee - \psi_\xsmall^\vee 
	\text{ near } (x,y) = (0,\infty),
	\\
	\nabla \psi_{\infty,\infty}^\vee 
	&= \psi_\xbig^\vee - \psi_\xsmall^\vee 
	\text{ near } (x,y) = (\infty,\infty).
\end{align}
Note that $\psi_\xbig^\vee - \psi_\xsmall^\vee\propto d\log (xy)$ so the two-index primitives are simply functions of $xy$.
Patching-up $[\phi^\vee_{123}]_c^{(1)}$, we find the closed and compactly supported form
\begin{align} 
	[\phi^\vee_{123}]_c 
	= [\phi^\vee_{123}]_c^{(1)} 
	+ \d\theta(x) \wedge \d\theta(1-xy) \psi_{0,\infty}^\vee
	+ \d\theta(1/y) \wedge \d\theta(1/x)\psi_{\infty,\infty}^\vee. 
\end{align}
From this equation, we see that the intersection number localizes on the points $(0,\infty)$ and $(\infty,\infty)$
\begin{align} \label{eq:tri bulk res}
	\la \vphi^\vee_{123} \vert \vphi \ra
	&= \bigg\la \phi^\vee_{123} \bigg\vert \res_{123} \bigg[\frac{u}{u_{123}}\vphi\bigg] \bigg\ra
	\nn \\
	&= \res_{y=\infty}\res_{x=0} 
	\left[ 
		\psi_{0,\infty}^\vee\; 
		\res_{123} \bigg[\frac{u}{u_{123}}\vphi\bigg] 
	\right]
	\nn \\ & \quad 
	+ \res_{y=\infty}\res_{x=\infty}
	\left[ 
		\psi_{\infty,\infty}^\vee\; 
		\res_{123} \bigg[\frac{u}{u_{123}}\vphi\bigg] 
	\right].
\end{align}
This result is remarkably simple -- only the double primitives $\psi_{0,\infty}^\vee$ and $\psi_{\infty,\infty}^\vee$ are needed! 

The choice to break up the neighbourhood about $1-xy = 0$ into the regions $x>\epsilon$ and $x<\epsilon$ is not unique and is somewhat analogous to choosing a fibration ordering. Especially when dealing with spherical twists, choices such as these are needed and break the spherical symmetry -- the codimension-2 points where the intersection number localizes cannot be deduced from the twist alone.

Our work is not done yet. Equation \eqref{eq:tri bulk res} does not extract the triangle coefficients of boxes or pentagons since it is not compactly supported on the additional boundaries present in a box or pentagon integral.
Fortunately, it is much simpler to enforce compact support at the boundaries. The trick is to consider the restriction of $[\phi^\vee_{123}]_c$ to the boundary (say, $D_4$ for concreteness). 
Then, if we can find a primitive for the restriction of $[\phi^\vee_{123}]_c$, the Leray construction (see, \eqref{eq:c-leray} or \cite{Caron-Huot:2021xqj}) effectively solves our problem. For example, let $\psi_4$ be a (compactly-supported) primitive for $[\phi^\vee_{123}]_c\vert_{D_4=0}$.
Then, the following is cohomologous to $[\phi^\vee_{123}]_c$ above:
\be
	[\phi^\vee_{123}]_c 
	\simeq \theta_{D_4}[\phi^\vee_{123}]_c + \frac{u}{u\vert_{D_4=0}}
 	\d\theta_{D_4} \wedge \pi_*\psi_4^\vee + \O(D_4 \d\theta_{D_4})\,
 	\quad\mbox{if}\quad 
 	\nabla \psi_4^\vee = [\phi^\vee_{123}]_c\vert_{D_4=0}.
\label{eq:tri-box contribution}
\ee
This is readily verified to be closed and vanishes in a neighbourhood of $D_4=0$.
The $\O(D_4\ \d\theta_{D_4})$ terms can always be computed order-by-order in a Taylor series around the boundary.
However, for most physics applications the $\O(D_4\ \d\theta_{D_4})$ corrections to \eqref{eq:tri-box contribution} can be ignored. These corrections are only needed when one wants to intersect $[\phi^\vee_{123}]_c$ with a Feynman form that contains double or higher order poles at the boundary. 

With the addition of the boundary $D_4$, the residue formula \eqref{eq:tri bulk res}, becomes
\begin{align} \label{eq:triangle with bd}
	\la \vphi^\vee_{123} \vert \vphi \ra
	&= \res_{(x,y)=(0,\infty)} 
	\left[ 
		\psi_{0,\infty}^\vee\; 
		\res_{123} \bigg[\frac{u}{u_{123}}\vphi\bigg] 
	\right]
	\nn \\ & \quad 
	+ \res_{(x,y)=(\infty,\infty)} 
	\left[ 
		\psi_{\infty,\infty}^\vee\; 
		\res_{123} \bigg[\frac{u}{u_{123}}\vphi\bigg] 
	\right]
	\nn \\ & \quad 
	+ \sum_\alpha \res_\alpha 
	\bigg[ 
		\psi_{4,\alpha}^\vee \; 
		\res_{1234} \bigg[\frac{u}{u_{1234}}\vphi\bigg] 
	\bigg]
\end{align}
where $\alpha$ denotes the singular points on the boundary $D_4$ (generically, there will be one or two finite singularities in addition to the singularity at infinity). 
In the case where we do not care about Feynman forms with higher order poles, the primitives $\psi_{4,\alpha}$ are easily computed since this boundary is simply a $\mathbb{C}^1$. 
We have tested this formula extensively on various examples with massive triangles as well as dimension shifting identities, as detailed below; it will also be used in section \ref{sec:4dbubble} for analyzing 4-dimensional limits of bubble coefficients.

This procedure can then be repeated one boundary at a time. For each iteration, we obtain additional twist-boundary contributions. When there are two or more additional boundaries (like for the pentagon), we also find boundary-boundary contributions to \eqref{eq:triangle with bd}. 
Proceeding in this way, one can derive residue formulas for triangle coefficients with an arbitrary number of boundaries.

\section{Compact support map for higher degree forms ($p>1$) \label{sec:higher form c-map}}

In this section we present an algorithm for constructing the compactly supported versions for multi-variate $(p>1)$ dual forms.

In its most basic form the compactly supported version of $\vphi^\vee$ is obtained by adding exact terms which remove support in tubular neighbourhoods of $\{\ell_\perp^2=0\}$ and $\{D_i=0\}$. For 1-forms, this is captured by equation \eqref{eq:c-map}. While we arrived at \eqref{eq:c-map} by slapping step functions in front of $\vphi^\vee$ and patching up to ensure that the resulting 1-form is closed, one can check that the difference $[\vphi^\vee]_c - \vphi^\vee$ is exact. That is, our construction of the compactly supported 1-forms coincides with the construction of \cite{matsuo1998}. 
Explicitly,
\begin{align} \label{eq:c-map2}
	[\vphi^\vee]_c = \Theta\ \vphi^\vee + \sum_\alpha \d\th_\alpha \psi_\alpha,
	\qquad 
	\vphi^\vee - [\vphi^\vee]_c = \nabla \Big( \sum_\alpha \bar{\theta}_\alpha \psi_\alpha \Big),
\end{align}
where $\Theta = \prod_\alpha\th_\alpha = 1 - \sum_\alpha \bar{\theta}_\alpha$, $\bar{\th}_\alpha = 1-\theta_\alpha$ and the $\th_\alpha$ are step functions with support outside the neighbourhood of the singular points indexed by $\alpha$. Here, the $\psi_\alpha$ are local primitives of $\vphi^\vee$ with respect to some connection $\omega^\vee$. 

Analogously, it is possible to construct compactly supported versions of $(p>1)$-forms by multiplying the original form by step functions with support outside tubular neighbourhoods about $\{\ell_\perp^2=0\}\cup\{D_i=0\}$ and patching up with terms $\wedge_{j\in J} \d\theta_i \psi_J$ such that the resulting $p$-form is closed at each nested boundary $J$ \cite{matsuo1998}.
While this method works well for low $p$, the multi-variate problem rapidly becomes computationally expensive since one needs a $\psi_J$ for each way of approaching each boundary. To avoid this combinatorial build up, it is often more efficient to break the problem into a series of $p=1$ problems. This is achieved by a procedure called fibration \cite{Mizera:2019gea, Frellesvig:2019kgj, Frellesvig:2019uqt}. While fibration plays an important role in practical intersection computations, we stress that it is just one particular way to compute the compact isomorphism $[\vphi^\vee]_c$. 

\begin{figure}
	\centering
	\includegraphics[width=.8 \textwidth]{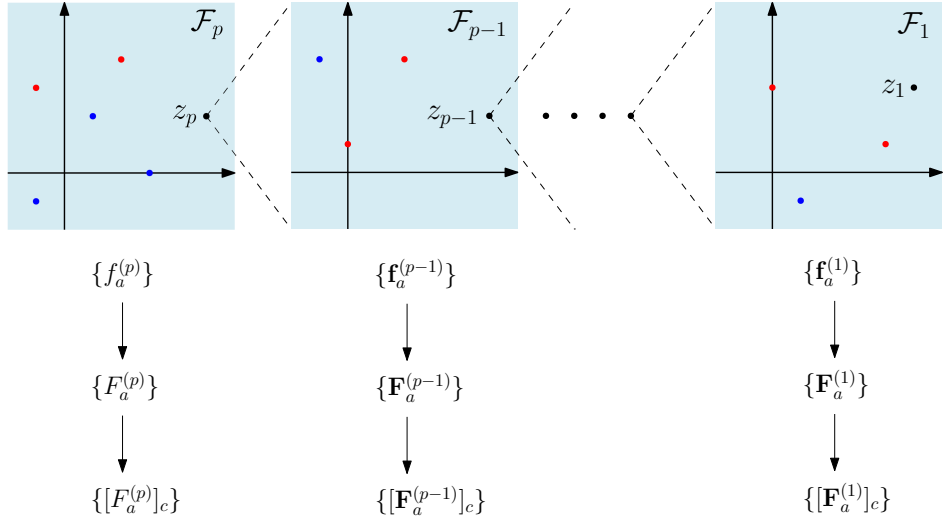}
	\caption{ \label{fig:fib sum}
	The process of the fibration c-map is illustrated. A $p$-dimensional manifold is split into a stack of 1-dimensional manifolds (called fibres). The $H^p$ of the original manifold is split into a series of $H^1$'s on each of the fibres (boundaries are blue dots while singular points are red dots). A basis $\{\bs{f}^{(i)}_a\}$ is chosen for each of the $H^1$, which in turn defines a connection on the fibre below it. The basis forms are then modified ($\bs{f}^{(i)}\to\bs{F}^{(i)}$) to have ``better'' commutation properties with the covariant derivative. Finally, each of the basis sets are compactified via the 1-form c-map.
	}
\end{figure}

Before moving on to the details, we summarize the fibration based multi-variate compactifying procedure here (see also figure \ref{fig:fib sum}):
\begin{itemize}
\item
Suppose that $\vphi^\vee$ is an algebraic $(p{>}1)$-form in $\{z_i\}_{i=1}^p$ and we wish to know its compactly supported version.

\item 
We choose an order of fibration, such that $z_i$ is fibred over $z_{j>i}$: $z_1$ is the first variable to be integrated out and the last base is $z_p$.

\item 
On the $i^\text{th}$ fibre, we choose a basis of 1-forms $\{f^{(i)}_a\}$. Here, the $f^{(i)}_a$ are a 1-forms in $\d z_i$ that are independent of $z_{j<i}$ and are generically vector valued (we will use bold typeface to denote vector-valued forms: $\bs{f}^{(i)}_a$).
The original $p$-form can be decomposed in terms of the fibre basis elements 
\begin{align}
	\vphi^\vee 
	\simeq \bs{f}^{(1)} \cdot \underline{\bs{f}}^{(2)} \cdots \underline{\bs{f}}^{(p)} \cdot \bs{n} 
\end{align}
where, generically, $\bs{f}^{(1)} = \bigoplus_a f^{(1)}_a$ is a vector-valued 1-form and the $\underline{\bs{f}}^{(i)} = \bigoplus_a \bs{f}^{(i)}_a$ are matrix-valued 1-forms. Here, $\bs{n}$ is a constant vector that picks out the correct combination of components with coefficients such that the above equality holds.\footnote{As the astute reader may have guessed, the elements of $\bs{n}$ can be computed using intersection numbers. We will however choose the $\{\bs{f}^{(i)}_a\}$ for a given $\vphi^\vee$ such that the components of $\bs{n}$ are obvious.} 
Note that there is an implicit wedge product in the vector/matrix multiplication above. 

\item Commuting the covariant derivative across the basis elements $\bs{f}^{(i)}_a$, modulo total derivatives, determines the connection on
the $(i+1)^\text{th}$ fibre ( $\mat{\omega}^{(i+1)}$).
We define a new form $\bs{F}^{(i)}$ that differs from $\bs{f}^{(i)}$ by an IBP form $\bs{V}^{(i)}$ so as to remove the total derivatives and get a \emph{strict equality}:
\be
 \nabla^{(i)} \bs{F}^{(i)}\wedge(\cdots) =
-\bs{F}^{(i)} \wedge \nabla^{(i+1)}(\cdots)\,. \label{F strict eq}
\ee
The decomposition of the original top form is unchanged:
\begin{align}
	\vphi^\vee 
	= \bs{F}^{(1)} \cdot \underline{\bs{F}}^{(2)}
		\cdots \underline{\bs{F}}^{(p)} \cdot \bs{n} .
\end{align}
\item 
The compactly supported version of $\phi^\vee$ is given by replacing the $\underline{\bs{F}}^{(i)}$ with their compactly supported versions (with respect to $z_i$) via the 1-form c-map
\begin{align} \label{eq:phic}
	\boxed{
		[\vphi^\vee]_c = [ \bs{F}^{(1)} ]_c 
		\cdot [ \underline{\bs{F}}^{(2)} ]_c 
		\cdots  [ \underline{\bs{F}}^{(p)}]_c \cdot \bs{n} 
		\simeq \vphi^\vee.
	}
\end{align}
Thanks to the commutation relation \eqref{F strict eq},
showing that $[\vphi^\vee]_c$ is cohomologous to $\vphi^\vee$ is relatively straightforward.  
\end{itemize}
While the additional vector and matrix structure may look complicated, our algorithm is still efficient since all connections are (block) lower triangular due to that fact that we are using relative cohomology rather than the parametric deformation as in \cite{Mizera:2019gea, Mizera:2019vvs, Frellesvig:2019uqt}.  Note that only the top anti-holomorphic term in $[\vphi^\vee]_c$ will contribute
to the intersection with a Feynman integrand since Feynman integrands are a top holomorphic forms. 

In section \ref{sec:fibration-gen}, we describe the fibration process that allows one to turn the multi-variate problem into a series of single-variate problems. Then, we illustrate how to compactly the fibration of $\vphi^\vee$ one variable at a time in section \ref{sec:compactifying-gen}. This compactification scheme leads to a simple recursive formula for the computation of any intersection number.

\subsection{Fibration \label{sec:fibration-gen}}

We would like to write $\vphi^\vee$ schematically as 
\begin{align}
	\vphi^\vee = f^{(1)} \wedge f^{(2)} \cdots \wedge f^{(p)}
\end{align} 
where $f^{(i)}$ is a 1-form in $\d z_i$ that is independent of $z_{j<i}$. Since only $f^{(1)}$ depends on $z_1$, it is useful to split the action of the covariant derivative into a piece that acts on $z_1$ and a piece that acts on $z_{i>1}$: $\nabla^\vee = \nabla^\vee_{\F_1} + \nabla^\vee_{\B_1}$. Here, the first term is the covariant derivative on the $1^\text{st}$ fibre ($\mathcal{F}_1 = z_1 \in \mathbb{C} \setminus \text{poles}(\omega^\vee)$)
\begin{align}
	\nabla^\vee_{\F_1} = \d_{z_1} + \omega_{\F_1}^\vee \wedge,
	\qquad 
	\omega_{\F_1}^\vee = (\omega^\vee \vert_{\d z_{i>1} = 0})
\end{align}
while the second term is the covariant derivative on the $1^\text{st}$ base $\B_1 = (z_2, \dots, z_p) \in \mathbb{C}^p \setminus \text{poles}(\omega^{(1)})$ where $\omega^{(1)}$ is some undetermined connection resulting from the splitting into fibre and base. To determine the connection on the first base, we need to specify a basis on the fibre cohomology
\begin{align}
	\{f_a^{(1)}\} \in H_{\F_1}^1 = H^1(\F_1, \cup_{z_1 \subset D_j} \{D_j=0\}; \omega_{\F_1}^\vee)
\end{align}
where $f^{(1)} \equiv f^{(1)}_1$. Then, the covariant derivative of any basis element can be expressed in terms of the same basis
\begin{align}
	\label{eq:f-base-connection}
	\nabla^\vee f^{(1)}_a 
	\simeq - f^{(1)}_b \wedge \omega^{(1)}_{ba}
\end{align}
where the coefficients $\omega_{ab}^{(1)}$ are 1-forms in the $z_{i>1}$. Note that since the base connection is matrix-valued, the elements of the base cohomology are vector-valued differential forms. 

It would be preferable if \eqref{eq:f-base-connection} was an exact equality on the level of differential forms rather than an equivalence of cohomology classes. Therefore, we define $F^{(1)}_a = f^{(1)}_a - V^{(1)}_a$ such that 
\begin{align}
	\label{eq:F-base-connection}
	\nabla^{(1)} F^{(1)}_a = - F^{(1)}_b \wedge \omega^{(1)}_{ba}
\end{align}
where $V^{(1)}$ is an IBP 1-form on $\B_1$ (0-form on $\F_1$) that is independent of $\d z_1$. Note that if $\phi^\vee_{\B_1,a}$ is any vector-valued form on the base, 
\begin{align}
	\label{eq:dF}
	\nabla^\vee F^{(1)}_a \wedge \phi^\vee_{\B_1} 
	= - F_b^{(1)} \wedge (\delta_{ba} d_{\B_1} + \omega^{(1)}_{ba} \wedge ) \phi_{\B_1}
	\equiv - F_a^{(1)} \wedge \nabla_{\B_1} \phi_{\B_1} \,.
\end{align}
Moreover, $F^{(1)}_a \wedge \phi_{\B_1} = f^{(1)}_a \wedge \phi_{\B_1}$ if $\phi_{\B_1}$ is an holomorphic top form, since
$V_a^{(1)} \wedge \phi_{\B_1} = 0$.

\begin{figure}
	\centering
	\includegraphics[scale=.4]{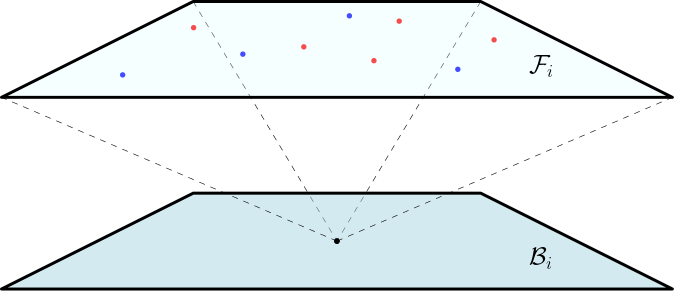}
	\caption{
	Each fibre $\F_i$ sits above a point on the base $x \in \B_i$. 
	The red dots are the singular points of the twist, while the blue dots are boundaries. 
	As $x$ moves around in $\B_i$ the red and blue dots in $\F_i$ move.
	}
\end{figure}

Repeating this process, we define the 1-dimensional fibres and the accompanying $(p-i)$-dimensional bases
\begin{align}
	\F_i &= z_i \in \mathbb{C} \setminus \{\text{poles}(\omega^{(i-1)})\} ,
	\\
	\B_i &= (z_{i+1},\cdots,z_p)\in \mathbb{C}^{p-i} \setminus \{\text{poles}(\omega^{(i)})\},
\end{align}
where $\omega^{(0)} \equiv \omega^\vee$ is the connection on the total space, $\B_0$. Then, for each fibre, we choose a basis, $\{\bs{f}^{(i)}_a\}$, for the fibre cohomology $H_{\F_i}^1 \equiv H^1(\F_i,\mathcal{R}_i; \mat{\omega}^{(i)}_{\F_i})$ where $\mathcal{R}_i$ is the set of $z_i$ boundaries and $\mat{\omega}_{\F_i}^\vee = \mat{\omega}^{(i-1)}\vert_{\d z_{j>i}=0}$. Note that $\bs{f}^{(i)}_a$ is a vector-valued form if the cohomology $H^1_{\B_{i-1}}$ is more than one-dimensional.
As a result, on the subsequent bases we get vector-valued version of the above:
\begin{align}
	\nabla^{(i)} \bs{f}^{(i)}_a  \simeq - \bs{f}^{(i)}_b \wedge \omega^{(i)}_{ba},\quad\Rightarrow\quad
	\nabla^{(i)} \bs{F}^{(i)}_a  = - \bs{F}^{(i)}_b \wedge \omega^{(i)}_{ba} \label{eq:nabla_commutation}
\end{align}
where $\bs{F}^{(i)}_a$ is a completion of $\bs{f}^{(i)}_a$ by adding IPB vectors proportional to $dz_{j\geq i}$.

The original top form $\vphi^\vee$ can then be expressed as a general linear combination
\begin{align}
	\vphi^\vee 
	= \bs{F}^{(1)} \cdot \underline{\bs{F}}^{(2)} \cdots 
	\underline{\bs{F}}^{(p)} \cdot \bs{n}\,.
\end{align}
The vector of forms, $\bs{F}^{(1)} \cdot \underline{\bs{F}}^{(2)} \cdots \underline{\bs{F}}^{(p)}$, span the un-fibred cohomology $H^p(\mathbb{C}^p\setminus\{\ell_\perp^2=0\}, \cup_i \{D_i=0\};\omega^\vee)$.\footnote{The dimension of the $H^p$ is equal to or less than the number of columns in $\bs{F}^{(1)} \cdot \underline{\bs{F}}^{(2)} \cdots \underline{\bs{F}}^{(p)}$. Columns can be cohomologous to other columns (massless limits) or vanish (when incompatible propagators are cut).}
In general, the coefficients $n_{a}$ can be computed via the intersection pairing or IBP relations. However, one can choose fibration bases such that $\vphi^\vee$ is easily related to the $\bs{f}^{(i)}_a$.

With this notation, it is easy to see how $\nabla^\vee$ commutes across the $\bs{F}$'s and acts on an element of the base cohomology $\bs{\phi}_{\B_i} \in H_{\B_i}^1$
\begin{align}
	\nabla^\vee ( \bs{F}^{(1)} \cdot \underline{\bs{F}}^{(2)} \cdots \underline{\bs{F}}^{(i-1)} \cdot \bs{\phi}_{\B_i} )
	= (-1)^{i-1} \bs{F}^{(1)} \cdot \underline{\bs{F}}^{(2)} \cdots \underline{\bs{F}}^{(i-1)} \cdot \nabla_{\B_i}^\vee \bs{\phi}_{\B_i}.
\end{align}
This relation allows us to apply the 1-form compact support isomorphism to each $\bs{F}^{(i)}$ on $\F_i$ independently of the other $\F_j$.

\subsection{Boundaries}

The next piece of fibration technology we need is the twist on $\B_i$. Since the $i^\text{th}$ cohomology is vector-valued, the associated twist $\mat{u}^{(i)}$ is a matrix valued 0-form. The twist satisfies the following differential equations
\begin{align}
	\d_{\B_i} \mat{u}^{(i)} = \mat{\omega}^{(i)} \cdot \mat{u}^{(i)},
	\qquad
	\nabla^\vee_{\B_i} \left(\mat{u}^{(i)}\right)^{-1} = 0.
\end{align}
Formally, these are solved by the path ordered exponential
\begin{align}
	\mat{u}^{(i)} 
	= \mathcal{P} \exp \int \mat{\omega}^{(i)}. 
\end{align}
Fortunately, explicit expressions are not required, only the abstract properties of $\mat{u}^{(i)}$ are needed.

Lastly, we need to define the Leray coboundary for vector-valued forms. 
One way is to use the matrix-valued twist:
\begin{align} \label{eq:matrix leray}
	\bs{\delta}_{D_i}^{(i)}
	= \left(\mat{u}^{(i)}\right)^{-1} \cdot \left(\mat{u}^{(i)}\vert_{D_i}\right) \d\theta(D_i).
\end{align}
Then, the columns of the Leray $\delta_{D_i}$ can be used as basis forms for the boundaries. 
However, in practice, we find that it is easier to take a more abstract approach that does not involve the matrix-valued twist. 

It is easiest to illustrate the idea with an example. Suppose that there are two boundaries $D_a$ and $D_b$ on the first fibre. The basis of form on $\F_1$ will contain a bulk form $f^{(1)}_1$ and two boundary forms $f^{(1)}_2 = \delta_{A}(1)$ and $f^{(1)}_3 = \delta_{B}(1)$. 
A generic form on $\B_1$ is a 3-vector 
\be
	\bs{\phi}_{\B_1}^\vee = \begin{pmatrix} \phi^\vee_1 & \phi^\vee_2 & \phi^\vee_3 \end{pmatrix}^T
\ee
that when wedged with $\bs{F}^{(1)}$ produces a form on the total space
\be \label{eq:FdotphiB1}
	\bs{F}^{(1)} \cdot \bs{\phi}^\vee_{\B_1} 
	= f^{(1)}_1 \wedge \phi^\vee_1 + f^{(1)}_2 \wedge \phi^\vee_2 + f^{(1)}_3 \wedge \phi^\vee_3
	= f^{(1)}_1 \wedge \phi^\vee_1 + \delta_A(\phi^\vee_2) + \delta_B(\phi^\vee_3). 
\ee
Each component $\phi^\vee_a$ belongs to the twisted cohomology produced by the corresponding diagonal component of the connection $\omega^{(1)}_{aa}$. By applying the total space derivative to \eqref{eq:FdotphiB1}, the action of the $\B_1$ covariant derivative can be determined
\be \label{eq:dFdotphiB1}
	\nabla^\vee ( \bs{F}^{(1)} \cdot \bs{\phi}_{\B_1} ) 
	= - F^{(1)}_a \wedge ( \delta_{ab} d + \omega^{(1)}_{ab} \wedge ) \phi^\vee_b + \delta_{AB}(\phi^\vee_2\vert_4 - \phi^\vee_3\vert_2).
\ee
Since $\delta_A(\delta_B(\bullet)) = -\delta_B(\delta_A(\bullet))$, we have to identify 
\be
	\begin{pmatrix}0&\delta_B(1)&0\end{pmatrix}^T
	\simeq \begin{pmatrix}0&0&-\delta_B(1)\end{pmatrix}^T
\ee
to avoid over counting. Choosing to put all double boundary contributions in the $\phi^\vee_3$ position, equation \eqref{eq:dFdotphiB1} becomes
\be
	\bs{F}^{(1)} \cdot \bs{\phi}^\vee_{\B_1} = - \bs{F}^{(1)} \cdot	\nabla^{(1)} \bs{\phi}^\vee_{\B_1},
\ee
where the $\B_1$ covariant derivative is defined to be
\be
	\nabla^{(1)} \bs{\phi}^\vee_{\B_1} 
	= (d  + \mat{\omega}^{(1)} - \mat{\L}^{(1)}) \cdot \bs{\phi}^\vee_{\B_1}.
\ee
Here, 
\be \label{eq:leray matrix}
	\mat{\L}^{(1)} 
	= \begin{pmatrix} 
		0 & 0 & 0
		\\ 
		0 & 0 & 0
		\\
		0 & -\delta_A(\ \vert_B) & \delta_A(\ \vert_A)
	\end{pmatrix}
\ee
encodes the boundary information reproducing the double boundary term in \eqref{eq:dFdotphiB1} and $\delta_A(\ \vert_B) \wedge \phi \equiv \delta_A(\phi \vert_B)$. 
When constructing a basis for the second fibre, the columns of $\mat{\L}$ will be used in place of the columns of $\eqref{eq:matrix leray}$.

\subsection{Compactifying \label{sec:compactifying-gen}}

Most of the heavy lifting has already been done by fibration. Once $\vphi^\vee$ has been written in terms of the $\bs{F}^{(i)}$, the compact version of $\vphi^\vee$ is obtained by the replacement $\bs{F}^{(i)}_a \to [\bs{F}^{(i)}_a]_c$, following
eq.~\eqref{eq:phic}.

In the remainder of this section, we define $[\bs{F}^{(i)}]_c$, show that the above replacement is cohomologous to the original form $\vphi^\vee$ and describe how to compute the intersection number. 

Basis forms associated to boundaries are conveniently represented by Leray forms, which already have compact support. 
Explicitly, if $\bs{f}^{(i)}_a$ is a Leray basis form (a column of \eqref{eq:matrix leray}), then $\bs{F}^{(i)}_a = [\bs{F}^{(i)}_a]_c$.
Therefore, the remaining task is to compactify the basis elements that do not already have compact support (bulk forms or non-Leray forms).   

Suppose that $\bs{f}^{(i)}_{a_\bk}$ is a bulk form. It's compactly supported version, $[\bs{f}^{(i)}_{a_\bk}]_c$, is defined via the 1-form c-map \eqref{eq:c-map2} on the $i^\text{th}$ fibre
\be \label{eq:vector_primitives}
	\left[\bs{f}^{(i)}_{a_\bk}\right]_c 
	= \Theta^{(i)}  \bs{f}^{(i)}_{a_\bk} 
		+ \sum_\alpha  \d\theta^{(i)}_\alpha\ \bs{\psi}^{(i)}_{{a_\bk},\alpha}
\ee
where 
$\Theta^{(i)} = \prod_\alpha \theta^{(i)}_\alpha$,
$\theta^{(i)}_\alpha = \theta(|z_i - z_{i,\alpha}|>\epsilon)$, 
$\bar{\theta}^{(i)}_\alpha = 1-\theta^{(i)}_\alpha$, 
and, the $z_{i,\alpha}$ are the twisted singular \emph{and} untwisted boundary points on $\F_i$. Here,
\be
	\bs{\psi}^{(i)}_{{a_\bk},\alpha} 
	= \left(\nabla_{\F_i}^\vee\right)^{-1} \bs{f}^{(i)}_{a_\bk}
	\text{ near } z_i \approx z_{i,\alpha}
\ee
is a local primitive of $\bs{f}^{(i)}_{a_\bk}$. 
Then, $[\bs{F}^{(i)}_{a_\bk}]_c$ is defined to be 
\be \label{eq:Fc}
	\left[\bs{F}^{(i)}_{a_\bk}\right]_c 
	= \left[\bs{f}^{(i)}_{a_\bk}\right]_c 
		- \Theta^{(i)}\ \bs{V}^{(i)}_{a_\bk}
	= \Theta^{(i)}\ \bs{F}^{(i)}_{a_\bk} 	
		+ \sum_\alpha  \d\theta^{(i)}_\alpha \ \bs{\psi}^{(i)}_{{a_\bk},\alpha}.	 
\ee
The vector primitives \eqref{eq:vector_primitives} appearing in \eqref{eq:Fc} are computed in the same manner as their 1-dimensional counterparts. The only new complication is that matrix-valued fibre connections $\mat{\omega}_{\F_i}^\vee$ need to be inverted. For one-loop integrals, the connections are lower-triangular and this inversion is efficient.\footnote{Boundaries help keep the fibration connection simple. If all the boundaries were instead twisted singularities, the connection would be some generic matrix.}

Knowing how $\nabla_{\B_i}$ commutes past $[\bs{F}^{(i)}_{a_\bk}]_c$, is essential to proving that \eqref{eq:phic} is cohomologous to $\vphi^\vee$. Crucially, it is possible to define compact versions of $\bs{F}^{(i)}_{a_\bk}$ that also satisfy equation \eqref{eq:nabla_commutation}
\be \label{eq:Fc-commutation}
	\nabla^\vee_{\B_{i-1}} \left[\bs{F}^{(i-1)}_{a_\bk}\right]_c 
		= - [\bs{F}^{(i-1)}_b]_c \wedge \omega^{(i)}_{b{a_\bk}}.
\ee
However, to prove \eqref{eq:Fc-commutation}, we need to know how $\nabla^{(i)}$ acts on the primitives $\bs{\psi}^{(i)}$ which can be found in appendix \ref{app:compactifying-dets}.

We can now show that \eqref{eq:phic} is cohomologous to $\vphi^\vee$. Assume that we are in the middle of the compactifying procedure
\be
	\vphi^\vee \sim 
	\bs{F}^{(1)} \cdot \underline{\bs{F}}^{(2)} \cdots 
		\underline{\bs{F}}^{(q)} \cdot [\underline{\bs{F}}^{(q+1)}]_c \cdots [\underline{\bs{F}}^{(p)}]_c \cdot \bs{n}. 
\ee
To convert $\underline{\bs{F}}^{(q)}$ into $[\underline{\bs{F}}^{(q)}]_c$, we subtract the total derivative 
\begin{align}
	&\nabla^\vee \left( (-1)^{q-1}
		\bs{F}^{(1)} \cdots \underline{\bs{F}}^{(q-1)} 
		\cdot \sum_{\alpha} (1-\theta^{(q)}_\alpha) \underline{\bs{\psi}}^{(q)}_{\alpha}  
		\cdot [\underline{\bs{F}}^{(q+1)}]_c \cdots [\underline{\bs{F}}^{(p)}]_c
		\cdot \bs{n}
		\right) 
	\nn \\ &\quad 
	= \bs{F}^{(1)} \cdots \underline{\bs{F}}^{(q-1)}
		\cdot \nabla^{(q-1)} 
		\left( \sum_{\alpha} (1-\theta^{(q)}_\alpha) \underline{\bs{\psi}}^{(q)}_{\alpha}  
		\cdot [\underline{\bs{F}}^{(q+1)}]_c \cdots [\underline{\bs{F}}^{(p)}]_c \right)
		\cdot \bs{n}
	\nn\\&\quad
	= \bs{F}^{(1)} \cdots \underline{\bs{F}}^{(q-1)}
		\cdot \left( \underline{\bs{F}}^{(q)} - [\underline{\bs{F}}^{(q)}]_c \right)
		\cdot [\underline{\bs{F}}^{(q+1)}]_c \cdots [\underline{\bs{F}}^{(p)}]_c
		\cdot \bs{n}.
\end{align}
Repeating this procedure for each fibre, one finds that $\vphi^\vee$ is indeed cohomologous to $[\vphi^\vee]_c$.

Inserting \eqref{eq:phic} into the formula for the intersection number one finds 
\be \label{eq:multivariate intersection number}
	\langle \vphi^\vee \vert \vphi \rangle
	= \frac{1}{2 \pi i} \int_{\F_p}  \bs{n}^T \cdot [\mat{F}^{(p)}]_c^T \cdot \bs{\vphi}^{(p)}
	= n_a \langle f^{(p)}_{ab} \vert \vphi^{(p)}_b \rangle_{_{\F_p}}
\ee
where 
\begin{align} \label{eq:phi(1)}
	\bs{\vphi}^{(1)} &= \frac{1}{2 \pi i} \int_{\F_{1}} [\bs{F}^{(1)}]_c^T\ \wedge \vphi 
		= \langle f^{(1)}_a \vert \vphi \rangle_{_{\F_1}},
	\\ \label{eq:phi(i)}
	\bs{\vphi}^{(i)} &= \frac{1}{2 \pi i} \int_{\F_{i}} [\underline{\bs{F}}^{(i)}]_c^T \cdot \bs{\vphi}^{(i-1)}
		= \langle f^{(i)}_{ba} \vert \vphi^{(i-1)}_b \rangle_{_{\F_i}},
\end{align}
and $\vphi$ is a generic Feynman integral. After fibration, the intersection number is realized recursively as a vector product of 1-dimensional intersection numbers. The vector-valued Feynman forms $\bs{\vphi}^{(i)}$ are forms on the base $\B_i$ \emph{not} $\F_i$ and are obtained by projecting $\bs{\vphi}^{(i-1)}$ onto the basis of $H_{\F_i}^1$. This formula is analogous to those derived in \cite{Mizera:2019gea,Frellesvig:2019uqt} -- the difference being that we have chosen a different fibration order. 

More concretely, the intersection number can be expressed as a series of consecutive residues. The top anti-holomorphic part of the dual form contains $p$-$\d\theta$ functions each of which take a residue when integrated over
\begin{align}
	\langle \vphi^\vee \vert \vphi \rangle
	&= (-1)^\frac{p(p-1)}{2} \int \sum_{\alpha_1,\dots,\alpha_p} 
		\d\theta_{z_1 = z_{1,\alpha_1}} \wedge  \cdots \wedge \d\theta_{z_p = z_{p,\alpha_p}} \wedge 
		\left(
			\bs{\psi}^{(1)}_{\alpha_1} \cdot \underline{\bs{\psi}}^{(2)}_{\alpha_2} 
			\cdots \underline{\bs{\psi}}^{(p)}_{\alpha_p} \cdot \bs{n}
		\right)^T \vphi
	\nonumber \\
	&= \sum_{\alpha_1,\dots,\alpha_p} 
		\res_{z_p = z_{p,\alpha_p}}  \cdots \res_{z_1 = z_{1,\alpha_1}} 
		\left[
			\left(
				\bs{\psi}^{(1)}_{\alpha_1} \cdot \underline{\bs{\psi}}^{(2)}_{\alpha_2} 
				\cdots \underline{\bs{\psi}}^{(p)}_{\alpha_p} \cdot \bs{n}
			\right)^T \hat{\vphi}
		\right]
\end{align}
where $\hat{\vphi}$ is the differential stripped version of $\vphi$.

\section{$d$-dimensional bubble coefficients} \label{sec:cbub4}

In this section we discuss the dual forms which extract bubble coefficients for planar 4-point massless amplitudes, testing out simple examples of dimension shifting identities. 
The 5-point case is a strightforward generalization of the 4-point case detailed here. 
Application to gluon amplitudes are discussed in the next section.

Section \ref{sec:conventions} fixes our conventions for 4-point integrals. 
A fibration for the massless 4-point problem is worked out in detail in section \ref{sec:4propfib}. 
Once the relation between the fibration basis and the UT dual form basis is know, the compactly supported version of the UT dual form basis is straightforwardly obtained. 
In section \ref{sec:orthobasis}, we intersect the UT dual form basis with a basis of Feynman forms and discover that the UT dual boxes are dual to combinations of Feynman integrals, which are finite. 
As the fist application of the formalism of section \ref{sec:higher form c-map}, we extract the generalized unitarity coefficients of dimension shifted integrals in section \ref{sec:dimshift}.

\subsection{Conventions for massless 4-particle kinematics \label{sec:conventions}}

The propagators are defined to be
\begin{align} 
	\label{eq:D1}
	D_1 &= \ell_\perp^2 + \ell^2, 
	\\
	\label{eq:D2} 
	D_2 &= \ell_\perp^2 + (\ell + p_1)^2,
	\\
	\label{eq:D3}
	D_3 &= \ell_\perp^2 + (\ell + p_1 + p_2)^2,
	\\
	\label{eq:D4}
	D_4 &= \ell_\perp^2 + (\ell + p_1 + p_2 + p_3)^2.
\end{align}
We use the 4-point rational parameterization of Appendix \ref{app:4pt param} for the external momenta and cartesian coordinates for the loop momentum 
\begin{align} \label{eq:coords}
	\ell_\mu = \sqrt{-s} \begin{pmatrix} z_2 & z_4 & i z_3 & z_1 & \bs{z}_\perp  \end{pmatrix}.
\end{align}
Here, the $z_i$ are dimensionless variables and the strange numerical labeling is connected to the fibration of section \ref{sec:4propfib}. Moreover, defining $\ell_3$ to be purely imaginary ensures that the propagators do not contain explicit factors of $i$ but introduces a factor of $i$ into the integration measure of both dual and non-dual forms (this will cancel in the end).  

Specializing to the 13-bubble cut, sets $z_\perp^2 = \frac{1}{4} - z_2^2 - z_3^2 + z_4^2$ and $z_1 = 1/2$. Then, on this cut, the remaining boundaries become
\begin{align} \label{eq:13boundaries}
	D_2\vert_{13} = -s \left( \frac12 + z_2 \right),
	\quad
	D_4\vert_{13} = -s \left(\frac12 + (1+2x) z_2 -z_3 + (1+2x) z_4 \right),
\end{align}
where $x=t/s$ 
(recall that $s = -(p_1 +p_2)^2$ and $t=-(p_2+p_3)^2$ are the 4-point Mandelstam invariants). 
In these coordinates, the twist is hyperboloidal instead of spherical: 
\be \label{eq:u0}
	u^{(0)} \equiv u\vert_{13}= \left( s Q^{(0)} \right)^\vep,
	\qquad 
	Q^{(0)} = \frac{1}{4} - z_2^2 - z_3^2 + z_4^2 .
\ee
In order to avoid unnecessary square roots and factors of $i$, we keep the hyperboloidal twist instead of the spherical twist of \eqref{eq:tadpoles}-\eqref{eq:pentagon} and \cite{Caron-Huot:2021xqj}. With these conventions 13-bubble dual form becomes
\begin{align} \label{eq:bubdual}
	\vphi^\vee_{13} = \delta_{13} (\phi^\vee_{13}),
	\qquad
	\phi^\vee_{13} = -\frac{ic_2}{2} \frac{ \d z_2 \wedge \d z_3 \wedge \d z_4}{(\frac{1}{4} - z_2^2 - z_3^2 + z_4^2 )^2},
	\qquad 
	c_2 = 8 (1-\vep).
\end{align}

To complete the basis on the 13-cut, the 1234-box dual form must be included. On the 1234-cut, the twist factorizes into a product of hyperplanes
\begin{align}
	u\vert_{1234} = \left( s Q^{(0)} \vert_{24} \right)^\vep
	\qquad 
	Q^{(0)} \vert_{24} =  x (2z_4-1) (2(1+x)z_4-x).
\end{align}
Up to an overall kinematic factor, the box dual form is simply the volume form on the 1234-cut divided by some power of the twist. The exact kinematic factor can be obtained from \eqref{eq:boxes} by a change of coordinates and we define
\begin{align} \label{eq:boxdual}
	\vphi^\vee_{1234} = \delta_{1234}(\phi^\vee_{1234}),
	\qquad 
	\phi^\vee_{1234} 
		= ic_0 \frac{\d z_4}{(2z_4-1) (2(1+x)z_4-x)},
	\qquad
	c_0 = \frac{4}{\vep} 
\end{align}
as the 1234-box dual form.

\subsection{Massless 4-point fibration \label{sec:4propfib}}

In order to use the fibration c-map of Section \ref{sec:higher form c-map}, we need to first specify a fibration. The dimension of the fibre cohomology greatly depends on the choice of coordinates and ordering of fibres. For example, ${\rm dim}H_{\F_1}^1=3$ if $z_2$ is chosen as the first fibre (one bulk and two boundary basis forms) while ${\rm dim}H_{\F_1}^1=2$ if either $z_3$ or $z_4$ are chosen (one bulk and one boundary basis form). Alternatively, we could change coordinates such that $z_3$ no longer appears in the propagators. Then, ${\rm dim}H_{\F_1}^1=1$ if $z_3$ is chosen as the first fibre. Since such choices are specific to the problem at hand, we will proceed naively in our chosen coordinates and choose $z_2$ as the first fibre, $z_3$ as the second and $z_4$ as the third and last fibre.

\subsubsection{Fibre 1}
The first fibre is $\F_1 = z_2 \in \mathbb{C}\setminus\{u^{(0)}=0\}$ and we choose the $\d\log$ basis\footnote{While the Leray forms are not technically $\d\log$'s, we will still call them logarithmic since they are dual to $\d\log$'s.}
\begin{align} \label{eq:f1(1)}
	f_1^{(1)} 
	= \frac{2\vep (\frac{1}{4}-z_3^2+z_4^2)}{\frac{1}{4}-z_2^2-z_3^2+z_4^2} \d z_3,
	\quad
	f_2^{(1)} = \delta_2(1),
	\quad
	f_3^{(1)} = \delta_4(1),
\end{align}
for the fibre cohomology $H^1_{\F_1}$. Next, we determine how the covariant derivative of the total space acts on the fibre cohomology in order to obtain the connection on $\B_1$. The total space covariant derivative acts simply on the Leray forms:
\begin{align}
	\nabla^\vee f_2^{(1)} = -\delta_{2}(\omega^\vee \vert_2) 
	= -f_3^{(1)} \wedge \omega_{22}^{(1)},
	\qquad 
	\omega^{(1)}_{22} = \vep\ \d\log Q_{22}^{(1)},
	\qquad 
	Q^{(1)}_{22} = Q^{(0)}\vert_2
	\\
	\nabla^\vee f_3^{(1)} = -\delta_{4}(\omega^\vee \vert_4) 
		= -f_3^{(1)} \wedge \omega_{33}^{(1)},
	\qquad 
	\omega^{(1)}_{33} = \vep\ \d\log Q_{33}^{(1)},
	\qquad 
	Q^{(1)}_{33} = Q^{(0)}\vert_4
\end{align}
where the $Q_{aa}^{(1)}$ are the quadrics appearing in the boundary twists ($u^{(1)}_{aa} =  (Q_{aa}^{(1)})^\vep$, which are the diagonal components of the matrix twist $\mat{u}^{(1)}$). Since our boundaries are non-generic (massless external and internal kinematics), the boundary twist factorizes into a product of hyperplanes
\begin{align}
	Q_{22}^{(1)} &=  (-z_3 {+} z_4) (z_3 {+} z_4),
	\\
	Q_{33}^{(1)} &= 
	-\frac{
		(2z_3{-}1) (x(1{+}x) + (1{+}2x{+}2x^2) z_3 {-} (1{+}2x) z_4)
	}{(1+2x)^2}.
\end{align}

The covariant derivative of the bulk-from, $f_1^{(1)}$, is slightly more complicated since we must add the covariant derivative of an IBP-form in order to project $\nabla^\vee f_1^{(1)}$ onto the $H^1_{\F_1}$ basis:
\begin{align}
	\nabla^\vee f_1^{(1)} 
	\simeq \nabla^\vee f_1^{(1)} + \nabla^\vee V^{(1)}_1
	= - f_1^{(1)} \wedge \omega_{11}^{(1)} - f_2^{(1)} \wedge \omega_{21}^{(1)} - f_3^{(1)} \wedge \omega_{21}^{(1)}
\end{align}
where 
\begin{align}
	V^{(1)}_{1}
	= - \vep \frac{z_2\ \d Q_{11}^{(1)}}{\frac14-z_2^2-z_3^2+z_4^2},
	\qquad 
	Q_{11}^{(1)} = \frac14{-}z_3^2{+}z_4^2,
\end{align}
and 
\begin{align}
	\omega_{11}^{(1)} &= (\vep+1/2)\ \d\log Q_{11}^{(1)},
	\\	
	\omega_{21}^{(1)} &= V_1^{(1)} \vert_2 = -\frac{\vep}{2} \d\log Q_{22}^{(1)},
	\\ 
	\omega_{31}^{(1)} &= V_1^{(1)} \vert_4 = \frac{\vep}{2(1{+}2x)Q_{33}^{(1)} }
 		\bigg[ 
			\left( - 1 + 2 z_3 + 4(1+2x) z_3 z_4 - 4 z_4^2 \right)	\d z_3
		\nn\\&\hspace{12em} 
			-\left(2 z_3-1\right) \left( 1+2x + 2(1+2x+) z_3 - 2z_4 \right)  \d z_4
		\bigg].
\end{align}
While the off-diagonal connections $\omega^{(1)}_{1a}$ may not be $\d\log$, they share the same singularities as the corresponding diagonal component $\omega^{(1)}_{aa}$.\footnote{This is only true if the basis forms are normalized in a specific way. More on this later.} Alternatively, one can compute $\omega^{(1)}_{ab}$ using single variable intersection numbers on $\F_1$. However, we will not show this here as it involves introducing a space dual to the (dual form) fibre cohomology.

To understand how the basis bulk forms ($f^{(1)}_1$ in this case) are chosen, suppose that the twist  $u^{(0)}=Q^\vep$ is a function some generic quadric $Q^{(0)}$. Assuming that $z_2$ is still the fibre variable, we can construct a $\d\log$ form from the $z_2$-roots of $Q^{(0)}$:
\be \label{eq:bulk forms}
	f_1^{(1)} 
	= \frac{\vep}{2}\ \sqrt{Q^{(1)}_{11}}\ \d\log\left(\frac{z_2-r_1}{z_2-r_2}\right)
	= 2\vep\ \frac{Q^{(1)}_{11}\ \d z_2}{Q^{(0)}},
	\qquad 
	Q^{(1)}_{11} = \frac14 \disc_{z_2}Q^{(0)}
\ee
where $\text{Disc}_{z_2}Q^{(0)}$ is the discriminant of $Q^{(0)}$ with respect to the variable $z_2$ and $r_i$ are the $z_2$-roots of $Q^{(0)}$. Despite the apparent square root, equation \eqref{eq:bulk forms} is always algebraic since the square root cancels once the $\d\log$ is expanded. When chosen as a basis element, the bulk form \eqref{eq:bulk forms} produces $(\vep+\frac12)\ \d\log Q^{(1)}_{11}$ as its diagonal component in the base connection. The connection $\omega_{11}$ introduces singularities at the roots of $Q^{(1)}_{11}$ on the base $\B_1$. However, on the full manifold, these singularities are not allowed (except at $z_2=0$). 
Indeed, the wedge product $f^{(1)}_1 \wedge \omega^{(1)}_{11}$ is free of spurious singularities since \eqref{eq:bulk forms} contains a factor of $\disc_{z_2}Q^{(0)}$ in the numerator. For this reason, the normalization of \eqref{eq:bulk forms} is particularly natural.\footnote{This normalization is also natural from the perspective of single variable intersection numbers. Since the intersection of identical $\d\log$ forms is proportional to the inverse discriminant,  $\la f^{(1)}_1 \vert f^{(1)}_1 \ra_{\F_1} \propto 1/Q^{(1)}_{11}$, normalizing the dual-basis brings a factor of the discriminant into the numerator of $f^{(1)}_1$.}

While convenient, normalizing \eqref{eq:bulk forms} by the discriminant is not necessary. For example, $\d z_{2}/Q^{(0)}$ could be used as a basis form on $\F_1$ instead. However, the off-diagonal connections would then contain singularities that do not appear in the some of the diagonal connections. This makes it impossible to find primitives for a certain class of base forms. Therefore, base forms must contain enough zeros at the bulk singularities introduced by the off-diagonal connections such that a local primitive always exists.
 
The last step on $\F_1$ is to construct the improved basis forms by combining them with their corresponding IBP form:
\begin{align}
	\nabla^\vee F^{(1)}_1 = - F^{(1)}_1 \wedge \omega^{(1)}_{11},
	\qquad 
	F_1^{(1)}=f_1^{(1)}-V_1^{(1)}.
\end{align}
The remaining basis forms $f^{(1)}_{2,3}$ are already in the desired form: $F^{(1)}_{2,3}=f^{(1)}_{2,3}$. Summarizing, the improved basis satisfies
\be
	\nabla^\vee \bs{F}^{(1)} = - \bs{F}^{(1)} \cdot \mat{\omega}^{(1)}
\ee 
where 
\be
	\mat{\omega}^{(1)} = 
	\begin{pmatrix}
		(\vep{+}\frac12)\ \d\log Q^{(1)}_{11} & 0 & 0
		\\
		V^{(0)}_1\vert_2 & \vep\ \d\log Q^{(1)}_{22} & 0 
		\\
		V^{(0)}_1\vert_4 & 0 & \vep\ \d\log Q^{(1)}_{33}
	\end{pmatrix}.
\ee

\subsubsection{Fibre 2}
Now, we can specify the $\B_1$ cohomology. 

Following the discussion at the end of section \ref{sec:fibration-gen}, a generic form on $\B_1$ is a 3-vector where each component belongs to the twisted cohomology of the corresponding diagonal of the connection 
such that the wedge product with the vector valued form $\bs{F}^{(1)}$ produces a valid element of the full un-fibred cohomology. 
The connection on $\B_1$ is given by 
\be
	\nabla^{(1)} \bs{\phi}^\vee_{\B_1} 
	= (d  + \mat{\omega}^{(1)} - \mat{\L}^{(1)}) \cdot \bs{\phi}^\vee_{\B_1}.
\ee
where \eqref{eq:leray matrix} becomes 
\be
	\mat{\L}^{(1)} 
	= \begin{pmatrix} 
		0 & 0 & 0
		\\ 
		0 & 0 & 0
		\\
		0 & -\delta_2(\ \vert_4) & \delta_2(\ \vert_2)
	\end{pmatrix}.
\ee

Having understood the $\B_1$ covariant derivative, we can define the second fibre $\F_{2} = z_3 \in \mathbb{C} \setminus \text{poles}(\mat{\omega}^{(1)})$, which has covariant derivative $\nabla_{\F_2} = \nabla^{(1)}\vert_{\d z_2=0}$. The $\F_2$-cohomology is 4-dimensional: one boundary form generated by $\mat{\L}^{(1)}$ and three bulk forms generated by the diagonal elements of $\mat{\omega}^{(1)}$. For the basis of the boundary cohomology, we choose the Leray form
\be
	\bs{f}^{(2)}_4 =  \delta_2\left(\begin{pmatrix}0&0&1\end{pmatrix}^T\right). 
\ee
Acting with $\nabla^{(1)}$, we can read off the component $\omega^{(2)}_{44}$ of the $\B_2$ connection
\be
	\nabla^{(1)} \bs{f}^{2}_4 = - \bs{f}^{2}_4 \wedge \omega^{(2)}_{44},
	\qquad 
	\omega^{(2)}_{44} = \omega^{(1)}_{33}\vert_2 = \vep\ \d\log Q^{(2)}_{44},
	\qquad 
	Q^{(2)}_{44} = Q^{(0)}\vert_{24}.
\ee
Mimicking \eqref{eq:bulk forms}, we choose the $\d\log$ forms
\begin{align}
	\bs{f}^{(2)}_1 &= 2\left(\vep+\frac12\right) \frac{Q^{(2)}_{11} \d z_3}{Q^{(1)}_{11}} 
		\begin{pmatrix}1&0&0\end{pmatrix}^T,
	\qquad 
	Q^{(2)}_{11} = \frac14 \disc_{z_3}Q^{(1)}_{11},
	\\
	\bs{f}^{(2)}_2 &=  2\vep\ \frac{Q^{(2)}_{22} \d z_3}{Q^{(1)}_{22}}
		\begin{pmatrix}0&0&1\end{pmatrix}^T,
	\qquad 
	Q^{(2)}_{22} = \frac14 \disc_{z_3}Q^{(1)}_{22},
	\\
	\bs{f}^{(2)}_3 &=  2\vep\ \frac{Q^{(2)}_{33} \d z_3}{Q^{(1)}_{33}} 
		\begin{pmatrix}0&0&1\end{pmatrix}^T,
	\qquad 
	Q^{(2)}_{33} = \frac14 \disc_{z_3}Q^{(1)}_{33},
\end{align}
as a basis for the bulk cohomology. Using the IBP-forms 
\begin{align}
	\bs{V}^{(2)}_1 &= V^{(2)}_1 \begin{pmatrix}1&0&0\end{pmatrix}^T,
	\qquad
	V^{(2)}_1 = - \left(\vep+\frac12\right) 
		\frac{z_3\ \d Q^{(2)}_{11}}{Q^{(1)}_{11}}
	\\
	\bs{V}^{(2)}_2 &= V^{(2)}_2 \begin{pmatrix}0&1&0\end{pmatrix}^T,
	\qquad 
	V^{(2)}_2 = - \vep \frac{z_3\ \d Q^{(2)}_{22}}{Q^{(1)}_{22}}
	\\
	\bs{V}^{(2)}_3 &= V^{(2)}_3 \begin{pmatrix}0&0&1\end{pmatrix}^T,
	\qquad V^{(2)}_3 = - \vep \frac{(z_3{-}\frac12)\ \d Q^{(2)}_{33}}{Q^{(1)}_{33}},
\end{align}
it is fairly easy to check that 
\begin{align}
	\nabla^{(1)} \bs{f}^{(2)}_a 
	\simeq \nabla^{(1)} \bs{f}^{(2)}_a + \nabla^{(1)} \bs{V}^{(2)}_a 
	= - \bs{f}^{(2)}_b \wedge \omega^{(2)}_{ba}
\end{align}
where 
\begin{align}
	\mat{\omega}^{(2)}
	=\begin{pmatrix}
		(\vep{+}1)\ \d\log Q^{(2)}_{11} & 0 & 0 & 0 
		\\
		-\frac12 (\vep{+}\frac12)\ \d\log Q^{(2)}_{22} & (\vep{+}\frac12)\ \d\log Q^{(2)}_{22} & 0 & 0 
		\\
		-\frac12 (1{+}2x) (\vep{+}\frac12)\ \d\log Q^{(2)}_{33} & 0 & (\vep{+}\frac12)\ \d\log Q^{(2)}_{33} & 0 
		\\
		0 & V^{(2)}_2\vert_4 & V^{(2)}_3\vert_2 & \vep\ \d\log Q^{(2)}_{44} 
	\end{pmatrix}
\end{align}
Note that the $\d\log$ basis is somewhat special since the $\bs{V}^{(2)}_{1,2,3}$ are unit vectors multiplied by a 1-form. Generically, the other components of $\bs{V}^{(2)}_{1,2,3}$ are complicated if $\bs{f}^{(2)}_{1,2,3}$ is not a $\d\log$ form. The last step on $\F_2$ is to construct the improved basis forms
\begin{align}
	\bs{F}^{(2)}_{a=1,2,3} = \bs{f}^{(2)}_{a=1,2,3} - \bs{V}^{(2)}_{a=1,2,3},
	\qquad 
	\bs{F}^{(2)}_{4} = \bs{f}^{(2)}_{4},
\end{align}
such that $\nabla^{(1)} \bs{F}^{(2)}_{a} = -\bs{F}^{(2)}_{b} \wedge \omega^{(2)}_{ba}$.

The matrix structure of $\mat{\omega}^{(2)}$ describes 4 different scenarios. The first row represents localization on the twisted singularities of $\F_2$ after localizing on the twisted singularities of $\F_1$ while the second and third row represents localization on the twisted singularities of $\F_2$ after localizing on the boundaries $D_2\vert_{13}$ or $D_4\vert_{13}$ on $\F_1$. The last row represents localization on the double boundary $D_2\vert_{13}=D_4\vert_{13}=0$. 

\subsubsection{Fibre 3}
With $\mat{\omega}^{(2)}$ in hand, we define the second base which is also the third and last fibre $\B_2 = \F_3 = \mathbb{C}\setminus\{\text{poles}(\mat{\omega}^{(2)})\}$. Since $\mat{\omega}^{(2)}$ is a $4\times4$ matrix, the elements of $\B_2$ cohomology are 4-vectors whose $a^\text{th}$-component belongs to the twisted cohomology generated by $\omega^{(2)}_{aa}$. Given a generic element of the base cohomology $\bs{\phi}^\vee_{\B_2}$, the covariant acts as
\be
 	\nabla^{(2)} \bs{\phi}^\vee_{\B_2} = (d + \mat{\omega}^{(2)} \wedge ) \cdot \bs{\phi}^\vee_{\B_2}
\ee
since there are no boundaries on $\B_2$. 

Having understood the action of the covariant derivative on the last fibre, we can choose a basis for the fibre cohomology $H^1_{\F_3}$. Since there are no boundaries, the basis forms are bulk forms generated by the diagonal components of $\mat{\omega}^{(2)}$. However, the diagonal components $\omega^{(2)}_{aa}$ for $a=2,3$ are degenerate since $\disc_{z_4}Q^{(2)}_{aa} = 0$. Since $\omega^{(2)}_{aa} \propto \d\log(\text{linear polynomial of } z_4)^2$ for $a=2,3$ there is not enough singularities to generate a non-trivial twisted cohomology class. Physically, this corresponds to the vanishing of triangle integrals in the massless limit. The triangle-dual components ($a=2,3$) of $\F_3$ form can always be reduced to the box-dual component ($a=4$). For example, consider the form on the 123-triangle cut
\be	\label{eq:tritobereduced}
	\d z_4 \begin{pmatrix}0&1&0&0\end{pmatrix}^T.
\ee 
Subtracting the total derivative 
\be
	\nabla^{(2)} \d z_4 
	\begin{pmatrix}
		0 
		& \frac{z_4}{2(\vep+1)}
		& 0
		& \frac{(1{+}2x)(1{+}\vep)+4(1+x)\vep z_4}{16(1+x)^2(1+\vep)(1+2\vep)}
	\end{pmatrix}^T
\ee
from \eqref{eq:tritobereduced}, one finds that \eqref{eq:tritobereduced} is cohomologous to 
\be
	- \frac{1+2x+2x^2+\vep+4x\vep+4x^2\vep}{16x(1+x)^2(1+\vep)(1+2\vep)} 
	\bs{f}^{(3)}_2
	.
\ee
where $\bs{f}^{(3)}_2$ is defined below. Thus, $\text{dim}H^1_{\F_3}=2$ and we choose the basis 
\begin{align}
	\bs{f}^{(3)}_1 &= 2(\vep+1) \frac{Q^{(3)}_{11}\ \d z_4}{Q^{(2)}_{11}} \begin{pmatrix}1&0&0&0\end{pmatrix}^T, 
	\qquad 
	Q^{(3)}_{11} = \frac14 \disc_{z_4} Q^{(2)}_{11} = - \frac14,
	\\ 
	\bs{f}^{(3)}_2 &= 2\vep \frac{Q^{(3)}_{44}\ \d z_4}{Q^{(2)}_{44}} \begin{pmatrix}0&0&0&1\end{pmatrix}^T, 
	\qquad 
	Q^{(3)}_{44} = \frac14 \disc_{z_4} Q^{(2)}_{44} = x^2.
\end{align}

\subsubsection{Basis for full cohomology}
A basis for the full cohomology on the 13-bubble cut is given by the columns of the vector 
\begin{align} 
	\bs{\vphi}^\vee_f 
	&=\bs{F}^{(1)} \cdot \underline{\bs{F}}^{(2)} \cdot \underline{\bs{F}}^{(3)}
	\nn \\
	&= \begin{pmatrix}
		-\vep(1+\vep)(1+2\vep)		
		\frac{\d z_2 \wedge \d z_3 \wedge \d z_4}{Q^{(0)}}
		&\quad
		-2\vep x^2 \delta_{24}\left(\frac{\d z_4}{Q^{(0)}\vert_{24}}\right)
	\end{pmatrix}
	\nn \\
	&= \begin{pmatrix}
		-\vep(1+\vep)(1+2\vep)	
		\frac{\d z_2 \wedge \d z_3 \wedge \d z_4}{Q^{(0)}}
		&\quad
		\frac{2ix\vep}{c_0} \vphi^\vee_{1234}
	\end{pmatrix}.
	\label{eq:fibbasis}
\end{align}
While the second column is directly proportional to the box dual form, the logarithmic basis form in the first column still needs to be related to the 13-bubble dual form. Using integration-by-parts, we find that 
\begin{align} \label{eq:UTduals}
	\begin{pmatrix}
		\vphi^\vee_{13}
		&
		\vphi^\vee_{1234}
	\end{pmatrix}
	=  \bs{\vphi}^\vee_f \cdot \mat{N}
\end{align}
where
\begin{align}
	\label{eq:N}
	\mat{N} 
	= \begin{pmatrix}
		\frac{ic_2}{(1+\vep)\vep(\vep-1)} 
			& 0 \\
		-\frac{ic_2}{4x\vep^2(\vep-1)} 
			& -\frac{ic_0}{2x\vep} 
	\end{pmatrix}
	.
\end{align}
The conversion $\mat{N}$ is non-diagonal since we are projecting the non-logarithmic uniform transcendental bubble \eqref{eq:bubdual} onto the logarithmic fibration basis \eqref{eq:fibbasis}. While choosing a non-$\d\log$ fibration basis could help diagonalize $\mat{N}$ it also makes the fibration more complicated.

The intersection numbers of a given Feynman integrand $\vphi$ with the basis of uniform transcendental dual forms (\eqref{eq:bubdual} and \eqref{eq:boxdual}) is simply
\begin{align} \label{eq:UTint}
	\begin{pmatrix}
		\la \vphi^\vee_{13} \vert \vphi \ra
		&
		\la \vphi^\vee_{1234} \vert \vphi \ra
	\end{pmatrix}
	= \la \bs{\vphi}^\vee_f  \vert \vphi \ra \cdot \mat{N} \ .
\end{align}
Moreover, the compactly supported versions of our basis on the 13-cut are 
\begin{align}
	\begin{pmatrix}
		[\vphi^\vee_{13}]_c
		&
		[\vphi^\vee_{1234}]_c
	\end{pmatrix}
	=  [\bs{\vphi}^\vee_f]_c \cdot \mat{N},
	\qquad 
	[\bs{\vphi}^\vee_f]_c
	= [\bs{F}^{(1)}]_c \cdot [\underline{\bs{F}}^{(2)}]_c 
		\cdot [\underline{\bs{F}}^{(3)}]_c.
\end{align}

\subsection{Intersection matrix on 13-cut \label{sec:orthobasis}}

To extract the coefficients of the bubble and box Feynman integrals, the intersection matrix of the fibration basis \eqref{eq:fibbasis} with a basis of Feynman forms is needed. 
We choose our initial basis of Feynman forms to be the UT basis of \cite{Henn:2014qga}
\begin{align}
	\label{eq:phi naive}
	\bs{\vphi}_{\rm naive} 
	= \begin{pmatrix} \vphi_{\text{bub}_s} & \vphi_\text{box} \end{pmatrix},
	\quad
	\vphi_{{\rm bub}_s} = \vep (2\vep{-}1)  \frac{\d^4\ell \wedge \d\ell_\perp^2}{\ell_\perp^2 D_1 D_3},
	\quad 
	\vphi_{{\rm box}} = st\vep^2 \frac{\d^4\ell \wedge \d\ell_\perp^2}{\ell_\perp^2 D_1 D_2 D_3 D_4}, 
\end{align}
and will discover that the UT dual forms (\eqref{eq:bubdual} and \eqref{eq:boxdual}) extract the coefficients corresponding to linear combinations of \eqref{eq:phi naive} such that the new box integral is finite!\footnote{Note that $(2\vep{-}1) \vep G_{1,0,1,0}$ is the uniform transcendental bubble and is cohomologous to $s \vep G_{1,0,2,0}$ of \cite{Henn:2014qga}.}

We begin by computing the intersection matrix for the dual form fibration basis \eqref{eq:fibbasis} and Feynman integrand basis \eqref{eq:phi naive}
\be
	(\mat{C}_{\rm fibre})_{ab} \equiv \la (\vphi^\vee_f)_a \vert \vphi_b \ra.
\ee
These intersection numbers are computed recursively starting from \eqref{eq:phi(1)}. Projecting $\bs{\vphi}_{\rm naive}$ onto the $\F_1$-basis, we find
\begin{align}
	\bs{\vphi}^{(1)}_{\rm naive}  &= 
	\begin{pmatrix}
		\vphi^{(1)}_{{\rm bub}_s}
		&
		\vphi^{(1)}_{{\rm box}}
	\end{pmatrix}
	\nn \\
	&= \begin{pmatrix}
		\frac{i}{2} \vep (2\vep-1)
		& \frac{ix\vep^2}{(1+2x)^2}
		\frac{
				 (1{+}x {-} z_3 - 2(1{+}2x) z_3^2 {+} (1{+}2x) z_4 {+} 2(1{+}2x) z_4^2)
			}{
				4Q^{(1)}_{22}Q^{(1)}_{33}
			}
		\\
		0 & -\frac{isx\vep^2}{2 Q^{(1)}_{22} D_4\vert_2}
		\\ 
		0 & \frac{isx\vep^2}{2 Q^{(1)}_{33} D_2\vert_4}
	\end{pmatrix} \d z_3 \wedge \d z_4 .
	\label{eq:phi0}
\end{align}
Then, following \eqref{eq:phi(i)}, the columns of $\bs{\vphi}^{(1)}_{\rm naive}$ are projected onto the $\F_2$-basis 
\begin{align}
	\mat{\vphi}^{(2)}_{\rm naive}  &= 
	\begin{pmatrix}
		\bs{\vphi}^{(2)}_{{\rm bub}_s}
		&
		\bs{\vphi}^{(2)}_{{\rm box}}
	\end{pmatrix}
	 = \begin{pmatrix}
		\frac{i(2\vep+1) (2\vep-1) Q^{(2)}_{11} }{4} & 0 
		\\
		0 & -\frac{ix \vep^2(-x{+}(1{+}2x) z_4)}{2Q^{(2)}_{44}}
		\\
		0 & -\frac{ix \vep^2 (1{+}2x{+}4x^2{-}2(1{+}2x)^2 z_4)}{4 (1{+}2x) Q^{(2)}_{44}}
		\\
		0 & \frac{ix\vep^2}{2 Q^{(2)}_{44}}
	\end{pmatrix} \d z_4
\end{align}
Then the intersection matrix is obtained by projecting the columns of $\mat{\vphi}^{(2)}_{\rm naive}$ onto the $\F_3$-basis 
\begin{align}
	\mat{\vphi}^{(3)}_{\rm naive}  &= 
	\begin{pmatrix}
		\bs{\vphi}^{(3)}_{{\rm bub}_s}
		&
		\bs{\vphi}^{(3)}_{{\rm box}}
	\end{pmatrix}
	 =
	 \begin{pmatrix}
		\frac{i(1+\vep)\vep}{16} & 0 
		\\
		0 & \frac{ix \vep^2}{2}
	\end{pmatrix}
	=\begin{pmatrix}
		\la \vphi^\vee_{f,1} \vert \vphi_{{\rm bub}_s} \ra  
			& \la \vphi^\vee_{f,1} \vert \vphi_{{\rm box}} \ra  
		\\
		\la \vphi^\vee_{f,2} \vert \vphi_{{\rm bub}_s} \ra  
			& \la \vphi^\vee_{f,2} \vert \vphi_{{\rm box}} \ra 
	\end{pmatrix}
	=\mat{C}_{\rm naive}.
	\label{eq:Cnaive}
\end{align}
With $\mat{C}_{\rm naive}$ in hand, the UT bubble/box \eqref{eq:phi naive} coefficients can be computed for any Feynman integrand $\vphi$: 
\begin{align}
	\begin{pmatrix} c_{{\rm bub}_s}[\vphi] \\ c_{{\rm box}}[\vphi] \end{pmatrix}
	= \mat{C}_{\rm  fibre}^{-1} \cdot \bs{\vphi}^{(3)}_{{\rm naive}}.
\end{align}
Note that it is surprising that the intersection matrix above is diagonal since the double pole in the bubble dual form (\eqref{eq:bubbles} or equivalently \eqref{eq:bubdual}) was needed to establish duality with the standard UT basis of Feynman forms \cite{Caron-Huot:2021xqj}. However, in the massless degeneration, the bubble dual forms are dual to a linear combination of Feynman forms instead. 


To find the basis of Feynman forms dual to \eqref{eq:bubdual} and \eqref{eq:boxdual}, we look for a basis of integrands, which we refer to as $\bs{\vphi}$, such that 
\begin{align} \label{eq:cmat norm}
	C_{ij}  
	= \la \vphi^\vee_i \vert \vphi_j \ra 
	= N^T_{ik} \la (\vphi^\vee_{f})_k \vert \vphi_j \ra
	= \delta_{ij}.
\end{align}
It follows that the integrands dual to \eqref{eq:bubdual} and \eqref{eq:boxdual} are
\begin{align}
	\label{eq:phi finite}
	\bs{\vphi} 
	= \bs{\vphi}_\text{nice}
	\equiv \begin{pmatrix}
		\vphi_{13}
		\\
		\vphi_{ 1234}
	\end{pmatrix}
	= \left(\mat{C}_\text{fibre}^T \cdot \mat{N}\right)^{-1} \cdot \bs{\vphi}_{\rm naive}   
	= \begin{pmatrix}
		2 \vphi_{{\rm bub}_s} \\
		\vphi_{{\rm box}} + 2 \vphi_{{\rm bub}_s}
	\end{pmatrix},
\end{align}
up to forms with vanishing $13$-residue. Performing the analogous calculation on the $24$-cut completely fixes $\vphi_{1234}$ (see \eqref{eq:4pt box}).

Interestingly, the UT basis of dual forms has picked a particularly nice basis of Feynman integrals where $\vphi_{1234}$ integrates to something finite
(by ``finite'', we mean that the integrated expression is not more singular than the integrand).
Even without knowing the part of $\vphi_{1234}$ that vanishes on the $13$-cut, we can check that the discontinuity of $\mathscr{I}[\vphi_{1234}]$ across the $13$-cut is finite
\begin{align}
	\disc_{13}\mathscr{I}[\vphi_{1234}]
	& = \disc_{13} \Big[ 
		\mathscr{I}[\vphi_{\text{box}}] 
		+ \mathscr{I}[2\vphi_{\text{bub}_s}] 
	\Big]
	= 4\vep^2 \log(t/s).
\end{align}
where 
\begin{align}
	\disc_{13}\mathscr{I}[\vphi_{\text{bub}_s}]
	& = \vep - \vep^2 \log(-s) + \mathcal{O}(\vep^3),
	\\
	\disc_{23}\mathscr{I}[\vphi_{\text{box}}]
	& = -2\vep + 2\vep^2 \log(-t) + \mathcal{O}(\vep^3),
\end{align}
and $\disc_{13}$ is $\frac{1}{2\pi i}$ times the discontinuity across the $13$-channel.
This can also be verified by taking the 13-channel discontinuity of \eqref{eq:Ibox massless} directly.


The corresponding $\bs{\vphi}$ coefficients are simply 
\begin{align}
	\begin{pmatrix} c_{13}[\vphi] \\ c_{1234}[\vphi] \end{pmatrix}
	= \mat{C}^{-1} \cdot \la \bs{\vphi}^\vee \vert \vphi \rangle
	= \mat{C}^{-1} \cdot \mat{N}^T \cdot \la \bs{\vphi}^\vee_f \vert \vphi \ra
	= \mat{C}^{-1} \cdot \mat{N}^T \cdot \bs{\vphi}^{(3)}
\end{align}
for any given FI $\vphi$. Since we only know the compactly supported versions of the fibration dual basis, $\bs{\vphi}^\vee$ has to be expressed in terms of $\bs{\vphi}^\vee_f$.

%

\subsection{Simple check: dimension shifting identities\label{sec:dimshift}}

\begin{table}
	\centering
	\bgroup
	\def\arraystretch{1.5}
	\begin{tabular}{c|cccc}
	\hline
		\multicolumn{5}{c}{
				$\vphi = \frac{(\ell_\perp^2)^{m-1} \d\ell_\perp^2 \wedge \d^{4}\ell}{D_1 D_3}$
			} 
		\\[.4em]
		\hline 
		$4+2m$ & 4 & 6 & 8 & 10
		\\
		\hline
		$c_{4+2m,4}[\vphi^\vee_{\text{bub}_s},\vphi]$ 
			& 1 
			& $\frac{s}{2(3-2\vep)}$ 
			& $\frac{s^2 \vep (\vep-1)}{4(5-2\vep)(3-2\vep)}$ 
			& $\frac{s^3 \vep (\vep-1) (\vep-2)}{8 (7- 2 \vep ) (5 - 2 \vep) (3-2 \vep)}$
		\\
		$c_{4+2m,4}[\vphi^\vee_{\text{box}},\vphi]$ 
			& 0  
			& 0 
			& 0 
			& 0
	\end{tabular}
	\egroup 
	\caption{ \label{tab:hdbub}
		The projection of the naive (non-UT) $s$-channel bubble in 4, 6, 8 and 10 dimensions onto 
		the 4-dimensional UT basis.
	}
\end{table}

\begin{table}
	\centering
	\bgroup
	\def\arraystretch{1.5}
	\begin{tabular}{c|cccc}
		\hline
		\multicolumn{5}{c}{
				$\vphi = \frac{(\ell_\perp^2)^{m-1} \d\ell_\perp^2 \wedge \d^{4}\ell}{D_1 D_2 D_3 D_4}$
			} 
		\\[.4em]
		\hline
		$4+2m$ & 4 & 6 & 8 & 10
		\\
		\hline
		$c_{4{+}2m,4}[\vphi^\vee_{\text{bub}_s},\vphi]$ 
			& 1 
			& $\frac{-1}{s (1{+}x) \vep^2 (2\vep{-}1)}$ 
			& $\frac{-x + (1+2x) \vep }{2 (1+x)^2 \vep^2 (\vep{-}1) (2\vep{-}1) (2\vep{-}3)}$
			& $\frac{
					-s \left(
						2 x^2
						- (1 + 4 x + 6 x^2 ) \vep 
						+ (1 + 3 x + 3 x^2 ) \vep^2
					\right)
				}{
					4 (1+x)^3 \vep^2 (\vep-1) (\vep-2) (2\vep-1) (2\vep-3) (2\vep-5) 
				}$
		\\
		$c_{4+2m,4}[\vphi^\vee_{\text{box}},\vphi]$ 
			& 1 
			& $\frac{-1}{2s(1+x)\vep^2(2\vep-1)}$ 
			& $\frac{x}{4 (1+x)^2 \vep^2 (2\vep-1) (2\vep-3)}$
			& $\frac{-s x^2}{8(1+x)^3 \vep^2 (2\vep-1) (2\vep-3) (2\vep-5)}$
	\end{tabular}
	\egroup	
	\caption{ \label{tab:hdbox}
		The projection of the naive (non-UT) box in 4, 6, 8 and 10 dimensions onto 
		the 4-dimensional  UT basis.	
		}
\end{table}

As a simple application of the newly obtained compactly supported bubble and box dual forms, we consider the projection of higher-dimensional bubbles and boxes onto the 4-dimensional bubble and box. 

The relationship between FI's of different dimensions was first noted in \cite{Tarasov_1996}. After writing a FI in parametric form, it is clear that the spacetime dimension only enters through the exponent of a homogeneous polynomial $P(\alpha)^{-d/2}$ in the Schwinger parameters $\{\alpha_i\}$. All propagators are given formal masses that can be set to zero later and $P(\alpha)$ is independent of the masses. Then, one constructs a differential operator out of $\partial_{m_i^2}$ such that its action on a FI gives the same integral except with an extra power of $P$ in the integrand effectively changing $d \to d-2$. The corresponding differential operator is simply obtained by the replacement $\alpha_i \to \partial_{m_i^2}$ in $P$. Using standard IBP techniques, the action of $P(\partial_{m_i}^2)$ on any FI can be decomposed into a know basis of $d$-dimensional integrals. 

For one-loop integrals, \cite{Tarasov_1996} gives compact expressions relating $(d+2)$-dimensional integrals to $d$-dimensional integrals. Suppose that we are considering integrals with $\leq n$ points. For the maximum topology, we have the propagators $D_{i=1,\dots,n-1} = (\ell + \sum_{j=1}^i p_j)^2 + m_i^2$ and $D_n = \ell^2 + m_n^2$. We label all other $m \leq n$-point integrals using the convections set by the $n$-point integral. Then, for some subset $J \subset {1,\dots,n}$,  
\begin{align} \label{eq:tarasov}
	& \left[
		\sum_{a \in J} \nu_a [(J) \cdot (j_1 \cdots j_{a-1}\ 0\ j_{a+1} \cdots j_{|J|})]
		- \left(d - \sum_{a \in J}^n \nu_a \right) (0J)^2 
	\right] \mathscr{I}_J^{(d)} 
	\nn \\
	&= - \sum_{a\in J} [(J) \cdot (j_1 \cdots j_{a-1}\ 0\ j_{a+1} \cdots j_{|J|})] \mathscr{I}_{J,a}^{(d-2)}
	+ 2 (J)^2 \mathscr{I}_J^{(d-2)}
\end{align}
where $\nu_a = 0$ if $a \notin J$, $I_{J}^{(d)}$ is the $d$-dimensional FI corresponding to the subset $J$ and $I_{J,a}^{(d)} = I_{J}^{(d)}\vert_{\nu_a\to\nu_a-1}$ is the same integral with a reduced power of the $a^\text{th}$ propagator. Here, the $(\bullet)\cdot(\bullet)$ corresponds to minors of the Gram determinant defined in section 2.5  of \cite{Caron-Huot:2021xqj}.

We can derive analogous dimension shifting identities using intersections numbers. First, recall that the $(\dint-2\vep)$-dimensional FI associated to the form $\vphi$ is 
\begin{align}
	\mathscr{I}^{(\dint-2\vep)}[\vphi ]	&= \C_{\dint} \int u(\ell_\perp^2)\ \vphi 
		= \C_{\dint} \int \left(\ell_\perp^2\right)^{-\vep}\ \vphi .
\end{align}
Then, note that the FI associated to the form $(\ell_\perp^2)^m \vphi$ corresponds to the same integral but in $(\dint+2m-2\vep)$-dimensions
\begin{align}
	\mathscr{I}^{(\dint-2\vep)} [(\ell_\perp^2)^m \vphi]
	&= \C(\dint-2\vep) \int \left(\ell_\perp^2\right)^{-(\vep-m)}\ \vphi 
		=\frac{\C_{\dint}}{\C_{\dint+2m}} \mathscr{I}^{(\dint + 2m - 2\vep)}[\vphi] .
\end{align}
Integrals in $(\dint+2m-2\vep)$-dimensions can be projected onto the $(\dint-2\vep)$-dimensional integral basis by intersecting the $(\dint-2\vep)$-dimensional dual forms with $(\ell_\perp^2)^m \vphi$
\begin{align}
	\mathscr{I}^{(\dint + 2m - 2\vep)}[\vphi] 
	&= \sum_i c_{\dint+2m,d}[\vphi^\vee_i,\vphi]\ \mathscr{I}^{(\dint - 2\vep)}[\vphi_i],
	\\
	c_{\dint+2m,d}[\vphi^\vee_i,\vphi] 
	&= \frac{\C_{\dint+2m-2\vep}}{\C_{\dint-2\vep}}  C^{-1}_{ij} \la \vphi^\vee_j \vert (\ell_\perp^2)^m \vphi \ra
\end{align}
where $\{\vphi_i\}$ is a basis of $(\dint-2\vep)$-dimensional Feynman integrands, $\{\vphi^\vee_i\}$ is a basis for the corresponding dual space and $C_{ij} = \la \vphi^\vee_i \vert \vphi_j \ra$ is the associated intersection matrix. The projection of higher-dimensional bubbles and boxes onto 4-dimensional bubbles and boxes is collected in Tables \ref{tab:hdbub} and \ref{tab:hdbox}, which agrees with equation \eqref{eq:tarasov}.

\section{Some gluon amplitudes in generic dimension \label{sec:gluon amplitudes}}

In this section we compute one-loop 4-point and 5-point gluon amplitudes using the 
the method detailed in section \ref{sec:cbub4} to extract bubble and box coefficients.
Our main goal is to see how the $d$-dimensional coefficients contrast with their four-dimensional limit,
and how intermediate steps are affected by coordinate choices.

Since our dual forms all localize to at least a bubble cut, we only need the helicity integrands on cuts. These cut integrands are obtained by gluing massive tree-level helicity amplitudes. To reproduce the rules of dimensional regularization, these are supplemented by $d_g-5$ scalars running in the loop (see appendix \ref{app:4pt helicity integrands} for details).
As is conventional, we record the spacetime dimension $d_g$ used in numerator algebra separately from that spanned by the loop momenta.

In section \ref{sec:4ptGU}, we compute the generalized unitarity coefficients of the one-loop 4-point helicity amplitudes and construct the integrated expressions. 
To make contact with traditional generalized unitarity, we also compactify the bubble form in the strict $d=4$ limit avoiding fibration all together in section \ref{sec:4dbubble}. 
The generalized unitarity coefficients of 5-point gluon amplitudes are extracted in section \ref{sec:5ptGU}.

\subsection{4-gluon scattering \label{sec:4ptGU}}

\begin{figure}
\centering
\includegraphics[scale=.2]{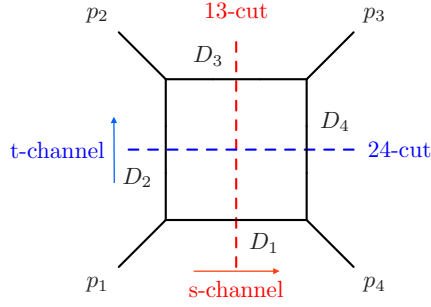}
\caption{ \label{fig:4pt conventions}
	Momentum labeling conventions. Note that the 13-cut corresponds to the $s$-channel and the 24-cut corresponds to the $t$-channel. 
}
\end{figure}

Projecting the integrands defined by \eqref{eq:4pt phi cut} onto the dual form basis yields the generalized unitarity coefficients for the one-loop helicity amplitudes.
Since not many kinematic variables are involved, the calculation can be done exactly so we record the exact dependence for each of the independent helicity choices: $(++++),(-+++),(--++),(-+-+)$.  
For each choice, we divide the amplitude by a dimensionless factor $L^\bullet$ that contains the little group weight (see equation \eqref{eq: 4pt little group factors}).

The 13-bubble (s-channel; figure \ref{fig:4pt conventions}) coefficients are
\begin{align} \label{4pt first coeff}
	c_{13}[\vphi^{++++}]
	&= -\frac{(d_g-2)}{4(2\vep-3) (2\vep-1) (1+x)}
	\\
	c_{13}[\vphi^{-+++}]
	&= - \frac{ (d_g-2) }{4(\vep-1) (2\vep-3) (2\vep-1) } \frac{x^2 (3+x-\vep) }{(1+x)}
	\\
	c_{13}[\vphi^{--++}]
	&= \frac{1}{\vep^2} - \frac{(d_g-2) x}{4(2\vep-3) (2\vep-1) (1+x)}
	\label{eq: 13 --++}
	\\
	c_{13}[\vphi^{-+-+}]
	&= -\frac{1 + x - 2\vep}{\vep^2 (2 \vep -1) (x+1)}
	+ \frac{(d_g-2) x (-2 +5 x + x^2 - x(x+6)\vep + x\vep^2) x}{(\vep -1) \vep  (2 \vep -3) (2 \vep -1) (x+1)^3}
\end{align}
Since all but the $(--++)$ little group factors $L^\bullet$ are symmetric in $s$ and $t$, three out of the four 24-bubble (t-channel; figure \ref{fig:4pt conventions}) coefficients can be obtained by exchanging $s$ and $t$
\begin{align}
	c_{24}[\vphi^{++++}]
	= c_{13}[\vphi^{++++}]\vert_{x\to x^{-1}},
	&\qquad
	c_{24}[\vphi^{-+++}]
	= c_{13}[\vphi^{-+++}]\vert_{x\to x^{-1}},
	\nn \\
	c_{24}[\vphi^{-+-+}]
	&= c_{13}[\vphi^{-+-+}]\vert_{x\to x^{-1}}.
\end{align}
The 24-bubble coefficient of the $(--++)$ integrand must be calculated separately
\begin{align}
	c_{24}[\vphi^{--++}]
	&= -\frac{1}{\vep^2 (2\vep-1)} - \frac{(d_g-2) (1+x-x\vep)}{4(1+x)\vep(2\vep-3)(2\vep-1)}.
\end{align}
We note that while our variable conventions are well suited to extracting the 13-bubble coefficients, these specific
formulas do not directly apply for the 24-bubble coefficients as the fibration axes align with some boundaries. We deal with this simply by making a rotation that leaves the twist unchanged and forces the boundaries into generic positions.

Lastly the box coefficients can be obtained from either the 13- or 24-cut integrands
\begin{align}
	c_{1234}[\vphi^{++++}]
	&= -\frac{(d_g-2) (\vep -1)  x}{4\vep  (2 \vep -3) (2 \vep -1) (x+1)^2},
	\\
	c_{1234}[\vphi^{-+++}]
	&= \frac{(d_g-2) (\vep -2) x}{4\vep  (2 \vep -3) (2 \vep -1) (x+1)^3},
	\\
	c_{1234}[\vphi^{--++}]
	&= \frac{1+x- 2\vep}{\vep ^2 (2 \vep -1) (x+1)}
		-\frac{(d_g-2) (\vep -1) x^2}{4\vep  (2 \vep -3) (2 \vep -1) (x+1)^2},
	\\
	c_{1234}[\vphi^{-+-+}]
	&= \frac{ 1 + x^2 - 2(1+x+x^2)\vep }{\vep ^2 (2 \vep -1) (x+1)^2} 
		-\frac{(d_g-2) x^2 (\vep -3) (\vep -2)}{4\vep ^2 (2 \vep -3) (2 \vep -1) (x+1)^4}. \label{4pt last coeff}
\end{align}
Note that fibration computes the box coefficient in two different ways (once starting from the 13-cut and once starting from the 24-cut), this provides important cross checks in more complicated calculations (5-points).

To obtain integrated amplitudes we multiply the above coefficients by the transcendental functions listed in our basis in eqs.~\eqref{eq:Ibub} and \eqref{eq:Ibox massless}
\begin{align}
	A^\bullet = \left( 
			c_{13}[\vphi^\bullet] \mathscr{I}[\vphi_{13}]
			+ c_{24}[\vphi^\bullet] \mathscr{I}[\vphi_{24}]
			+ c_{1234}[\vphi^\bullet] \mathscr{I}[\vphi_{1234}] 
		\right) L^\bullet,
\end{align}
where the little group weights $L^\bullet$ are defined in \eqref{eq: 4pt little group factors}.
Since the boxes are finite, all divergences in the $(--++)$ and $(-+-+)$ amplitudes appear explicitly in the above bubble coefficients, whereas the box contribution is always finite. 
On the other hand, the $(++++)$ and $(-+++)$ coefficients are finite since the corresponding tree-level amplitudes vanish. 

We find the finite amplitudes 
\begin{align}
	\A_4^{\text{one-loop}}(1^+,2^+,3^+,4^+) &
		= \frac{st}{3\la12\ra\la23\ra\la34\ra\la41\ra}
		+ \mathcal{O}(\vep),
	\\
	\A_4^{\text{one-loop}}(1^-,2^+,3^+,4^+)  &
		= \frac{\la13\ra^2[34]}{\la23\ra^2\la34\ra}  \frac{(s+t)^2}{3st} 
		+ \mathcal{O}(\vep), 
\end{align}
and the divergent amplitudes 
\begin{align} \label{mmpp}
		\frac{\A_4^{\text{one-loop}}(1^-,2^-,3^+,4^+)}{\A_4^{\text{tree}}(1^-,2^-,3^+,4^+)} = &
		-\frac{4}{\vep^2} 
		+ \frac{1}{\vep} 
			\left[ 
				-\frac{11}{3} 
				+ 2 \log\bigg(\frac{-s}{\mu^2}\bigg) 
				+ 2 \log\bigg(\frac{-t}{\mu^2}\bigg)
		 	\right]
		\nn\\&
		-\frac{64}{9} 
		- \frac{\delta_R}{3}
		+ \frac{11}{3} \log\bigg(\frac{-t}{\mu^2}\bigg)
		+ 2\zeta_2 
		- \log^2\left(\frac{-s}{\mu^2}\right) - \log^2\left(\frac{-t}{\mu^2}\right)
		\nn\\&
		+ \left(\log^2\bigg(\frac{t}{s}\bigg) + \pi^2\right)
		+ \mathcal{O}(\vep),
\\
\frac{\A_4^{\text{one-loop}}(1^-,2^+,3^-,4^+)}{\A_4^{\text{tree}}(1^-,2^+,3^-,4^+)}  
	=& \left[ 
		\mbox{ eq.~\eqref{mmpp}} 
		- \frac{11}{3} \log\left(-\frac{t}{\mu^2}\right) 
	\right]	 
	\nn\\&
	-\frac{st}{(s+t)^2}
	+ \frac{s t(s t -2(s+t)^2)}{(s+t)^4} 
		\left[\log^2\bigg(\frac{t}{s}\bigg) + \pi^2\right]
	\nn\\&	
	+ \left[
		\frac{t(14s^2+19st+11t^2)}{3(s+t)^3}
		\log\left(\frac{-s}{\mu^2}\right) 
		+ (s\leftrightarrow t)
	\right]
	+ \mathcal{O}(\vep).
\end{align}
Note that all we have also divided these amplitudes by a one-loop factor $g^2N_c(\mu^2)^{\vep}e^{-\vep\gamma_E}/(4\pi)^{2-\vep}$. 
The divergences (infrared and ultraviolet) are as expected from general results and going to the SCET hard function simply cancels out the first line \cite{Becher:2014oda,Feige:2014wja}.

Comparing with known results for the one-loop 4-point helicity amplitudes \cite{Bern:1991aq}, we find agreement.\footnote{Agreement is found after performing UV renormalization.} 
One can also check that there are no spurious u-channel singularities. 
Here, $d_g = 4-2\vep \delta_R$ where $\delta_R$ is 0 in the FDH scheme and 1 in the 't Hooft-Veltman scheme.

\subsection{4-dimensional limit of bubble coefficients (using null variables) 
\label{sec:4dbubble}}

It is clear from the above that $d$-dimensional coefficients package a lot of information that
cancels in the four-dimensional limit (compare eqs.~\eqref{4pt first coeff}-\eqref{4pt last coeff} with the subsequent amplitudes).
Thus, it is instructive to try to extract the relevant terms in the $d\to 4$ limit directly at the level of dual forms.
By examining this limit, we will discover how to construct primitives that extract 4-dimensional information such as rational terms.

To extract 4-dimensional information, we expand the $d$-dimensional ($\vep$-dependent) integral coefficients in terms of 4-dimensional ($\vep$-independent) coefficients
\begin{align}
	c_{13} (\vep)
	&= \frac{ c_{s}^{(\text{tri})} }{\vep^2}
		+ \frac{ c_{s}^{(\text{bub})} }{\vep}	
		+ c_{s}^{(\text{rat})}
		+ \O(\vep),
	\nn\\
	c_{24} (\vep)
	&= \frac{ c_{t}^{(\text{tri})} }{\vep^2}
		+ \frac{ c_{t}^{(\text{bub})} }{\vep}	
		+ c_{t}^{(\text{rat})}
		+ \O(\vep),
	\nn\\
	c_{1234}(\vep) &= \frac{c_\text{box}}{\vep^2} + \O(\vep^{-1}).
\end{align}
Since the $d$-dimensional coefficients have at most double poles as $\vep\to0$ and the box integral starts at $\mathcal{O}(\vep^2)$, six $\vep$-independent coefficients are needed to describe the finite part of a scattering amplitude. Since each of the $\vep$-independent coefficients, multiplies a linearly independent function in the amplitude (see \eqref{eq:A4 decomp} below), we anticipate that the 4-dimensional cohomology is in fact larger than the $d$-dimensional cohomology. For example, when $\vep\neq0$ the triangle and bubble integrals are cohomologous. This is no longer true in the $\vep=0$ limit, which explains the reappearance of the 4-dimensional triangle coefficients $c_\bullet^{(\text{tri})}$.

Discarding all terms that vanish as $\vep\to0$ yields the following decomposition of the amplitude
\begin{align} \label{eq:A4 decomp}
	\mathcal{A}_4^{\text{one-loop}}
	&= c_{13}(\vep) \mathscr{I}[\vphi_{13}]
	+ c_{24}(\vep) \mathscr{I}[\vphi_{24}]
	+ c_{1234}(\vep) \mathscr{I}[\vphi_{1234}]
	\nn\\&
	= -2\left(c_{s}^{(\text{rat})} + c_{t}^{(\text{rat})}\right)
	 -2 c_{s}^{(\text{bub})}\left(\frac{1}{\vep} + \log(-s)\right) -2c_t^{(\text{bub})}\Big(\bullet\Big)-c_{\text{box}} \left(\pi^2 + \log^2(s/t) \right)
\nn\\&\quad -c_{s}^{(\text{tri})}\left(\frac{2}{\vep^2} -\frac{2\log(-s)}{\vep} + \log^2(-s)-\zeta_2\right) -2c_t^{(\text{tri})}\Big(\bullet\Big)
\end{align}
where $\bullet$ stands for a trivial relabeling of the preceding term.
While massless triangles (which contain weight-two $\log^2(s)$ terms) disappear from the basis at finite $\vep$,
they effectively re-appear in the $d\to 4$ limit as the most singular part of bubble coefficients.
Looking at the finite part of \eqref{eq:A4 decomp}, it is clear that in the 4-dimensional limit the 3 master integrals
effectively span 6 independent functions (3 $\log^2$'s, 2 $\log$'s and $1$).
Understanding in general how the size of a basis jumps in the limit is left for future work.
For now, we will be satisfied by obtaining the specific combinations of primitives that extract each distinct 4-dimensional coefficients: $c_{s}^{(\text{tri})}, c_{s}^{(\text{bub})}, c_{s}^{(\text{rat})}$, etc.

A simple way to apply the null variables used for 2-forms in \sec{sec:ctri} to the three-dimensional bubble cut, is to first integrate out a variable orthogonal to all (cut and uncut) propagators. (This simplifies calculations but is probably not essential.)
We make the transformation\footnote{This transformation just brings us to the most natural projection of $\ell$ onto the external momenta. For example, $z_2^\prime$ is simply $\ell$ projected onto the direction defined by $p_1-p_2$. }
\begin{align} \label{eq:rot 4dlim}
	\begin{pmatrix}
		z_2
		\\
		z_3 
		\\ 
		z_4
	\end{pmatrix}
	=
	\begin{pmatrix}
		-\frac{1-y^2}{4y} 
		& -\frac{i(1+y^2)}{4y} 
		& 0 
		\\
		\frac{y}{2(1-y^2)}
		& \frac{iy}{2(1+y^2)}
		& -\frac{i(1+y^4)}{2(1-y^4)}
		\\
		\frac{1+y^4}{4y(1-y^2)}
		&-\frac{i(1+y^4)}{4y(1+y^2)}
		&-\frac{iy^2}{1-y^4}
	\end{pmatrix}
	\cdot 
	\begin{pmatrix}
		z_2^\prime
		\\
		z_3^\prime
		\\ 
		z_4^\prime
	\end{pmatrix}
\end{align}
where $y=\sqrt{x}+\sqrt{1+x}$ rationalizes the square roots that appear in the rotation above. 
While there are many rotations that make the boundaries free of $z_4$, \eqref{eq:rot 4dlim} simplifies the expression for the primitives. 
Applying \eqref{eq:rot 4dlim}, the boundaries become free of $z_4$
\begin{align}
	D_2\vert_{13} = -s \left( 
			\frac12 
			- \frac{1-y^2}{4y} z_2^\prime 
			- \frac{i(1+y^2)}{4y} z_3^\prime 
		\right),
	\nn\\
	D_4\vert_{13} = -s \left( 
			\frac12 
			+ \frac{1-y^2}{4y} z_2^\prime 
			- \frac{i(1+y^2)}{4y} z_3^\prime 
		\right).
\end{align}
and the bubble dual form \eqref{eq:bubdual} becomes
\begin{align} \label{eq:4d bubdual}
	\vphi^\vee_{13} = \delta_{13} (\phi^\vee_{13}),
	\qquad
	\phi^\vee_{13} = 
	\frac{ 
		-8i(\vep-1) 
		\d z_2^\prime \wedge \d z_3^\prime \wedge \d z_4^\prime
	}{
		(1 + (z_2^\prime)^2 + (z_2^\prime)^2 +  (z_4^\prime)^2 )^2
	}.
\end{align} 

We now integrate out $z_4^\prime$ first since propagators are free of it ({\it ie.} we treat it as the first fibre as in eq.~\eqref{eq:bulk forms}). The intersection of a Feynman form $\vphi$ with the bubble dual form becomes 
\begin{align}
	\la \vphi^\vee_{13} \vert \vphi \ra
	= \la \phi^\vee_{13} \vert \phi \ra
	= \big\la 
		\la \phi^\vee_{13} \vert f \ra_\F 
	\big\vert 
		\la f^\vee \vert \phi \ra_\F 
	\big\ra_{\B}
	= \big\la 
		(\phi^\vee_{13})_\B
	\big\vert 
		\phi_\B
	\big\ra_{\B}
\end{align}
where $\phi = \res_{13}[\vphi]$, $(\vphi^\vee_{13})_\B$ is the projection of the bubble dual form onto the base and $\phi_\B$ is the projection of the Feynman form onto the base. 
Here, we have chosen the following basis on the fibre
\begin{align}
	f^\vee = \frac{Q^{(1)}\ \d z_4^\prime}{Q^{(0)}},
	\qquad
	f = \frac{2\vep\ \d z_4^\prime}{Q^{(0)}},
	\qquad 
	\la f^\vee \vert f \ra_\F = 1,
\end{align}
where 
\begin{align}
	Q^{(0)} = 1 + (z_1^\prime)^2 + (z_2^\prime)^2 + (z_3^\prime)^2,
	\qquad 
	Q^{(1)} = 1 + (z_1^\prime)^2 + (z_2^\prime)^2 .
\end{align}
This choice of fibre basis yields the following base connection: 
\begin{align}
	\omega_\B^\vee = \left(\vep+\tfrac12\right) \d\log Q^{(1)}.	
\end{align}

Projecting the bubble dual onto the $z_2',z_3'$ base, for example
\begin{align}
	(\phi_{13}^\vee)_\B
	= \frac{4i(2\vep-1)\ \d^2z^\prime}{(1+z_1^2+z_2^2)^2}
\end{align}
Now that we have a 2-form, we can use the methods of \sec{sec:ctri} to compactify $(\phi_{13}^\vee)_\B$. 
The double primitives of $(\phi_{13}^\vee)_\B$ need to be computed to $\O(\vep)$ since the leading term of $\phi_\B$ scales as $1/\vep$. 
The $1/\vep$ terms of $\phi_\B$ are present if and only if $\phi$ has a simple pole at $\ell_\perp^2\vert_{13}=Q^{(0)}=0$. 
The higher order terms in $\phi_\B$ come from poles at $z_4^\prime=\infty$ (we assume that $\phi$ does not have higher order poles in $Q^{(0)}$).  
Since the leading term of the double primitive also scale as $1/\vep$ (at the collinear singularity of a massless triangle), we also need to expand $\phi_\B$ up to $\O(\vep)$. 
However, as we will see later, the $\O(\vep)$ term of $\phi_\B$ is orthogonal to the $1/\vep$ term of the double primitives. 

Changing to light cone variables, 
\begin{align}
	z_2^\prime = \frac{z_+ + z_-}{2} 
	\quad 
	z_3^\prime = \frac{i(z_+ - z_-)}{2},
\end{align}
the connection and bubble dual form become
\begin{align}
	\omega^\vee_\B = \left(\vep+\frac12\right) \d\log(1+z_+z_-)
	\quad
	(\phi_{13}^\vee)_\B = \frac{2(2\vep-1)\ \d z_+ \wedge \d z_-}{(1+z_+z_-)^2}.
\end{align}
In these coordinates, there exists almost global primitives for the bubble dual form
\begin{align}
	\psi_1^\vee = -\frac{4\ \d z_+}{z_+(1+z_+z_-)},
	\qquad
	\psi_2^\vee = \frac{4\ \d z_-}{z_-(1+z_+z_-)}.
\end{align}
The difference of these almost global primitives depends only on the combination $w=z_+z_-$
\begin{align}
	\psi_2^\vee -\psi_1^\vee = \frac{4\ \d(z_+z_-)}{(z_+z_-)(1+z_+z_-)}.
\end{align}
Thus, the bulk double primitive can be computed in terms of $w$ about the twisted points $w=-1,\infty$ ($(z_+,z_-)\to(0,\infty),(\infty,\infty)$). Computing this double primitive order by order in $\vep$, yields
\begin{align}
	\psi_{0,\infty}^\vee
	&= -\frac{8\ \text{arctanh}(\sqrt{1+w})}{\sqrt{1+w}}
		+ \frac{
			8\left( \Li_2(\sqrt{1+w}) - \Li_2(-\sqrt{1+w}) \right)
		}{\sqrt{1+w}} \vep
		+ \O(\vep^2)
\\
	\psi_{\infty,\infty}^\vee
	&= -\frac{4(1-W^2)}{(1+W^2)} 
	\bigg(
		2\log(W) 
		+ \vep \left( \Li_2(1-W^4) - 4 \Li_2(1-W^2) \right)
	\bigg)	
	+ \O(\vep^2)
\end{align}
where $W=\sqrt{w^{-1}}+\sqrt{1+w^{-1}}$. 

To make the bubble dual form compactly supported about the boundaries $D_2$ and $D_4$
\begin{align}
	D_2 = -\frac{sy}{4} \left( z_+ - \frac{z_-}{y^2} + \frac{2}{y} \right),
	\qquad 
	D_4 = -\frac{s}{4y} \left( z_+ - y^2 z_- + 2y \right),
\end{align}
we also need primitives for the restriction of $\psi_1^\vee$ to the boundaries
\begin{align}
	\psi_1\vert_{D_2} = \frac{4y^2\ \d z_-}{(y-z_-)^2(2y-z_-)},
	\qquad 
	\psi_1\vert_{D_4} = \frac{4y\ \d z_-}{(1-yz_-)^2(2-yz_-)}.
\end{align}
We will specialize to the $D_2$ boundary since the $D_4$ is obtained by simply inverting $y$. On the $D_2$ boundary, there are twisted singularities at $z_- = y,\infty$. In these neighbourhoods, the primitives are
\begin{align} 
\label{eq:psi D2 finite}
	\psi^\vee_{D_2,y}
	&= \frac{2y}{z_--y} 
	\left[
		\frac{1}{\vep}
		- 2 \log\left( 1-\frac{z_--y}{y} \right) 
		- 4 \Li_2\left(\frac{z_--y}{y}\right) \vep
	\right]
	+ \O(\vep^2)
\\
\label{eq:psi D2 infty}
	\psi^\vee_{D_2,\infty }
	&= \frac{4y}{y-z_-} 
	\bigg[
		\log\left(\frac{2y-z_-}{y-z_-}\right)
		+\bigg(
			\log^2\left(1-\frac{y}{z_-}\right)
			+ 2 \Li_2\left(\frac{y}{z_-}\right)
		\nn\\&\qquad 
			- 4 \Li_2\left(\frac{2y}{z_-}\right)
			+ \Li_2\left(-\frac{4y(y-z_-)}{z_-^2}\right)
		\bigg)\vep
	\bigg]
	+ \O(\vep^2).
\end{align}
To make the bubble dual form truly with compact support a primitive for $\psi_1\vert_{D_2}$ near the intersection $D_2 \cap D_4$ and $\O(D_2)$ corrections to \eqref{eq:psi D2 finite} and \eqref{eq:psi D2 infty} are needed. However, such corrections are only needed when the Feynman forms have propagators raised to higher powers. The $D_4$ primitives are obtained from the $D_2$ primitives by the replacement $y\to1/y$.

The boundary primitive restricted to the 123-triangle cut contains poles at $\vep=0$ and at $z_- =y$. Mathematically, the $z_-=y$ pole simply comes from the twist on the triangle cut. Physically, the point $z_-=y$ corresponds to the collinear singularity in the massless triangle. From this point of view, it is natural that the $z_-=y$ and $\vep=0$ poles are coupled. 

Putting everything together, we obtain the following residue formula for divergent and finite terms of 13-bubble coefficient
\begin{align} \label{eq:4d cbub}
	c_{13}[\vphi] 
	&= \big\la 
		(\phi^\vee_{13})_\B
	\big\vert 
		 \phi_\B
	\big\ra_{\B}
	= \sum_{(z_-^*,z_+^*)} \res_{z_-=z_-^*} \res_{z_+=z_+^*}
		\bigg[\psi^\vee_{z_+^*,z_-^*}\phi_\B\bigg]
	\\
	(z_+^*,z_-^*) &\in \Big\{ \{0,\infty\},\{\infty,\infty\},
		\{D_2,y\},\{D_2,\infty\} ,\{D_4,1/y\},\{D_4,\infty\} \Big\}
\end{align}
where $\phi_\B = \la f^\vee \vert \res_{13}[\vphi] \ra$. Note that this formula is only valid when the propagators appear as simple poles. To accommodate higher order poles, one cannot ignore the parallel transport factor $u\vert_\bullet/u$ and the $\O(D_{2,4})$ corrections to the primitives. 

The residues that contribute to each order in $\vep$ are summarized in table \ref{tab:res chart}. From this table formulas for the 4-dimensional generalized unitarity coefficients can be extracted from \eqref{eq:4d cbub}. 

The first thing to note is that while the $\O(\vep)$ terms of $\psi^\vee_\bullet$ are required, the $\O(\vep)$ terms of $\phi^\vee_\B$ are not needed. 
The $\O(\vep)$ term of $\phi_\B$ simply has too many powers of $Q^{(1)}$ in the numerator to have a residue at the collinear singularity of massless triangles. 
After restricting to $D_2$, the twist $Q^{(1)}\vert_{D_2}\propto (z_--y)^2$ has a double zero at the collinear singularity canceling the pole in $\psi^\vee_{D_2,y}$. A similar argument reveals that the $\O(\vep^0)$ term of  $\phi_\B$  is orthogonal to the $\O(\vep^{-1})$ term of $\psi^\vee_\bullet$.
Moreover, the residues in $c_{ij}^{(\text{bub})}$ and $c_{ij}^{(\text{rat})}$ are only non-zero when a subset of the loop momentum components go to infinity. More specifically, $c_{ij}^{(\text{bub})}$ receives contributions from residues at $\ell_\perp^2 = 0$ and $\ell_i=\infty$ while $c_{ij}^{(\text{rat})}$ receives contributions from residues at $\ell_\perp^2 = \infty$ and $\ell_i = \infty$.
These formulas highlight two important physical facts. First, the rational term only receives contributions from the infinite momentum limit. 
Secondly, the bubble term does not receive contributions form the collinear singularity of the massless triangle.

While an algorithm for extracting the rational terms of one-loop amplitudes already exists \cite{Badger:2008cm,Giele:2008ve}, an interesting aspect of our method is its
systematic nature and thus expect it to be useful for higher loop generalizations.
We see some technical differences: in \cite{Badger:2008cm,Giele:2008ve}, the rational terms are generated exclusively by the $\ell_\perp^2=\infty$ limit,
while here we see contributions from both $\ell_\perp^2=0$ and $\ell_\perp^2=\infty$.
It would be interesting to better understand the relation between the formulas. 
The differences may be due to their use of better-adapted coordinate choices,
or to taking residues in $\ell_\perp^2$ last rather than first. 
In any case, these differences suggest that the above formulas can be improved.

\begin{table}
\def\myxmark{}
	\centering
	\bgroup
	\def\arraystretch{1.5}
	\begin{tabular}{c|c|c|c|c|c|c|c}
	 \multicolumn{2}{c|}{} & $(0,\infty)$ & $(\infty,\infty)$ & $(D_2,y)$ & $(D_2,\infty)$ & $(D_4,y^{-1})$ & $(D_4,\infty)$
	\\\hline
	 \multirow{2}{*}{$\vep^{-2}$} 
	 	& $Q^{(0)}=0$ & 
		\myxmark & \myxmark & \cmark & \myxmark & \cmark & \myxmark
	\\
		& $Q^{(0)}=\infty$ & 
		\myxmark & \myxmark & \myxmark & \myxmark & \myxmark & \myxmark
	\\\hline
	 \multirow{2}{*}{$\vep^{-1}$} 
	 	& $Q^{(0)}=0$ & 
		\myxmark & \cmark & \myxmark & \cmark & \myxmark & \cmark
	\\
		& $Q^{(0)}=\infty$ & 
		\myxmark & \myxmark & \myxmark & \myxmark & \myxmark & \myxmark
	\\\hline
	 \multirow{2}{*}{$\vep^{0}$} 
	 	& $Q^{(0)}=0$ & 
		\myxmark & \cmark & \myxmark & \cmark & \myxmark & \cmark
	\\
		& $Q^{(0)}=\infty$ & 
		\myxmark & \cmark & \myxmark & \cmark & \myxmark & \cmark
	\end{tabular}
	\egroup 
	\caption{  \label{tab:res chart}
		Residues that contribute to \eqref{eq:4d cbub} at each order in $\vep$.
		Since $Q^{(0)}$ is $\ell_\perp^2$ restricted to the 13-cut, the $Q^{(0)}=0$ residues
		pick up the leading 4-dimensional terms while residues at $Q^{(0)}=\infty$,
		corresponding to integrals with a $\ell_\perp^2$ in the numerator and 
		contribute to the rational term only.
	}
\end{table}

To illustrate this procedure, we calculate $c_{13}^{(\text{rat})}$ for the 4-point $(--++)$ amplitude
\begin{align}
	\phi
	= \res_{13}[\vphi^{--++}] 
	= \frac{is^2 (1-y^2)^2}{128 y^2} 
	\frac{
		 P\ \d z_2^\prime \wedge \d z_3^\prime \wedge \d z_4^\prime
	}{
		 Q^{(0)} D_2 D_4
	}
\end{align}
where 
\begin{align}
	P &= 1 
		- 6 Q^{(0)}
		+ (1-\vep \delta_R) \left( Q^{(0)} \right)^2
\end{align}
is a polynomial in the twist ($Q^{(0)}=\ell_\perp^2\vert_{13}$). 
Projecting onto the base, we find
\begin{align}
	\phi_\B\vert_{\vep^{-1}}
	&= -\frac{ (1-y^2)^2}{64 y^2}  
		\frac{\d z_+ \wedge \d z_-}{(D_2 D_4)/s^2 },
	\\
	\phi_\B\vert_{\vep^{0} }
	&= \frac{ (1-y^2)^2}{384 y^2}  
		\frac{ (1+z_+z_-) (-11+z_+z_-) \d z_+ \wedge \d z_-}{(D_2 D_4)/s^2 }
\end{align}
where $\vert_{\vep^n}$ denotes the coefficient of $\vep^n$. 
From \eqref{eq:4d cbub} and table \ref{tab:res chart}, we find that 
\begin{align} \label{eq: 13 rat}
	c_{13}^{(\text{rat})}[\vphi^{--++}] 
	&= \res_{z_-=\infty}\res_{D_2}\bigg[
			\psi^\vee_{D_2,\infty}\vert_{\vep^{0} }\ \phi_\B\vert_{\vep^{0} }
		\bigg]
		+\res_{z_-=\infty}\res_{D_4}\bigg[
			\psi^\vee_{D_2,\infty}\vert_{\vep^{0} }\ \phi_\B\vert_{\vep^{0} }
		\bigg]
	\nn\\&
	= -\frac{x}{6(1+x)},
\end{align}
where only two residues contribute. 
Comparing with the $\O(\vep^0)$ term of \eqref{eq: 13 --++}, we find agreement. 

In this section, we have presented a compact formula for extracting the rational term of 4-point integrals \eqref{eq:4d cbub}. 
While some work is needed to extend our algorithm to any multiplicity at one-loop, our method is systematic and we expect that it will be useful for higher loop generalizations.

\subsection{5-particle examples \label{sec:5ptGU}}

We now illustrate how the above calculations, using the fibration method, generalize to higher points, focusing on one-loop 5-gluon amplitudes.

In the massless limit, the basis of dual forms (\eqref{eq:tadpoles}-\eqref{eq:pentagon}) shrinks to five bubble duals
\begin{align}
	\label{eq:degen bubbles}
	\vphi^\vee_{ab} = c_{2}\,\ \delta_{ab} \left( \th\ \frac{ r_{ab}\ \d^3 k_{ab} }{ (r_{ab}^2 + k^2_{ab})^2 } \right),
	&\quad  (ab) \in \text{cyclic perms}(1,3),
\end{align}
five box duals 
\begin{align}
	\label{eq:degen boxes}
	\vphi^\vee_{abcd} = c_0\, \delta_{abcd} \left( \th\ \frac{ r_{abcd}\ \d^1 k_{abcd} }{ r_{abcd}^2 + k^2_{abcd} } \right),
	&\quad (abcd) \in \text{cyclic perms}(1234),
\end{align}
and one pentagon dual
\begin{align}
	\label{eq:degen pentagon}
	\vphi^\vee_{12345} = c_{-1}\, \delta_{12345} \left( r_{12345} \right) .
\end{align}

The duality between the basis of dual forms (\eqref{eq:degen bubbles}, \eqref{eq:degen boxes} and \eqref{eq:degen pentagon}) and FIs (\eqref{eq:Ibub}, \eqref{eq:5ptFI boxes} and \eqref{eq:5ptFI pentagon}) can be seen in two ways. The straightforward way is to simply compute the intersection matrix and note that it is diagonal. Alternatively, one can note that the massless degeneration of \eqref{eq:oddOmega} and \eqref{eq:evenOmega} is equal to the minus transpose of the FI differential equations.

The corresponding integrals were described in \eqref{eq:Ibub} and \eqref{eq:5ptFI boxes}.
Recall that in our basis, all infrared and ultraviolet divergences are pushed into the bubble coefficients; the box contribution is a finite weight-two transcendental function, and the pentagon contributes at $\O(\vep)$.
Due to the nature of the 5-point problem, the full $\vep$-dependence of the generalized unitarity coefficients is too complicated to present here.
Expressions for the 5-point generalized unitarity coefficients to all orders in $\vep$ can be found in the ancillary file.  

Since the $(+++++)$ and $(-++++)$ helicity amplitudes have no four-dimensional cuts (see  appendix \ref{sec:unitary_cuts}), they are finite and at order $\O(\vep^0)$ come entirely from the bubble coefficients. 
In fact, the finite part of the $(+++++)$ and $(-++++)$ amplitudes are proportional to the sum of all the bubble coefficients (in the $\vep\to0$ limit) reproducing the standard result \cite{Bern:1993mq}
\begin{align}
	\A^{\text{one-loop}}_{5}(1^+,2^+,3^+,4^+,5^+)
	&= \frac{s_{12}s_{23}+s_{23}s_{34}+s_{34}s_{45}+s_{45}s_{51}+s_{51}s_{12}+\trfive}{6\; \la12\ra\la23\ra\la34\ra\la45\ra\la51\ra},
	\\
	\A^{\text{one-loop}}_{5}(1^-,2^+,3^+,4^+,5^+)
	&= \frac{1}{3 [12]\la23\ra\la34\ra\la45\ra[51]}
	\bigg[ 
		(s_{23}{+}s_{34}{+}s_{45})[25]^2
		{-} [24]\la43\ra[35][25]
	\nn\\
		{-} \frac{[12][15]}{\la12\ra\la15\ra}
		&
			\bigg(
				 \frac{\la12\ra^2 \la13\ra^2[23]}{\la23\ra}
				{+} \frac{\la13\ra^2 \la14\ra^2[34]}{\la34\ra}
				{+} \frac{\la14\ra^2 \la15\ra^2[45]}{\la45\ra}
			\bigg)
	\bigg].
\end{align}
In this case, the sum of all contributions is particularly nice since it directly corresponds to a well-defined piece of an observable amplitude. On the other hand, contributions from individual bubble coefficients are complicated and contain spurious $\trfive$ singularities (which only cancel in the sum).

For other helicity configurations, the bubble contributions diverge like $1/\vep^2$ and there are additional finite contributions coming from boxes.
Somewhat like the $(\pm++++)$ amplitudes the now divergent sum over all bubble coefficients is still simple since it controls the universally divergent part of the amplitude 
\begin{align}
	\frac{\A_{5}^\text{one-loop}(1^-,2^\mp,3^\pm,4^+,5^+)}{\A_{5}^\text{tree}(1^-,2^\mp,3^\pm,4^+,5^+)}
	&=\frac{5}{\vep^2} 
		+ \frac{1}{\vep}
		\bigg( 
			\frac{11}{3}
			- \sum_{i} \log\left(\frac{-s_{ii+1}}{\mu^2}\right)
		\bigg)
		\nn\\& \qquad
		+ H^{-\mp\pm++}
	 + \mathcal{O}(\vep).
\end{align}
Here
\begin{align}
	H^{\bullet} = \sum_{i=0}^2 H^{\bullet}_i,
\end{align}
is the finite part of the amplitude (hard function in SCET) organized by transcendental weight $i=0,1,2$.
We present the finite part of the $(--+++)$ amplitude here while the corresponding expressions for the $(-+-++)$ amplitude can be found in appendix \ref{app:-+-++} (see also the ancillary file).

After using a dilog identity, the  weight 2 piece of the $(--+++)$ finite term is
\begin{align}
	H_2^{--+++}
	&= - \frac{5\pi^2}{4}+ \frac{1}{2} \sum_{i} \log^2\left(\frac{-s_{i,i+1}}{\mu^2}\right)
	- \sum_{i} \log\left(\frac{s_{i,i+1}}{s_{i+1,i+2}}\right) \log\left(\frac{s_{i+2,i+3}}{s_{i+3,i+4}}\right).
\end{align}
Moving onto the weight 1 term, we find
\begin{align}
	H_1^{--+++} &= 
	-\frac{\la14\ra\la23\ra[34]+\la15\ra\la24\ra[45]}{6\la12\ra(s_{23}-s_{51})}
		\bigg( 
			11 
			+ 2 \frac{\la14\ra\la15\ra\la23\ra\la24\ra[34][45]}{\la12\ra^2 (s_{23}-s_{51})^2}
		\bigg)
	\log\left(\frac{s_{23}}{s_{51}}\right) 
	\nn\\&
	-\frac{11}{6} \log\left(\frac{s_{23}}{\mu^2}\right)
	-\frac{11}{6} \log\left(\frac{s_{51}}{\mu^2}\right)
	+\O(\vep).
\end{align}
Lastly, the weight 0 term is
\begin{align}
		H_0^{--+++} 
		&= \frac{\la 23 \ra \la 24 \ra \la 41 \ra \la 51 \ra (s_{23}+s_{51}) [ 34 ] [ 45 ] (\la 23 \ra \la 41 \ra [ 34 ]+\la 24 \ra \la 51 \ra [ 45 ])}{6 \la12\ra^3 s_{23} (s_{23}-s_{51})^2 s_{51}}
		\nn\\&
		-\frac{\la 23 \ra \la 51 \ra \left(\la 24 \ra \la 41 \ra [ 23 ] [ 34 ] [ 45 ] [ 51 ]+2 s_{23} s_{51} [ 35 ]^2\right)}{6 \la12\ra^2 s_{23} s_{51} [ 23 ] [ 51 ]}
		\nn\\&
		+\frac{\la 23 \ra \la 35 \ra \la 51 \ra [ 35 ]^3}{3 \la12\ra^3 [ 12 ] [ 23 ] [ 51 ]}
		+\frac{\delta_R}{3}
		+\frac{64}{9}.
\end{align}
Comparing with the known results of \cite{Bern:1993mq}, we find agreement. Moreover, one can check that there are no spurious poles in the Mandelstams. 

Generically, the full $\vep$-dependence of the generalized unitarity coefficients becomes extremely complicated. Understanding how to keep exactly the minimal number of terms in the $\vep$-expansion of these coefficients will be important for future applications. Working with a finite basis has helped tame the $\vep$-dependence of the coefficients; divergences appear only in bubble coefficients. Moreover, the residues of the $\vep$ poles are simple because they are physical -- there can be no cancelations with the higher weight boxes and pentagon. On the other hand, the finite terms $\O(\vep^0)$ of the bubble coefficients contain spurious poles that cancel when combined with the box coefficients.

\section{Conclusions \label{sec:conclusion}}

In \cite{Caron-Huot:2021xqj}, we identified the space Poincar\'e dual to the vector space of Feynman integrals (twisted cohomology) as a certain twisted relative cohomology. Its elements, called dual forms, appear simpler than their Feynman counterparts since they are localized to generalized unitarity cuts.
Intuitively, dual forms are integrals in $(\dint+2\vep)$-dimensions that are forced to vanish near the zero locus of the propagators ($\{D_i=0\}$) instead of having poles.
The wedge product of a dual form and Feynman form can thus be meaningfully integrated, yielding an algebraic invariant called the intersection number. 

Since the intersection number is invariant under adding integration-by-part identities, it allows to reduce complicated integrals to a minimal basis without generating such identities \cite{Mastrolia:2018uzb, Frellesvig:2019kgj, Frellesvig:2019uqt, Mizera:2019vvs, Mizera:2020wdt, Frellesvig:2020qot}. 
Our work differs from previous intersection-based approaches in two ways. 
First, we insist on working directly in loop momentum space rather than an auxiliary parameter space in order to exploit the factorization of amplitudes on their cuts. 
Second, we do not modify the non-generic (integer) propagator exponents that arise naturally in Feynman integrals. 
The second condition allows dual forms to localize on maximal generalized unitarity cut, thus providing a systematic method for computing
integral coefficients for arbitrary sub-topologies and away from the 4-dimensional limit. 

Dual forms are best characterized by algebraic representatives, which are holomorphic top forms supported on various cuts.
The main ingredient to compute intersection numbers is the $c$-isomorphism, which maps the algebraic version of a form
to a compactly supported version by adding certain ``$d\theta$'' anti-holomorphic terms (see section \eqref{sec:distributions}).
We provide two methods for computing the $c$-map of multi-variate dual forms: fibrations in Cartesian and light-cone variables.
Since the $c$-map is only an isomorphism in cohomology,
different coordinate choices generally yield distinct residue-like formulas that compute the same integral coefficients.
While fibration is perhaps the most systematic algorithm at the moment and applicable to $p$-forms of any degree,
it requires one to compute the $c$-map for the entire basis of dual cohomology at once and is currently hard to apply to a single dual form on is own.
For 2-forms, light-cone variables provide a direct and efficient solution.

As an example of our formalism, one-loop 4- and 5-point gluon amplitudes were extracted in different ways from $d$-dimensional cuts.
While these amplitudes have long been known, we focused here on validating the method and also
to determine how coordinate choices affect intermediate steps.  It would be interesting to see if these would further simplify using other choices
or integration orders, for example those used in \cite{Badger:2008cm}.
An irreducible issue is that $\vep$-dependant generalized unitarity coefficients at higher points admit intrinsically
bulky analytic expressions. 
With this in mind, we took first steps in section \ref{sec:4dbubble} to analyze the physically relevant $d\to 4$ limit,
focusing on 4-gluon scattering, and exploiting the fact that the Laurent series around $d=4$ of (the $c$-image of) dual forms can be computed directly under
the integration sign. Amusingly, the duals of lower-transcendental integrals turn out to be pure higher transcendental forms: 
the dual of a rational term contains a dilogarithm primitive, etc.
Since the method is systematic, we expect that it could prove useful at higher loop orders. 
Other fibration strategies, such as a loop-by-loop approach, could streamline the calculation of multi-variate intersection numbers. 
Such considerations are left for future work. 

\section{Acknowledgements}
The authors would like to thank Sebastian Mizera, Mathieu Giroux, Brent Pym, Kale Coville and Samuel Abreu for helpful discussions at various stages of this work, and the participants of the 2019 Padova workshop ``Intersection theory \& Feynman Integrals''. 
This work (S.C.-H.) was supported by the National Science and Engineering Council of Canada, the Canada Research Chair program,
the Fonds de Recherche du Qu\'ebec - Nature et Technologies, and the Sloan Foundation.
A.P. is grateful for support provided by the National Science and Engineering Council of Canada and the Fonds de Recherche du Qu\'ebec - Nature et Technologies.

\appendix

\section{Closed form primitives for the 1234-box \label{app:closed form box prim}}

Instead of deriving the primitives as a Laurent series about the various singular points and boundaries, one can try to directly integrate the formal expression 
\be
	\psi^\vee_{z_i} [\vphi^\vee]
	= \nabla_{-\omega}^{-1} \vphi^\vee \;\; (\text{near } z=z_i)
	= \int_{z_i}^z \frac{u(z,\{s_a\})}{u(z^\prime,\{s_a\})} \vphi^\vee(z^\prime,\{s_a\}).
\ee
in order to obtain a closed form expression for the primitives. Here, the $z_i$ are the singular and boundary points and $s_a$ are the kinematic variables. 

It is indeed possible to find closed form expressions for the primitives of 1-forms with a quadratic twist. For the one-loop 1234-box form, we find closed form primitives
\begin{align}
	\psi^\vee_+ &= -\frac{ 2^{\vep-1} c_0 r_{1234}^{\vep}\ (r_{1234} + \ell_{1234})^{\vep} }{\vep} 
		\;_2F_1 \left( 1-\vep,\vep,1+\vep,\frac{r_{1234}-\ell_{1234}}{2r_{1234}} \right),
\\
	\psi^\vee_- &= \frac{ 2^{\vep-1} c_0 r_{1234}^{\vep}\ (r_{1234} - \ell_{1234})^{\vep} }{\vep} 
		\;_2F_1 \left( 1-\vep,\vep,1+\vep,\frac{r_{1234}+\ell_{1234}}{2r_{1234}} \right),
\\
	\psi^\vee_\infty &= \frac{c_0\ r_{1234} }{(1-2\vep)\ell_{1234}} 
	\left(\frac{\ell_{1234}^2}{\ell_{1234}^2-r_{1234}^2}\right)^{\vep}
	{}_2F_1\left( \frac{1}{2}-\vep, 1-\vep, \frac{3}{2}-\vep, \frac{r_{1234}^2}{\ell_{1234}^2}\right),
\\
	\psi^\vee_0 &= \frac{c_0}{2}
	\left( \frac{4 r_{1234}^2}{r_{1234}^2-\ell_{1234}^2} \right)^\vep
	\left( 
		B\left(\frac{r_{1234}+\ell_{1234}}{2r_{1234}},\vep,\vep\right) 
		- B\left(\frac{ \ell_0 +r_{1234}}{2r_{1234}},\vep,\vep\right) 
	\right),
\end{align}
where $B$ is the incomplete beta function. The primitives are indexed as in Sec.~\ref{sec:cbox} ($\pm$ for the finite roots of the twist, $\infty$ for the point at infinity and $0$ for any boundaries).

\section{Compactifying forms: details \label{app:compactifying-dets} }

Using the formal expression
\be
	\psi^{(1)}_{{a_\bk},\alpha} = \int_{z_{1,\alpha}}^{z_1} \frac{u^{(0)}\vert_{z_1=z_{1,\alpha}}}{u^{(0)}} f_a^{(1)},
\ee
it follows that 
\begin{align}
	\nabla^\vee \psi^{(1)}_{{a_\bk},\alpha}
	&= F^{(1)}_{a_\bk} + \psi^{(1)}_{b_\bk,\alpha}\ \omega^{(1)}_{b_\bk a_\bk} 
		+ \frac{[u^{(0)} F^{(1)}_{a_\bk}]\vert_\alpha}{u^{(0)}}
	= F^{(1)}_{a_\bk} + \psi^{(1)}_{b,\alpha}\ \omega^{(1)}_{b{a_\bk}},
\end{align}
where the boundary components of the connection are simply the restriction of $F^{(1)}_{a_\bk}$ and we have defined the primitive of the boundary/Leray forms on $\F_1$ to be $u^{(0)}\vert_{\alpha = b_\bd}/u^{(0)}$ or equivalently $\d\theta_{\alpha = b_\bd}$ striped Leray. 
Explicitly, $\omega^{(1)}_{b_\bd a_\bk} = F^{(1)}_{a_\bk}\vert_{\alpha = b_\bd}$ and $\psi^{(1)}_{{a_\bk},\alpha=b_\bd} = u^{(0)}\vert_{\alpha = b_\bd}/u^{(0)}$. 
Note that if $\alpha$ corresponds to a twisted singularity, $u^{(0)}\vert_{\alpha} = 0$.

Similarly, the vector valued analogue is 
\be
\bs{\psi}^{(i)}_{{a_\bk},\alpha} 
	= \int_{z_{i,\alpha}}^{z_i} \left(\mat{u}^{(i-1)}\right)^{-1} 
		\cdot \left(\mat{u}^{(i-1)}\vert_{z_i=z_{i,\alpha}}\right) f_{a_\bk}^{(i)}.
\ee
It follows that
\begin{align}
	\nabla^{(i-1)}\bs{\psi}^{(i)}_{a_\bk,\alpha}
	&= \bs{F}^{(i)}_{a_\bk} 
		+ \bs{\psi}^{(i)}_{b_\bk,\alpha}\ \omega^{(i)}_{b_\bk a_\bk} 
		+ \left[\mat{u}^{(i-1)}\right]^{-1} \cdot 
		\left.\left[
			\mat{u}^{(i-1)} \cdot \bs{F}^{(i)}_{a_\bk}
		\right]\right\vert_\alpha,
	\nn\\
	&= \bs{F}^{(i)}_{a_\bk} 
		+ \bs{\psi}^{(i)}_{b,\alpha}\ \omega^{(i)}_{b a_\bk}, 
\end{align}
where the primitive of the boundary/Leray forms on $\F_i$ are the corresponding column of $[\mat{u}^{(i-1)}]^{-1} \cdot [\mat{u}^{(i-1)} \cdot \bs{F}^{(i)}_{a_\bk}]\vert_{\alpha = b_\bd}$. 
One subtlety here is that the rows of $\bs{f}^{(i)}_{a_\bk}$ have to contain different $\theta$'s in order to construct compactly supported forms satisfying \eqref{eq:Fc-commutation}. 
While it is not correct to compactify each column with respect to the location of all twisted singularities, it does not hurt in situations where it can be done.
The extra terms produced by this over compactification do not contribute to the intersection simply because the projection of Feynman forms onto the $\B_i$ (\eqref{eq:phi(1)} and \eqref{eq:phi(i)}) never develop the singularities needed survive the corresponding residues. 
In such cases, each row is uniformly compactified with respect to all the twisted singularities of the connection.

\section{Rational parameterizations \label{app:ratparm} }

Analytic calculations of scatting amplitudes usually involves complicated functions of scalar products $p_i\cdot p_j$ and spinor products $\la ij \ra, [ij]$. For example, the region momenta $\{x_{i,i+1}^2, x_{i,i+2}\}$ forms a basis of Mandelstam-like variables for kinematic space in the case of planar scattering. However, due to momentum conservation and algebraic constraints (like the Schouten identity) these scalar products are not mutually independent. Instead, we work with momentum twistor variables, which automatically satisfy all identities (Schouten identity, momentum conservation, etc.). Moreover, momentum twistor variables are rational functions of the $x_{ij}$ (and vice versa) making conversation between these variables trivial. 

Momentum twistors are functions of the holomorphic spinor variables $\lambda_{i}$ and anti-holomorphic spinors $\mu_{i}$
\begin{align}
	Z_i = \begin{pmatrix} \lambda_{i} \\ \mu_{i} \end{pmatrix}. 
\end{align}
The usual anti-holomorphic spinor is defined in terms of the $\mu_{i}$ \cite{Elvang:2015rqa}
\begin{align}
	\tilde{\lambda}_{i} = \frac{\la i,i+1 \ra \mu_{i-1} + \la i+1, i-1 \ra \mu_{i} + \la i-1, i \ra \mu_{i+1}}{\la i,i+1 \ra \la i-1, i \ra}.
\end{align}
Thus, parametrizing the components of $Z=(Z_1 Z_2 \cdots )$ fixes our representation of the external kinematics. We can use the symmetries of $Z$ (Poincar\'e and a $U(1)$ little group scaling) to reduce the number of unfixed twistor components. For example, consider $n$-particle scattering. The momentum twistor $Z$ has $4n$ components but symmetries fix $n + 10$ components leaving $3n-10$ free components. 

While the momentum twistor parameterization may not seem very useful for 4-point scattering, it considerably simplifies 5-point kinematics. As a warm up, we the momentum twistor parametrization for 4-point scattering in section \ref{app:4pt param}. Then the 5-point parameterization is given in section \ref{app:5pt param}.

\subsection{4pt parametrization \label{app:4pt param} }

We choose the following representation for the 4-point momentum twistor\footnote{A different parameterization can be found in appendix A of \cite{Badger:2013gxa}.}
\begin{align}
	Z_4 = 
	\begin{pmatrix}
		1 & 0 & 1 & -\left(1+\frac{s}{t}\right) \\
		0 & 1 & 1 & -1 \\
		0 & 0 & 0 & \sqrt{-s} \\
		0 & 0 & \sqrt{-s} & 0
	\end{pmatrix}.
\end{align}
In our parameterization, the twistor variables are just the ordinary Mandelstam variables: $\la 12 \ra [21] = s$ and $\la 23 \ra [32] = t$. Twistor variables can be used to calculate any physical expression that is does not contain an overall helicity factor.

We will use the following cartesian coordinates for the loop momentum
\begin{align}
	\ell_\mu = \sqrt{-s} \begin{pmatrix} z_3 & z_4 & i z_2 & z_1 & \bs{z}_\perp  \end{pmatrix}.
\end{align}
The strange labeling of the dimensionless variables $z_i$ was chosen so that the fibration of section \ref{sec:cbub4} has simple labeling. Moreover, the factor of $i$ in front of $z_2$ ensures that there are no explicit factors of $i$ in the boundaries. Reconstructing the 4-momentum from the spinors,
and substituting into \eqref{eq:D1}-\eqref{eq:D4} yields the boundaries in terms of the $z_i$
\begin{align}
	D_1 &= -s\left( {-}z_\perp^2 {-} z_1^2 {+} z_2^2 {+} z_3^2 {-} z_4^2 \right),
	\\
	D_2 &= D_1 {-} s\left( z_1 {+} z_3 \right),
	\\
	D_3 &= D_1 {-} s\left( 2z_1 {-} 1 \right),
	\\
	D_4 &= D_1 {-} s\left( z_1 {-} z_2 + \left(1{+}2\frac{t}{s}\right) z_3 {+} \left(1{+}2\frac{t}{s}\right) z_4 \right).
\end{align}

\subsection{5pt parametrization \label{app:5pt param} }

The following momentum twistor parameterization for the 5-point kinematics was used to simplify the intermediate results of the intersection calculations.
\begin{align}
	Z_5 = 
	\begin{pmatrix}
		1 & 0 & 1 & 1+\frac{1}{x_1} & 1+\frac{1}{x_1} + \frac{1}{x_1 x_4} \\
		0 & 1 & 1 & 1 & 1 \\
		0 & 0 & 0 & \sqrt{-s}\ x_2 & -\sqrt{-s}\ \\
		0 & 0 & \sqrt{-s}\ & \sqrt{-s}\  x_3  & 0
	\end{pmatrix}
\end{align}
The generalized unitarity coefficients of the one-loop 5-point helicity amplitudes in the attached ancillary file are expressed in terms of these parameters.

\section{One-loop cut integrands from glueing trees\label{sec:unitary_cuts}}

In this appendix, bubble cuts of the one-loop $++++$, $-+++$, $--++$ and $-+-+$ helicity amplitudes in $(4-2\vep)$-dimensions are constructed by glueing \emph{massive} tree-level amplitudes. 
Here, the propagators are massive to account for the effective mass provided by the non-physical $\ell_\perp$ directions.

\subsection{Some massive tree amplitudes}
\label{ssec:trees}

We follow the conventions of \cite{Elvang:2013cua}.  
In brief: metric is mostly-plus and spinors satisfy $\ab{ij}\sb{ji}=-2p_i{\cdot}p_j=s_{ij}$ when the $p_i$, $p_j$ are massless and for a general set $I$, $s_I\equiv -(\sum_{i\in I} p_i)^2$.
Note that this implies a minus sign in the Clifford algebra: $\pslash_1\pslash_1 = -p_1^2 = +m_1^2$.
For massive vectors polarizations (boldface), left- and right- spinors are related as
$\keta{\b{1}}=\frac{\pslash_1}{m_1} \kets{\b{1}}$, normalized to $\ab{1^\dagger1}=2m$
\cite{Arkani-Hamed:2017jhn}.
Note that a minus sign is required for bras by consistency: $\braa{\b{1}}= -\bras{\b{1}}\frac{\pslash_1}{m_1}$.

To construct the one-loop cut integrands we employ factorization:
\be
 A \mapsto \sum_{\lambda_I} A_L(\lambda_I) \frac{-1}{s_I^2-m_I^2}  A_R(\bar\lambda_I)
\ee
where for a particle of spin $j$ the polarization sum satisfies:
$\sum_{\lambda} \ab{a \lambda}^{2j} \ab{\bar\lambda b}^{2j} = m^{2j} \ab{a b}^{2j}$.
Since the internal loop momentum can be massive, we need 3-,4- and 5-point tree amplitudes with two massive legs to construct all one-loop 4- and 5-point helicity amplitudes. 

We find concise formulas for amplitudes where the massless gluons all have the same helicity:
\be\begin{aligned} \label{eq:same h trees}
 A(\b{1},\b{2},3^+) &= \ab{\b{1}\b{2}}^2  \frac{\sb{3\mu}}{\braa{3}\pslash_1\kets{\mu}}\ , \\
 A(\b{1},\b{2},3^+,4^+) &= \frac{\ab{\b{1}\b{2}}^2}{(s_{23}-m^2)} \frac{\sb{34}}{\ab{34}}\ ,\\
 A(\b{1},\b{2},3^+,4^+,5^+) &= \frac{\ab{\b{1}\b{2}}^2\left( \bras{3}\pslash_2\pslash_1\!\kets{5}+m^2\sb{35}\right)}
 {(s_{15}-m^2)(s_{23}-m^2)\ab{34}\ab{45}}\,.
\end{aligned}\ee
The first one is the ``minimal coupling" from Ref.~\cite{Arkani-Hamed:2017jhn}; note that the ratio is independent of the reference $\mu$ thanks to the on-shell conditions. The other amplitudes
were obtained by applying BCFW recursion (shifting $\kets{3}$ and $\keta{4}$).
Same-helicity amplitudes involving a massive scalar (instead of vector) are obtained by the simple replacement:
$\ab{\b{1}\b{2}}^2\mapsto m^2$.
General formulas for massive scalar and $(n-2)$ positive-helicity gluons were given in ref.~\cite{Elvang:2011ub} and we expect that by multiplying these formulas by $\ab{\b{1}\b{2}}^2$ gives the corresponding amplitude for massive vectors. 

Amplitudes with non-identical helicity are slightly more complicated. For example, at 4-points \cite{Arkani-Hamed:2017jhn}:
\be
 A(\b{1},\b{2},3^-,4^+) = \frac{ \big(\ab{3\b{1}}\sb{\b{2}4}+\ab{3\b{2}}\sb{\b{1}4}\big)^2}{(s_{23}-m^2)s_{34}},
 \qquad
 A(\b{1},\b{2},3^-,4^+) = \frac{ \braa{3}p_1\kets{4}^2}{(s_{23}-m^2)s_{34}}\,.
\ee
The other helicity configurations at 5-points can be obtained from \eqref{eq:same h trees} by glueing 3-point and 4-point trees.

\subsection{Unitarity cuts for 4-point one-loop amplitudes \label{app:4pt helicity integrands}}

\begin{figure}
	\centering
	\includegraphics[width=.3\textwidth]{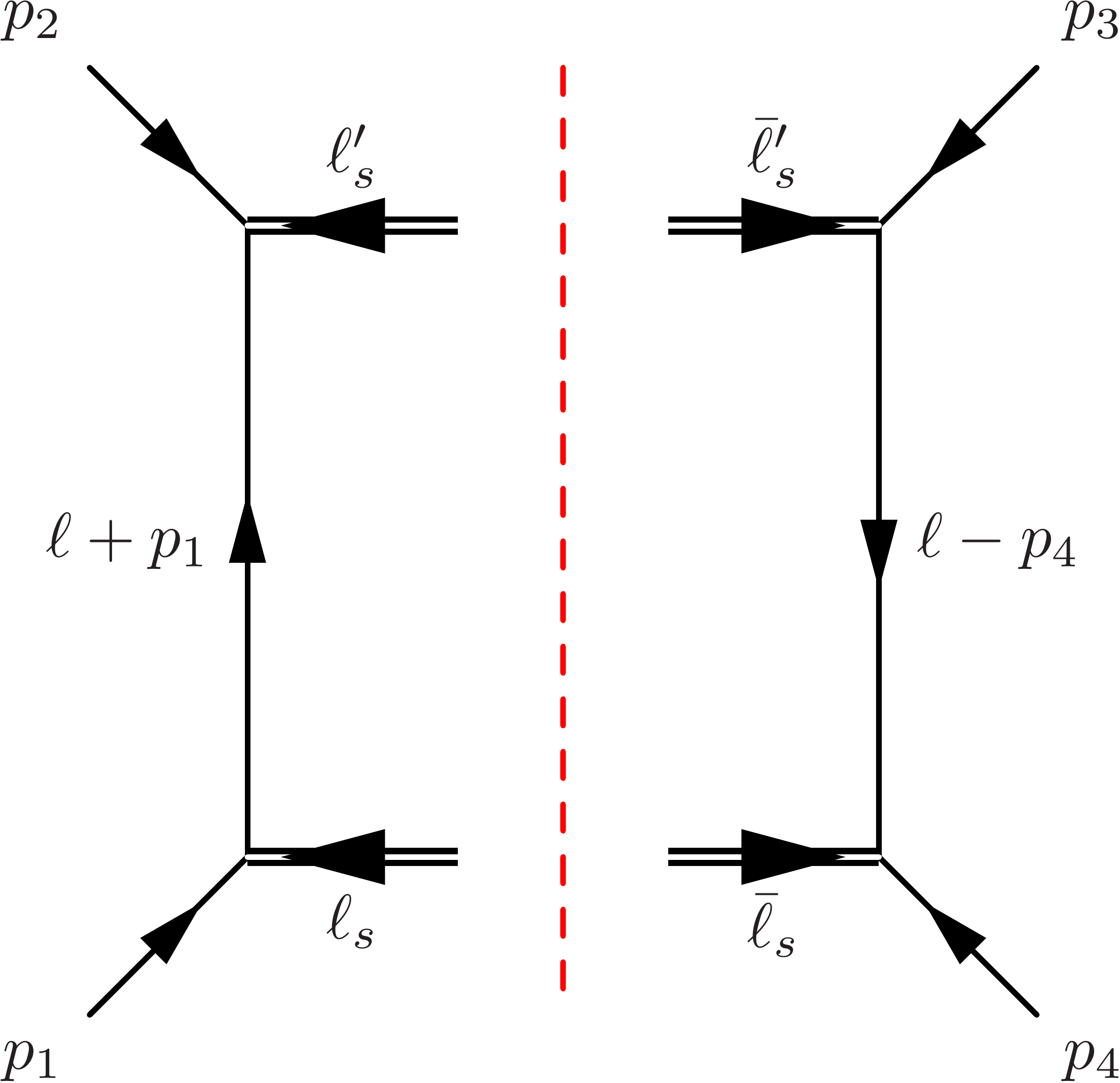}	
	\qquad
	\includegraphics[width=.3\textwidth]{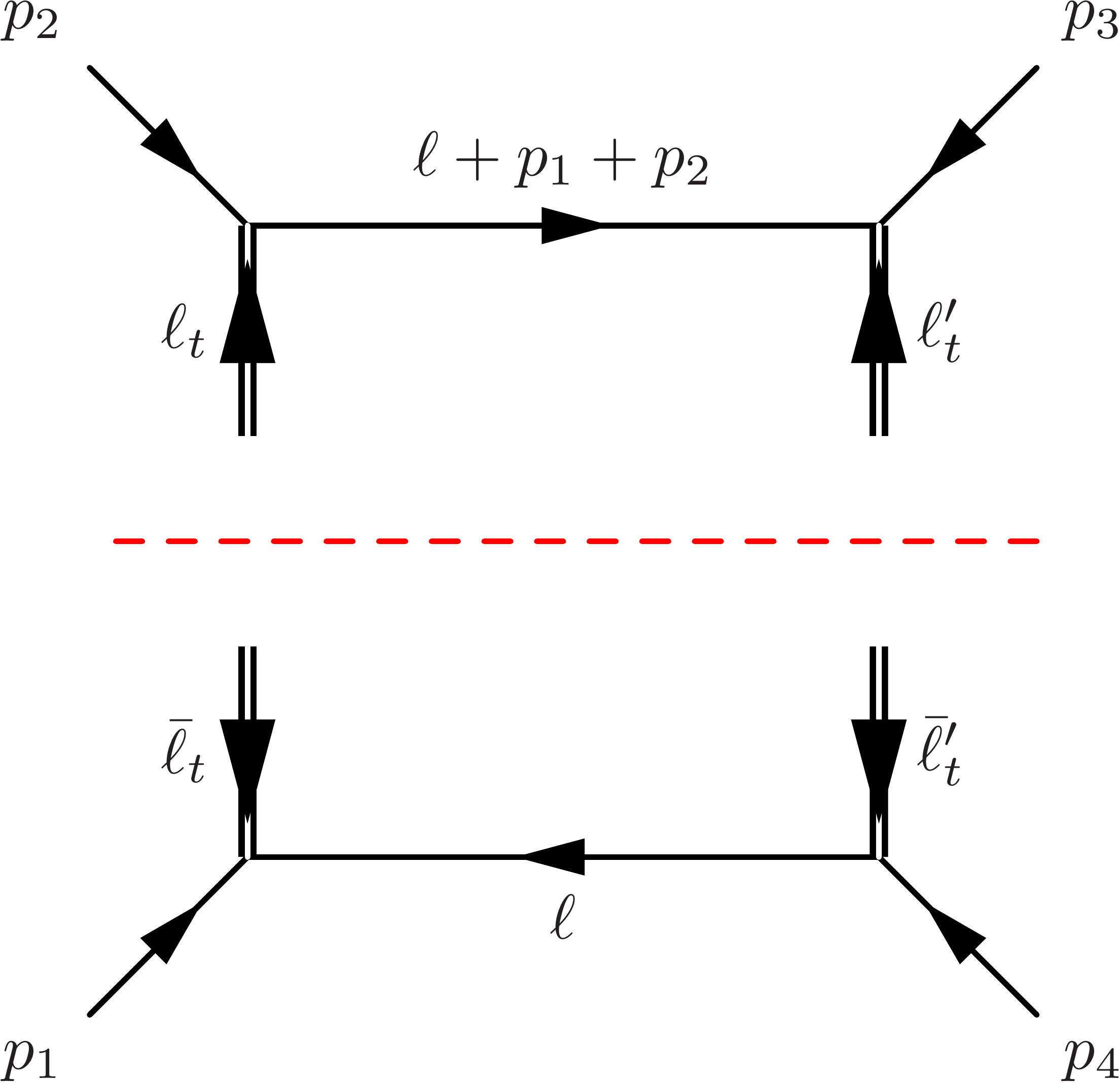}	
	\caption{\label{fig:s/t cuts}
		Conventions for the labeling of loop momentum. The $s$-channel
		momentum are $\ell_s = -\bar{\ell}_s = \ell$ and 
		$\bar{\ell}^\prime_s = -\ell^\prime_s = \ell+p_1+p_2$ while
		the $t$-channel momentum are $\ell_t = -\bar{\ell} = \ell + p_1$
		and $\bar{\ell}^\prime_t = - \ell^\prime_t = \ell - p_4$.  
		The double lines indicate a propagator with a mass squared equal 
		to $\ell_\perp^2$. 
	}
\end{figure}

Since we only consider planar amplitudes so there are $s=s_{12}$ and $t=s_{23}$ bubble cuts.
We treat each $d$-dimensional gluon as a minimally coupled vector of mass $m^2=\ell_\perp^2$.
Summing over all polarizations of the massive vector is expected to give the integrand of 
five-dimensional Yang-Mills.  To reduce to $d_g$ dimensions, we add $(d_g-5)$ scalars running 
inside the loop, where $d_g=\eta^\mu_\mu$ is the spacetime dimension that enters in numerator 
algebra ({\it ie.} $d_g-2$ is the number of degrees of freedom of the gluon).
That is, $d_g=4-2\vep \delta_R$ where $\delta_R=0$ in the FDH scheme and $\delta_R=1$ in the `t Hooft-Veltmann scheme.

Up to parity and cyclic rotations, there are four distinct helicity amplitudes: $A^{++++}$, $A^{-+++}$, $A^{--++}$ and $A^{-+-+}$.
Each has an $s$- and $t$- channel bubble cuts. See figure \ref{fig:s/t cuts} for the labeling conventions.

Lets start with the easiest helicity configuration: the $(++++)$ amplitude.  
The numerator gives a factor $\ab{{\bf \ell}{\bf \ell'}}^2\ab{{\bf \bar\ell'}{\bf \bar\ell}}^2$, 
which becomes $3\ell_\perp^4$ after summing over the massive vector helicities.  Adding the 
$(d_g-2)$ scalars turns the 3 into $(d_g-5)$, or:
\be
 {\rm Cut}_{s}^{++++} 
 = \frac{\sb{12}^2}{\ab{34}^2} 
 \frac{\ell_\perp^4 (d_g-2)}{(\ell_s{+}p_1)^2 (\ell_s{-}p_4)^2}\,.
\ee
More generally, we find the following identity useful for summing over vector polarizations:
\be
 \frac{1}{m^4}\sum_{\ell,\ell'} \braa{\ell}X\keta{\ell'}^2\braa{\bar\ell'}Y\keta{\bar\ell}^2 =
 \frac{{\rm tr}(X YXY)+{\rm tr}(XY)^2}{2}
\ee
where the trace is over $2\times 2$ Weyl matrices.  For example, with $X=Y=1$, the right-hand side gives 3,
in agreement with the 3 states of a spin-1 massive particle.


Multiplying the appropriate tree amplitudes and summing over polarizations we find:
\be\begin{aligned}
 {\rm Cut}_{s}^{++++} &= 
 \frac{\sb{12}^2}{\ab{34}^2} 
 \frac{\ell_\perp^4(d_g-2)}{(\ell_s{-}p_4)^2(\ell_s{+}p_1)^2}\,,
 &\quad
 {\rm Cut}_{t}^{++++} &= 
 \frac{\sb{23}^2}{\ab{14}^2} 
 \frac{\ell_\perp^4(d_g-2)}{(\ell_t{-}p_1)^2(\ell_t{+}p_2)^2}\,,
\\
 {\rm Cut}_{s}^{-+++}  &= 
 \frac{1}{\ab{34}^2} 
 \frac{\ell_\perp^2\braa{1}\ell_s\kets{2}^2(d_g-2)}{(\ell_s{-}p_4)^2(\ell_s{+}p_1)^2}\,,
 &\qquad
{\rm Cut}_{t}^{-+++}  &= 
\frac{1}{\ab{23}^2} 
\frac{\ell_\perp^2\braa{1}\ell_t\kets{4}^2(d_g-2)}{(\ell_t{-}p_1)^2(\ell_t{+}p_2)^2}\,.
\end{aligned}
\label{eq:4pt cuts1}
\ee
Note that these cuts vanish when $\ell_\perp^2\to 0$. 
The other ones take somewhat more algebra, but can all be written as a piece that is the 4d cut, plus terms that vanish with $\ell_\perp^2$:
\be\begin{aligned}
	{\rm Cut}_{s}^{--++} 
	&= \frac{\ab{12}\sb{34}}{\sb{12}\ab{34}}
	\frac{s^2-4s\ell_\perp^2+\ell_\perp^4(d_g-2)}{(\ell_s{-}p_4)^2(\ell_s{+}p_1)^2}\,,
\\
	{\rm Cut}_{t}^{--++} &= 
		\frac{1}{t^2(\ell_t{-}p_1)^2(\ell_t{+}p_2)^2}
		\left[\begin{array}{l}	
			\phantom{+}
   			\big(
				\langle 1\ell_t 3]\langle 2\ell'_t 4]
				- \ell_\perp^2\ab{12}\sb{34}
			\big)^2
			+ 2 t \ab{12}^2\sb{34}^2 \ell_\perp^2 
			\\
			+ \big(
				\langle 1\ell'_t 3]\langle 2\ell_t 4]
				- \ell_\perp^2\ab{12}\sb{34}
			\big)^2
			+ (d_g-4) \langle 1\ell_t 4]^2\langle 2\ell_t 3]^2
		\end{array}\right],
\\
	{\rm Cut}_{s}^{-+-+}  &= 
		\frac{1}{s^2(\ell_s{-}p_4)^2(\ell_s{+}p_1)^2}
		\left[\begin{array}{l}
			\phantom{+}
   			\big(
				\langle 1\ell_s 4]\langle 3\ell'_s 2]
				+ \ell_\perp^2\ab{13}\sb{24}
			\big)^2
			+ 2 s\ab{13}^2\sb{24}^2 \ell_\perp^2 
			\\
			+ \big(
			\langle 1\ell'_s 4]\langle 3\ell_s 2]
			+ \ell_\perp^2\ab{13}\sb{24}
		\big)^2
		+ (d_g-4) \langle 1\ell_s 2]^2\langle 3\ell_s 4]^2
		\end{array}\right],
\\
	{\rm Cut}_{t}^{-+-+}  &= 
		\frac{1}{t^2(\ell_t{-}p_1)^2(\ell_t{+}p_2)^2}
		\left[\begin{array}{l}
			\phantom{+}
   			\big(
				\langle 1\ell_t 2]\langle 3\ell'_t 4]
				- \ell_\perp^2\ab{13}\sb{24}
			\big)^2
			+ 2 t \ab{13}^2\sb{24}^2 \ell_\perp^2 
			\\
			+ \big(
				\langle 1\ell'_t 2]\langle 3\ell_t 4]
				- \ell_\perp^2\ab{13}\sb{24}
			\big)^2
			+ (d_g-4)\langle 1\ell_t 4]^2\langle 3\ell_t 2]^2
		\end{array}\right].
\\
\end{aligned}
\label{eq:4pt cuts2}
\ee
Note also that dimensional regularization is expensive. In contrast, the 4-dimensional cuts ($\ell_\perp=0$ nd $d_g=4$)
are much simpler:
\be\begin{aligned}
{\rm Cut}_{s,d=4}^{--++} &= 
\frac{\ab{12}^2\sb{34}^2}{(\ell_s{-}p_4)^2(\ell_s{+}p_1)^2}\ ,
&\quad
{\rm Cut}_{t,d=4}^{--++}&= 
\frac{\langle 1\ell_t 3]^2\langle 2\ell'_t 4]^2+\langle 1\ell'_t 3]^2\langle 2\ell_t 4]^2}{t^2(\ell_t{-}p_1)^2(\ell_t{+}p_2)^2}\ ,
\\
{\rm Cut}_{s,d=4}^{-+-+} &= 
\frac{\langle 1\ell_s 4]^2\langle 3\ell'_s 2]^2+\langle 1\ell'_s 4]^2\langle 3\ell_s 2]^2}{s^2(\ell_s{-}p_4)^2(\ell_s{+}p_1)^2}\ ,
&\quad
{\rm Cut}_{t,d=4}^{-+-+} &= 
\frac{\langle 1\ell_t 2]^2\langle 3\ell'_t 4]^2+\langle 1\ell'_t 2]^2\langle 3\ell_t 4]^2}{t^2(\ell_t{-}p_1)^2(\ell_t{+}p_2)^2}\ .
\end{aligned}\ee

Since the momentum twistor parameterization is only unambiguous for quantities that are little group invariant, we multiply each cut amplitude by a factor that absorbs any little group weight. That is, we actually compute the generalized unitarity coefficients of the little group invariant integrals
\begin{align}
	\vphi^\bullet_{s,t} = \frac{\text{Cut}_{s,t}^\bullet}{L^\bullet} \frac{\d^3\ell\vert_{s,t}}{\ell_\perp^2\vert_{s,t}}
	\label{eq:4pt phi cut}
\end{align}
where
\begin{align} \label{eq: 4pt little group factors}
	L^{++++}= \frac{[12][23][34][41]}{st}
	\qquad&\qquad 
	L^{-+++}= \frac{[34]^2[23]^2\la31\ra^2}{st(s+t)},
	\nn\\
	L^{--++}= \frac{\la12\ra^4}{\la12\ra\la23\ra\la34\ra\la41\ra},
	\qquad&\qquad
	L^{-+-+}= \frac{\la13\ra^4}{\la12\ra\la23\ra\la34\ra\la41\ra}.
\end{align}
When it exists, we factor our the tree level amplitude as done in the second line above. 
Also note that all but $L^{-+++}$ are symmetric in $s$ and $t$. This allows us to predict all but the $(-+++)$ $t$-channel coefficients from symmetry.



For planar five-gluon scattering, up to dihedral and parity symmetries, there are four inequivalent helicity configurations:
\be
(+++++),\quad
(-++++),\quad
(--+++),\quad
(-+-++).
\ee
Each has five distinct bubble cuts.
The corresponding integrands are obtained by multiplying the massive 4- and 5-point trees from section
\ref{ssec:trees}. While the trees are relatively concise, and the polarization sums would be straightforward in $d=4$ dimensions,
the result in $d$-dimensions (including a vector boson plus $d_g-5$ scalars as above) is rather too large to be quoted here.
After factoring out all little group scaling, the generalized unitarity coefficients were extracted using the algorithm of section \ref{sec:higher form c-map} and \ref{sec:cbub4}. 


\section{The finite terms of the $(-+-++)$ helicity amplitude \label{app:-+-++}}

In this appendix, we list the finite terms of the $(-+-++)$ helicity amplitude that were too long to fit into the main body of this work. The full $\vep$-dependence of these expressions can be found in the ancillary {\tt{Mathematica}} notebook.

The finite weight two term has a compact expression in terms of the finite box integrals
\begin{align}
	H_2^{-+-++} &= w_2^{--+++}
	+ \frac{\mathscr{I}[\vphi_{1235}]}{\vep^2}
	\bigg( 
		\frac{8 \la12\ra \la15\ra \la23\ra \la35\ra [25]^2}{\la13\ra^2 (s_{12}-s_{34}+s_{51})^2}
		- \frac{4 \la12\ra^2 \la15\ra^2 \la23\ra^2 \la35\ra^2 [25]^2}{\la13\ra^4 \la25\ra^2 (s_{12}-s_{34}+s_{51})^2}
	\bigg) 
	\nn\\&
	+ \frac{\mathscr{I}[\vphi_{2345}]}{\vep^2}
	\bigg(
		\frac{8 \la12\ra \la14\ra \la23\ra \la34\ra [24]^2}{ \la13\ra^2 (s_{23}+s_{34}-s_{51})^2}
		-\frac{4 \la12\ra^2 \la14\ra^2 \la23\ra^2 \la34\ra^2 [24]^2}{\la13\ra^4 \la24\ra^2 (s_{23}+s_{34}-s_{51})^2}
	\bigg). 
\end{align}
The finite wight one term is 
\begin{align}
	H_1^{-+-++} &= 
	\frac{11}{6} \log \left(\frac{-s_{34}}{\mu^2}\right)
	+ \frac{11}{6} \log \left(\frac{-s_{51}}{\mu^2}\right)
	\nn\\&
	+ \log\left(\frac{s_{23}}{s_{51}}\right) 
	\bigg(
		\frac{2 \la12\ra \la14\ra^3 \la23\ra^3 \la34\ra [24]^3}{3 \la13\ra^4 \la24\ra (s_{23}-s_{51})^3}
		-\frac{\la12\ra^2 \la14\ra^2 \la23\ra^2 \la34\ra^2 [24]^2}{\la13\ra^4 \la24\ra^2 (s_{23}-s_{51})^2}
		\nn\\&\quad
		+\frac{4 \la12\ra \la14\ra \la23\ra \la34\ra [24]^2}{\la13\ra^2 (s_{23}-s_{51}) (s_{23}+s_{34}-s_{51})}
		-\frac{2 \la12\ra^2 \la14\ra^2 \la23\ra^2 \la34\ra^2 [24]^2}{\la13\ra^4 \la24\ra^2 (s_{23}-s_{51}) (s_{23}+s_{34}-s_{51})}
	\bigg) 
	\nn\\&
	+ \log\left(\frac{s_{12}}{s_{34}}\right) 
	\bigg(
		\frac{2 \la12\ra^3 \la15\ra \la23\ra \la35\ra^3 [25]^3}{3 \la13\ra^4 \la25\ra (s_{12}-s_{34})^3}
		-\frac{\la12\ra^2 \la15\ra^2 \la23\ra^2 \la35\ra^2 [25]^2}{\la13\ra^4 \la25\ra^2 (s_{12}-s_{34})^2}
		\nn\\&\quad
		+\frac{4 \la12\ra \la15\ra \la23\ra \la35\ra [25]^2}{\la13\ra^2 (s_{12}-s_{34}) (s_{12}-s_{34}+s_{51})}
		-\frac{2 \la12\ra^2 \la15\ra^2 \la23\ra^2 \la35\ra^2 [25]^2}{\la13\ra^4 \la25\ra^2 (s_{12}-s_{34}) (s_{12}-s_{34}+s_{51})}	
	\bigg) 
	\nn\\&
	+ \log\left(\frac{s_{34}}{s_{51}}\right) 
	\bigg(
		\frac{2 \la12\ra^3 \la14\ra \la23\ra \la34\ra^3 [24]^3}{3 \la13\ra^4 \la24\ra (s_{34}-s_{51})^3}
		-\frac{\la12\ra^2 \la14\ra^2 \la23\ra^2 \la34\ra^2 [24]^2}{\la13\ra^4 \la24\ra^2 (s_{34}-s_{51})^2}
		\nn\\&\quad
		+\frac{4 \la12\ra \la14\ra \la23\ra \la34\ra [24]^2}{\la13\ra^2 (s_{34}-s_{51}) (s_{23}+s_{34}-s_{51})}
		-\frac{2 \la12\ra^2 \la14\ra^2 \la23\ra^2 \la34\ra^2 [24]^2}{\la13\ra^4 \la24\ra^2 (s_{34}-s_{51}) (s_{23}+s_{34}-s_{51})}
		\nn\\&\quad
		+\frac{2 \la12\ra \la15\ra^3 \la23\ra^3 \la35\ra [25]^3}{3 \la13\ra^4 \la25\ra (s_{34}-s_{51})^3}
		+\frac{\la12\ra^2 \la15\ra^2 \la23\ra^2 \la35\ra^2 [25]^2}{\la13\ra^4 \la25\ra^2 (s_{34}-s_{51})^2}
		\nn\\&\quad
		+\frac{4 \la12\ra \la15\ra \la23\ra \la35\ra [25]^2}{\la13\ra^2 (s_{34}-s_{51}) (s_{12}-s_{34}+s_{51})}
		+\frac{-11 \la12\ra \la34\ra [24]-11 \la15\ra \la23\ra [25]}{6 \la13\ra (s_{34}-s_{51})}
		\nn\\&\quad
		+\frac{-2 \la12\ra^2 \la15\ra \la23\ra \la34\ra^2 [25] [24]^2-2 \la12\ra \la15\ra^2 \la23\ra^2 \la34\ra [25]^2 [24]}{6 \la13\ra^3 (s_{34}-s_{51})^3}
		\nn\\&\quad
		-\frac{2 \la12\ra^2 \la15\ra^2 \la23\ra^2 \la35\ra^2 [25]^2}{\la13\ra^4 \la25\ra^2 (s_{34}-s_{51}) (s_{12}-s_{34}+s_{51})}
	\bigg) .
\end{align}
Lastly, the rational term is
\begin{align}
		&H_0^{-+-++} 
		= \frac{\delta_R}{12}  
			+ \frac{16}{9}
			+ \frac{\la12\ra^2 \la14\ra^2 \la23\ra \la34\ra [ 24 ]^3}{12 \la13\ra^4 \la24\ra [ 23 ] [ 34 ]}
				\frac{1}{s_{51}}
			+ \frac{\la12\ra \la15\ra \la23\ra^2 \la35\ra^2 [ 25 ]^3}{12 \la13\ra^4 \la25\ra [ 12 ] [ 15 ]}
				\frac{1}{s_{34}}
		\nn\\&\quad
			+ \frac{\la12\ra^3 \la15\ra \la23\ra \la35\ra^3 [ 25 ]^3}{12 \la13\ra^4 \la25\ra}
				\frac{(s_{12}+s_{34})}{s_{12} s_{34} (s_{12}-s_{34})^2}
			+\frac{\la 12 \ra \la 14 \ra^3 \la 23 \ra^3 \la 34 \ra [ 24 ]^3}{12 \la13\ra^4 \la24\ra}
				\frac{s_{23}+s_{51}}{s_{23} s_{51} (s_{23}-s_{51})^2}			
		\nn\\&\quad
			+ \frac{\la 12 \ra^3 \la 14 \ra \la 23 \ra \la 34 \ra^3[ 24 ]^3}{12 \la13\ra^4 \la24\ra}
				\frac{s_{34}+s_{51}}{s_{34} s_{51} (s_{34}-s_{51})^2}
			+ \frac{\la12\ra \la15\ra^2 \la15\ra \la23\ra^3 \la35\ra [ 25 ]^3}{12\la13\ra^4 \la25\ra}
				\frac{(s_{34}+s_{51})}{s_{34} s_{51} (s_{34}-s_{51})^2}
		\nn\\&\quad 
			+ \frac{\la12\ra \la15\ra \la23\ra \la34\ra [ 24 ]^2 [ 25 ]^2}{6 \la13\ra^4 [ 12 ] [ 15 ] [ 23 ] [ 34 ]}
				\frac{s_{34}s_{51}}{(s_{34}-s_{51})^2}
			+ \frac{\la12\ra \la15\ra \la23\ra \la34\ra [ 24 ] [ 25 ]}{24 \la13\ra^2}
					\frac{s_{34}}{s_{51} (s_{34}-s_{51})^2}
		\nn\\&\quad
			+ \frac{\la12\ra \la15\ra \la23\ra \la34\ra [ 24 ] [ 34 ] [ 25 ]}{24 \la13\ra^2 [ 34 ]}
				\frac{s_{51}}{s_{34} (s_{34}-s_{51})^2}
			- \frac{\la12\ra \la15\ra \la23\ra \la34\ra [ 24 ]^2 [ 25 ]^2}{12 \la13\ra^4 [ 12 ] [ 15 ] [ 23 ] [ 34 ]}
				\frac{s_{51}^2}{(s_{34}-s_{51})^2}
		\nn\\&\quad
			- \frac{\la12\ra \la15\ra \la23\ra \la34\ra [ 24 ] [ 25 ]}{12 \la13\ra^2}
				\frac{1}{(s_{34}-s_{51})^2 }
			- \frac{\la12\ra \la15\ra \la23\ra \la34\ra [ 24 ]^2 [ 25 ]^2}{12 \la13\ra^4 [ 12 ] [ 15 ] [ 23 ] [ 34 ]}
				\frac{s_{34}^2}{(s_{34}-s_{51})^2}
\nn\\&\quad
			- \frac{\la12\ra \la15\ra \la23\ra \la34\ra [ 24 ] [ 25 ]}{24 \la13\ra^3}
				\frac{\la 12 \ra \la 34 \ra [ 24 ] + \la 15 \ra \la 23 \ra [ 25 ]}{s_{34} (s_{34}-s_{51})^2}
		\nn\\&\quad
			- \frac{\la12\ra \la15\ra \la23\ra \la34\ra [ 24 ] [ 25 ]}{24 \la13\ra^3}
				\frac{\la 12 \ra \la 34 \ra [ 24 ] + \la 15 \ra \la 23 \ra [ 25 ]}{s_{51} (s_{34}-s_{51})^2}
		\nn\\&\quad
			+ \frac{\la 12 \ra^2 \la 14 \ra^2 \la 23 \ra^2 \la 34 \ra^2 [ 24 ]^2}{4\la 13 \ra^4 \la 24 \ra^2}
				\frac{s_{23}+s_{34}-2 s_{51}}{s_{51} (s_{23}-s_{51}) (s_{51}-s_{34})}
		\nn\\&\quad
			+ \frac{\la12\ra^2 \la15\ra^2 \la23\ra^2 \la35\ra^2 [ 25 ]^2 }{4\la13\ra^4 \la25\ra^2}
				\frac{s_{12}-2 s_{34}+s_{51}}{s_{34} (s_{12}-s_{34})(s_{34}-s_{51})}
			.
\end{align}

\bibliographystyle{JHEP}
\bibliography{physics}

\end{document}